\documentclass[1p,12pt,times,sort&compress]{elsarticle}
\usepackage[colorlinks]{hyperref}
\usepackage{amsmath}%to use \boldsymbol{}
\usepackage{multirow}
\journal{Mechanics of Materials}

\usepackage{float}%fixed figures
\biboptions{authoryear}
\begin{document}
\begin{frontmatter}
\title{PREPRINT:
Micromechanical and numerical analysis of shape and packing effects in elastic-plastic particulate composites}
\author[]{M. Majewski \corref{mycorrespondingauthor}}
%\orcidlink{0000-0001-9023-9194}
 \ead{mmajew@ippt.pan.pl}
\address{Institute of Fundamental Technological Research, Polish Academy of Sciences,\\
Pawi\'{n}skiego 5B, 02-106 Warsaw, Poland}
\cortext[mycorrespondingauthor]{Corresponding author, fax: +4822 8269815}
\begin{abstract}
The purpose of this study is to inspect the combined effect of reinforcement shape and packing on the macroscopic behaviour of particulate composites.
The introduced micromechanical approach modifies the Morphologically Representative Pattern scheme with the Replacement Mori-Tanaka Model.
The statistical volume elements have randomly placed inclusions with a selected shape.
Four shapes of inhomogeneities are studied: a sphere, a prolate spheroid, three prolate spheroids crossing at right angles, and a drilled spheroid.
The concentration tensors of non-ellipsoidal inhomogeneities are found numerically using simple simulations of a single particle.
The extension to the regime of non-linear material behaviour is performed by employing the tangent or secant incremental linearization of the material response.
The results are compared with the outcomes of numerical simulations and predictions of the classical mean-field models based on the Eshelby solution, e.g., the Mori-Tanaka model or the Self-Consistent scheme.
It is found that the proposed modification of the Morphologically Representative Pattern approach can be used as an alternative to computational homogenization in the case of elastic-plastic composites with different shapes and packings of particles.
\end{abstract}
\begin{keyword}
Micro-mechanics,
Composite materials,
Elastoplastic behavior,
Homogenization,
Shape of particles,
Packing of particles.
\end{keyword}
\end{frontmatter}
\section{Introduction\label{Sec:Intro}}
%slowo wstepu
Innovative processing techniques allow the manufacturing of materials with the designed microstructure, e.g., additive manufacturing of metallic components \citep{Debroy2018} or, as a specific example, materials design for electrodes of molten carbonate fuel cells \citep{Wejrzanowski2020}.
The design parameters can be, e.g., the spatial distribution or shape of microstructural components.
Therefore, the knowledge of the relationship between the effective behaviour of heterogeneous materials and morphological features of the microstructure is crucial for producing modern materials with desired overall properties.

%klasyczne modele i ich ograniczenia, jak modelowane jest upakowanie i ksztalt
The classical mean-field micromechanical models, e.g., the self-consistent (SC) scheme or standard Mori-Tanaka (MT) model, do not consider the influence of the distribution of components in the representative volume element on the macroscopic properties.
Furthermore, the classical mean-field micromechanical approaches usually account for only ellipsoidal shapes of heterogeneities since those models are based on Eshelby's solution (\citeyear{Eshelby1957}).
As a consequence, the quality of the classical mean-field models predictions decreases with a growing contrast in phase properties or an increasing volume fraction of phases \citep{Christensen1990}.

Some modifications of the existing formulations were proposed to improve the available mean-field approaches with regard to elastic composites. 
Among analytical models, one can distinguish the following extensions accounting for \textit{particle distribution}:
(i) the Morphologically Representative Pattern (MRP) approach \citep{Bornert1996}, e.g., in linear elasticity with the size effect \citep{Marcadon2007} or with shapes of inclusions \citep{Majewski2022}, in linear viscoelasticity \citep{Diani2013}, in the non-linear elastic-plastic response \citep{Majewski2020}, whose strategy for describing packing is followed in the present paper;
(ii) variational bounds with $n$-point correlation functions of elastic moduli \citep{Kanaun2008}, e.g., of cracked media \citep{PonteCastaneda1995}, or as an application to isotropic dispersions \citep{Torquato1998};
(iii)  $n$-site versions of mean-field models \citep{Chaboche2005} or far-field theories c.f. \citep{Sevostianov2019}, e.g. cluster models \citep{Kowalczyk2021}.

In the literature, e.g. \citep{Klusemann2012}, two analytical approaches for predicting the composite behaviour in the presence of \textit{non-elliptical heterogeneities} were distinguished.
The first one utilizes analytical procedures for estimating position-dependent Eshelby tensors,
like: the so-called effective self-consistent scheme \citep{Zheng2001} and its simplified version---the interaction direct derivative estimate \citep{Du2002}.
The second strategy is based on the Replacement Mori-Tanaka Method (RMTM) which is used and described in the further part of this paper.
Nogales and B{\"o}hm (\citeyear{Nogales2008}) used RMTM to predict the thermal conductivity of diamond-reinforced composites.

As an alternative to the discussed micromechanical models, we should mention analytical approaches for selected small-scaled elements like nanobeams, tubular structures, and circular cylindrical nanoshells.
Some recent works in this field can be mentioned as examples.
The authors  of \citep{Xuejie2021} have established a thorough model for appraisal of size-dependent thermoelastic vibrations of Timoshenko nanobeams by capturing small-scale effect on both structural and thermal fields.
The article \citep{Ming2022} provides a size-dependent generalized thermoelasticity model and closed-form solution for thermoelastic damping in cylindrical nanoshells.

%numeryczne analizy jako alternatywa do mikromechanicznego podejscia
In several papers, there are also fully numerical analyses of composites using: the Finite Element Method (FEM), e.g., studies of particles clustering in the plastic deformation \citep{NafarDastgerdi2018}, or particles spatial distribution on the tensile deformation \citep{Segurado2006}, or the fast Fourier transform method \citep{Escoda2016}.
The influence of particles shape on strength of metal matrix composites was studied in this fashion in \citep{Qing2015}.
The numerical approach is much more time consuming, especially for a non-linear response of the heterogeneous material, than the analytical methods.

%opis modelu MRP jak ewoluował 
The focus of the present study is on the morphologically representative pattern approach.
The MRP approach is based on the idea of describing microstructure with patterns.
The basis for the division of the microstructure is its distinct features, e.g., varied particle shapes or different packings of inclusions.
Initially, the MRP model \citep{Marcadon2007} was based on the formulation of the self-consistent scheme and its modification to the 3-phase Generalized Self-Consistent (3-GSC)  scheme \citep{Christensen1990}, hence it was possible to model the packing of spherical inclusions.
With the development of the 3-GSC into the $n$-phase configuration n-GSC by Herve and Zaoui \citep{Herve1993}, the MRP approach gained the capacity to model spherical inclusions with coatings.
By using coatings around the spherical particles, the effect of the size of inclusions on the effective properties of the composite can be estimated \citep{Majewski2017}.
Classic micromechanical models such as SC, n-GSC, and thus MRP allow estimating the properties of the composite governed by linear relationships, such as elasticity or heat conduction.
Generalization of the MRP model to nonlinear problems was performed in \citep{Majewski2020} using linearization proposed by Hill (\citeyear{Hill1965}).
Extension of the MRP model by taking into account inclusion shapes required a reformulation of the initial MRP model \citep{Majewski2022}:
the n-GSC pattern was replaced with the Replace Mori-Tanaka Method \citep{Klusemann2012}, and the SC pattern was changed to the classical Mori-Tanaka model.
The analysis of the influence of selected inclusion shapes on the effective elastic properties of a Metal Matrix Composite (MMC) was performed in \citep{Majewski2022}.
The present work goes one step further - it modifies the MRP model so that it can estimate the nonlinear elastic-plastic response of an MMC reinforced with ceramic particles, which have various shapes.
Compared to classical mean-field models, the proposed MRP model takes into account the geometry of inclusions (shape, orientation, packing) which improves micromechanical estimates in the case of moderate volume contents, especially for the non-linear response of the particulate composites.

The MMCs are used in many industries: antenna waveguides in Hubble space telescope, commercial satellites, automotive (e.g., cylinder liner and brake discs), aerospace (e.g., rotor blade sleeve in helicopters).
Because the influence of the geometry of the ceramic particle on the macroscopic response of the MMC is crucial \citep{Jarzabek2016}, we focus on the MMC reinforced with ceramic particle examples.
Nevertheless, the presented model can be used for other composites with a continuous reinforced/weakened dispersed phase matrix.

%kolejnosc publikacji co gdzie
The paper is organized as follows.
In the next section, a short summary is provided of the theoretical foundations of the developed MRP model:
(i) a description of the proposed micromechanical mean-field approach, (ii) presentation of the inclusions with complex shapes used in the model, (iii) extension of the MRP to the non-linear response.
In Section 3, numerical procedures are described for generating periodic Statistical Volume Elements (SVE) having random particle distributions with varying inclusion shapes and values of the packing ratio.
In Section 4, the results obtained using the mean-field and the FE methods are shown and discussed.
The last section summarizes the studies presented in this paper.

\section{Morphologically representative pattern-based approach\label{Sec:MRP}}
\subsection{Linear constitutive law}\label{SubSec:LinConLa}
The morphologically representative pattern-based approach proposes a subdivision of the composite microstructure, e.g., Representative Volume Element (RVE) of volume $V$, with some morphological properties into $M$ representative patterns $\alpha$.
In the previously mentioned papers \citep{Majewski2017, Majewski2022} the MRP framework for a linear constitutive law was described.
Here we shortly repeat the basic formulations.

The linear relationship between the auxiliary far-field strain $\mathbf{E}_0$ and the average strain $\boldsymbol{\varepsilon}_k^{\alpha}$ in the phase $k$ and the pattern $\alpha$ is described by the fourth-order concentration tensor $\mathbb{A}_k^{\alpha}$, that is
\begin{equation}\label{Eq:MRPCentral}
\boldsymbol{\varepsilon}_k^\alpha=\mathbb{A}_k^{\alpha}\cdot\mathbf{E}_0\,.
\end{equation}
The overall average strain of the composite $\mathbf{E}=\langle \boldsymbol{\varepsilon}\rangle_V$, where $\langle \cdot \rangle_V$ is defined as the volume averaging operation $1/V \int_V (\cdot)dV$, is identified that satisfies
\begin{equation}\label{Eq:MRPCentral2}
\boldsymbol{\varepsilon}_k^\alpha=\mathbb{A}_k^{\alpha}\left(\sum_{\beta=1}^Mc^{\beta}\sum_{j=1}^Nf_j^{\beta}\mathbb{A}_j^{\beta}\right)^{-1}\cdot\mathbf{E}=\bar{\mathbb{A}}_k^{\alpha}\cdot\mathbf{E}\,,\quad\textrm{where}\quad
\sum_{\alpha=1}^Mc^{\alpha}\sum_{k=1}^Nf_k^{\alpha}\bar{\mathbb{A}}_k^{\alpha}=\mathbb{I} \,.
\end{equation}
$f_k^{\alpha}$ is the volume fraction of the phase $k$ in the subvolume $V^{\alpha}$ occupied by the pattern $\alpha$ and $c^{\alpha}$ is the volume fraction of the pattern $\alpha$ in the representative volume element $V$.
Let us highlight that $f$ refers to the volume fraction of composite phases or their portions and $c$ to the volume fraction of patterns in the composite material's volume.

Using the formulation for the overall average stress $\boldsymbol{\Sigma}=\langle \boldsymbol{\sigma}\rangle_V$ and the constitutive law $\boldsymbol{\sigma}_k^\alpha=\mathbb{L}_k^{\alpha}\cdot\boldsymbol{\varepsilon}_k^\alpha$, the overall constitutive law is found as
\begin{equation}
\boldsymbol{\Sigma} 
=\left(\sum_{\alpha=1}^Mc^{\alpha}\sum_{k=1}^Nf_k^{\alpha}\mathbb{L}_k^{\alpha}\bar{\mathbb{A}}_k^{\alpha}\right)\cdot\mathbf{E}=\bar{\mathbb{L}}\cdot\mathbf{E} \,.
\end{equation}
The fourth-order tensor $\bar{\mathbb{L}}$ is the overall composite stiffness.

\begin{figure}[H]
\centering
\begin{tabular}{ccccc}\\
(a)&&(b) RMTM&&(c) MT*\\
RVE&&Effective composite inclusion&&remaining matrix\\
\includegraphics[angle=0,height=4cm]{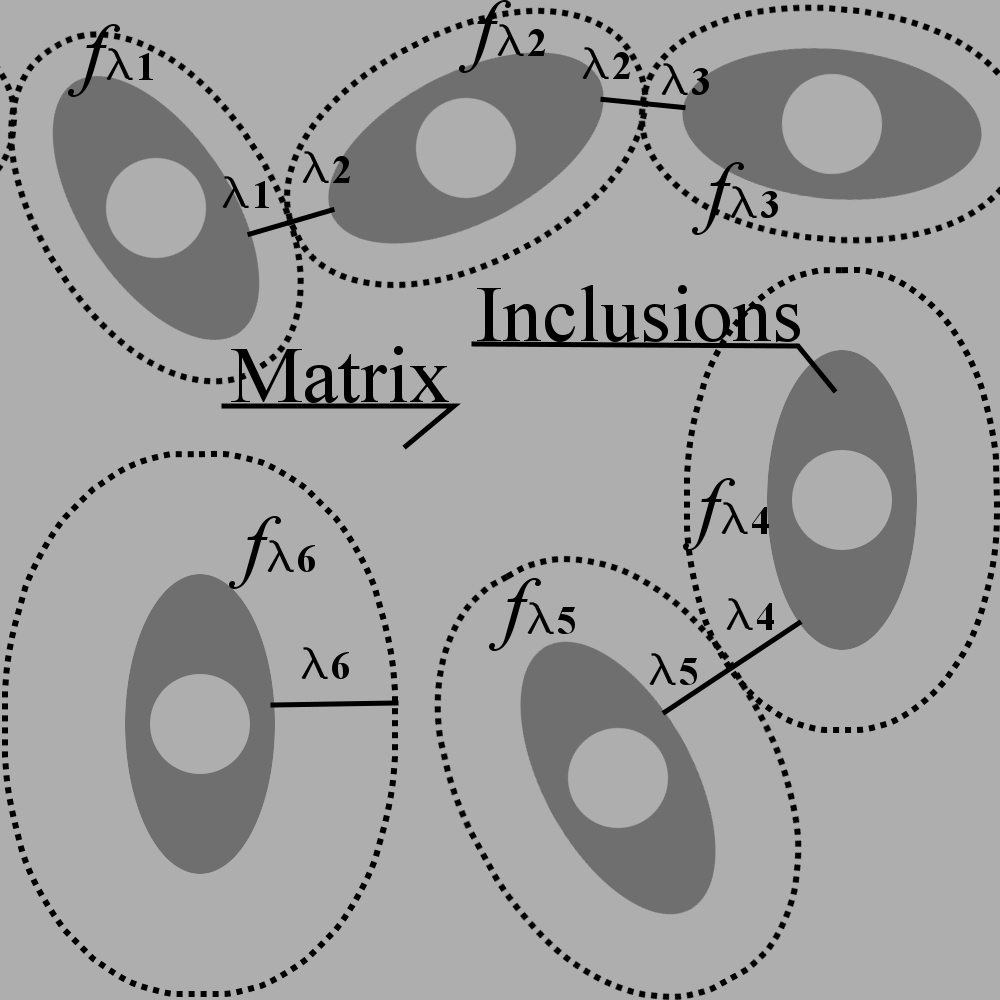}&
\includegraphics[angle=0,height=4cm]{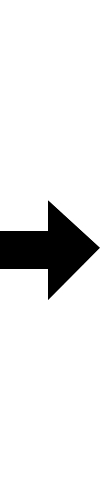}&
\includegraphics[angle=0,height=4cm]{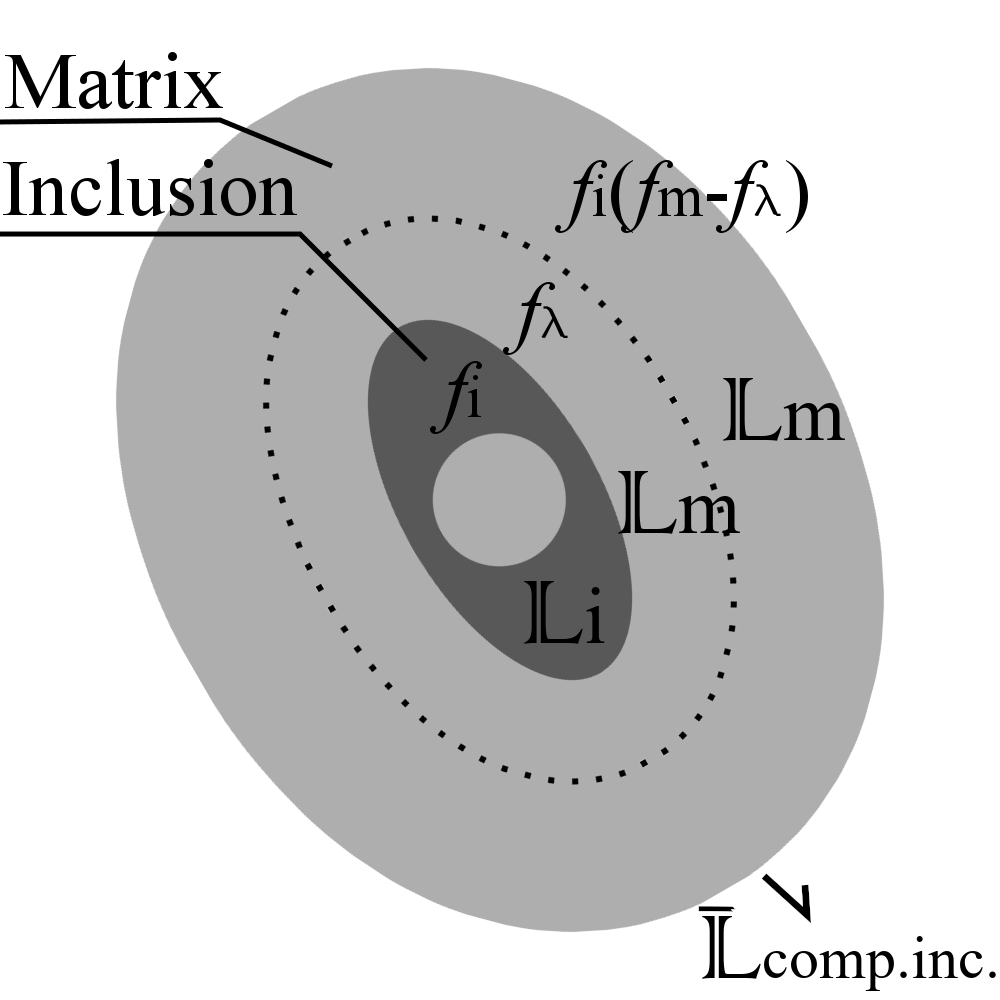}&
\includegraphics[angle=0,height=4cm]{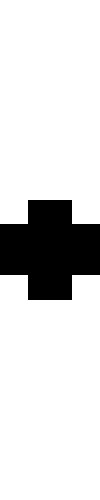}&
\includegraphics[angle=0,height=4cm]{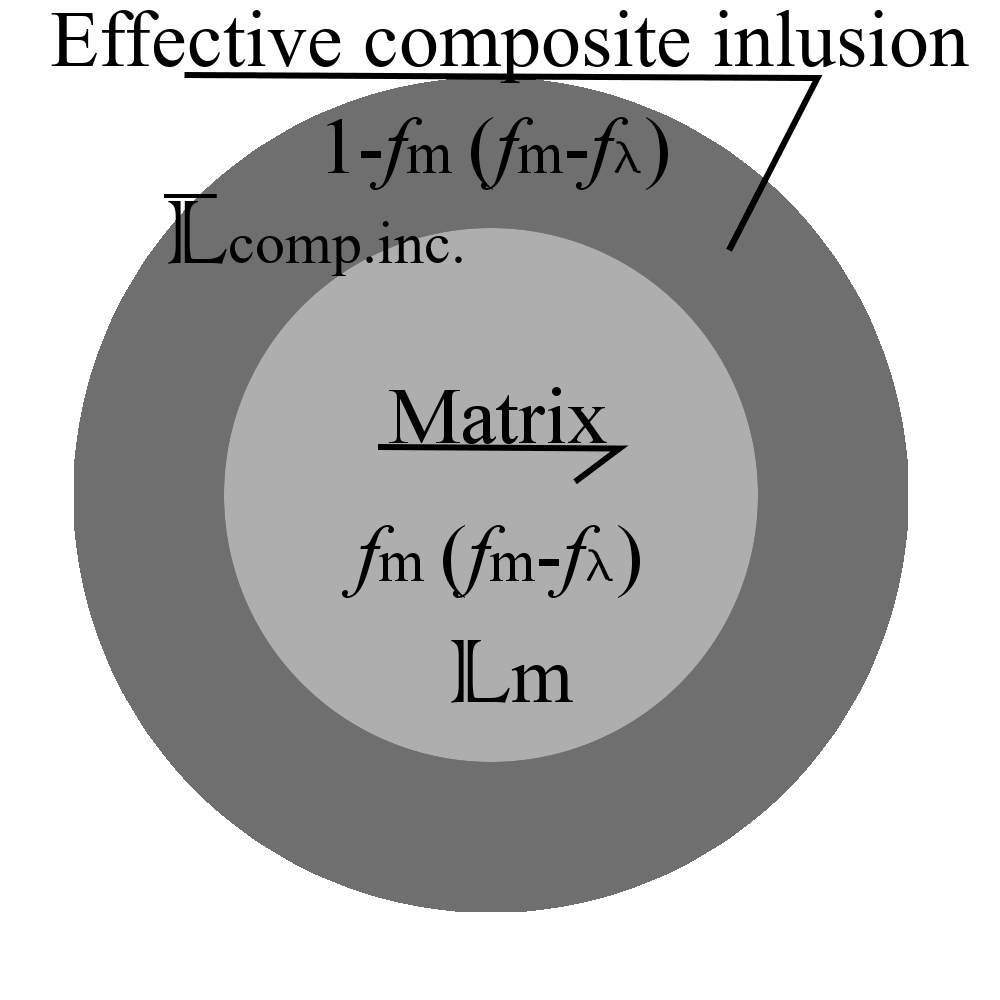}
\end{tabular}
\caption{
(a) The microstructure of a two-phase composite reinforced by identical drilled spheroids $k=1,\ldots,6$ represented in the Morphologically Representative Pattern-based (MRP) approach by two patterns: (b) "effective composite inclusion" modelled by the Replace Mori-Tanaka Method (RMTM), and (c) "remaining matrix" modelled by Mori-Tanaka (MT*).
$f_{\textup{i}}$ - the volume fraction of all inclusions,
$f_{\textup{m}}=1-f_{\textup{i}}$ - the volume fraction of the matrix,
$\lambda_k$ - half a distance between inclusion $k$ and its nearest neighbour,
$f_{\lambda,k}$ - the volume fraction of the matrix material included in $\lambda_k$,
$f_{\lambda}=\Sigma_{k=1}^{6} f_{\lambda,k}$ - the volume fraction of all inclusion coatings in the RVE.
$\mathbb{L}_{\textup{i}}$ and $\mathbb{L}_{\textup{m}}$ are the stiffness tensors of the inclusion and matrix phase, respectively.
$\bar{\mathbb{L}}_{\textup{comp.inc.}}$ is the overall stiffness tensor of the effective composite inclusion.
\label{Fig:Morph}
}
\end{figure}

\begin{figure}[H]
\centering
\begin{tabular}{c}
(a)\qquad\qquad\qquad\quad(b)\quad\qquad\qquad\qquad(c)\qquad\qquad\qquad\qquad\qquad(d)\qquad\qquad\qquad\qquad(e)\qquad\\
\includegraphics[angle=0,height=3cm]{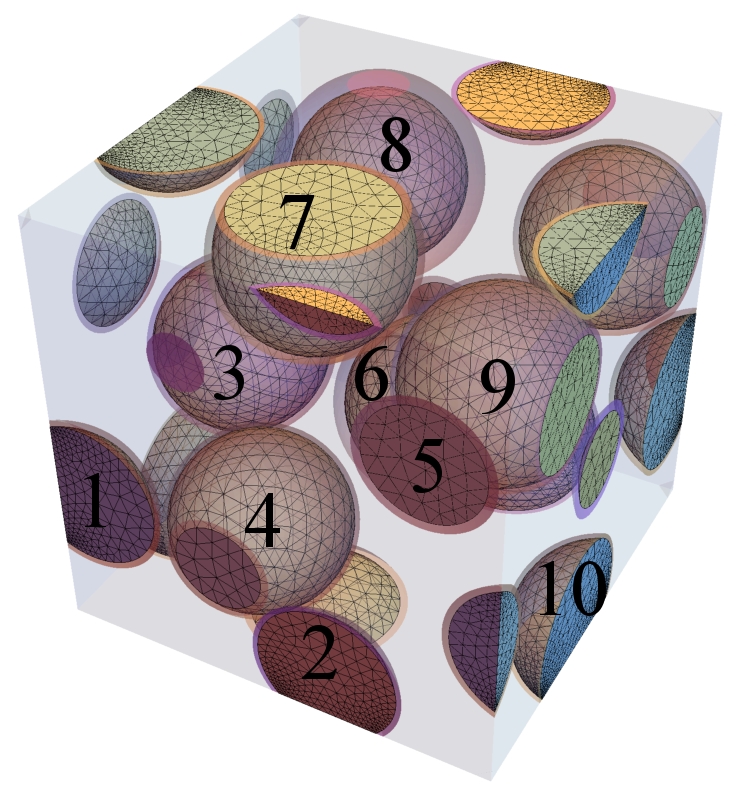}
\raisebox{.3\height}{\includegraphics[angle=0,height=2cm]{strz.jpg}}
\includegraphics[angle=0,height=3cm]{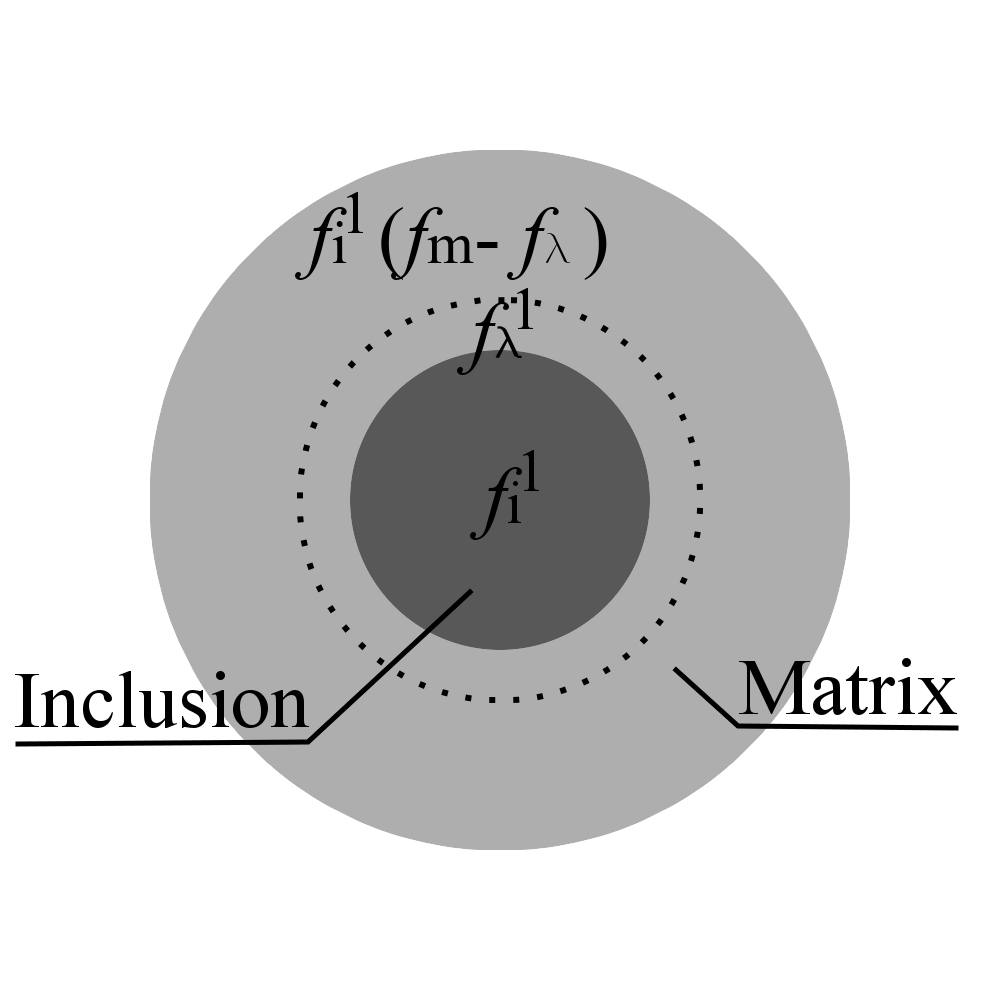}
\raisebox{.3\height}{\includegraphics[angle=0,height=2cm]{plus.jpg}}
\includegraphics[angle=0,height=3cm]{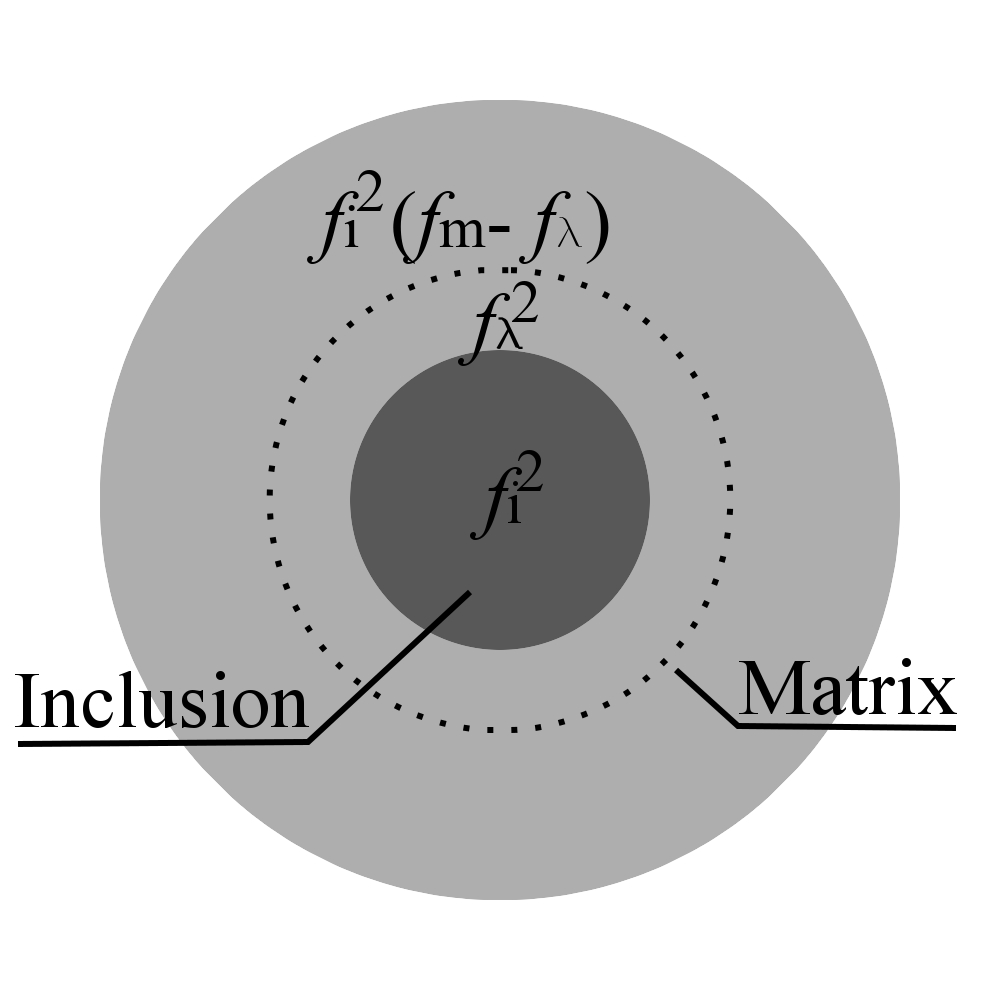}
\raisebox{.3\height}{\includegraphics[angle=0,height=2cm]{plus.jpg}}
\raisebox{.3\height}{\includegraphics[angle=0,height=2cm]{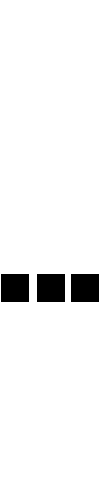}}
\raisebox{.3\height}{\includegraphics[angle=0,height=2cm]{plus.jpg}}
\includegraphics[angle=0,height=3cm]{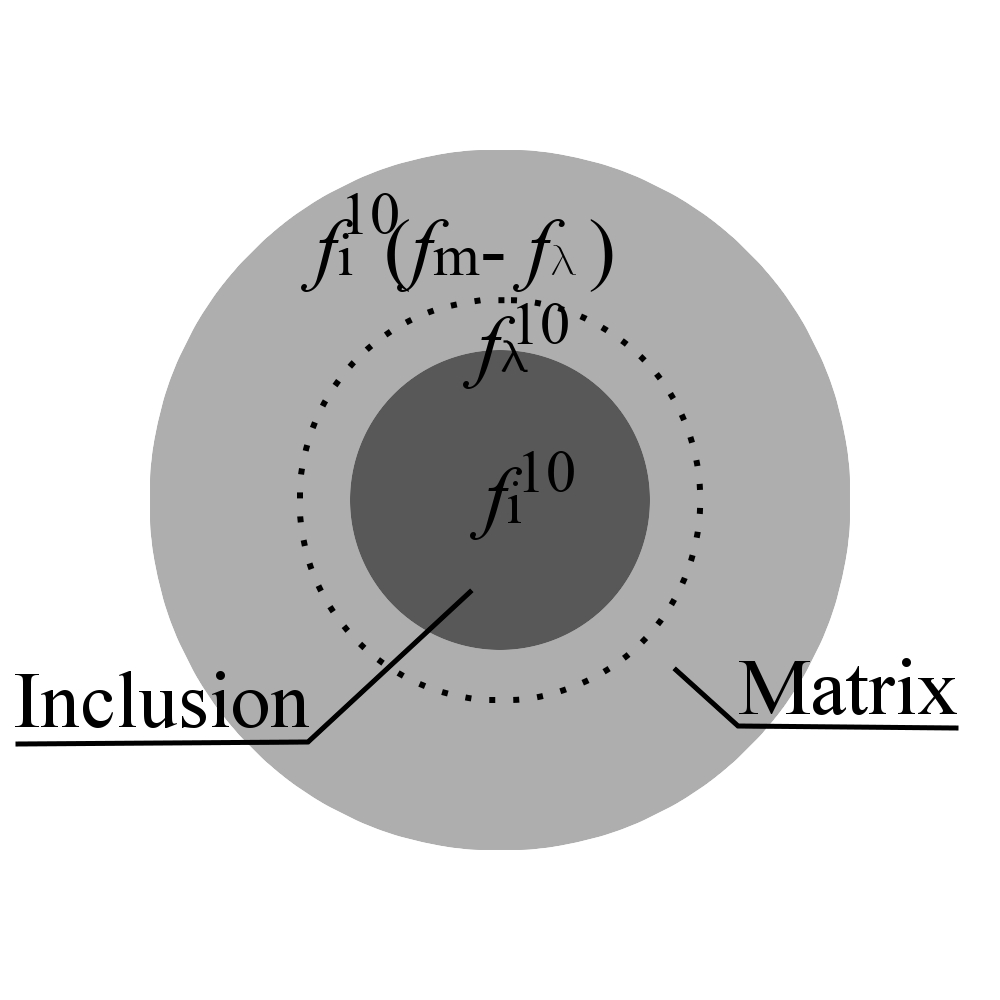}
\raisebox{.3\height}{\includegraphics[angle=0,height=2cm]{plus.jpg}}
\includegraphics[angle=0,height=3cm]{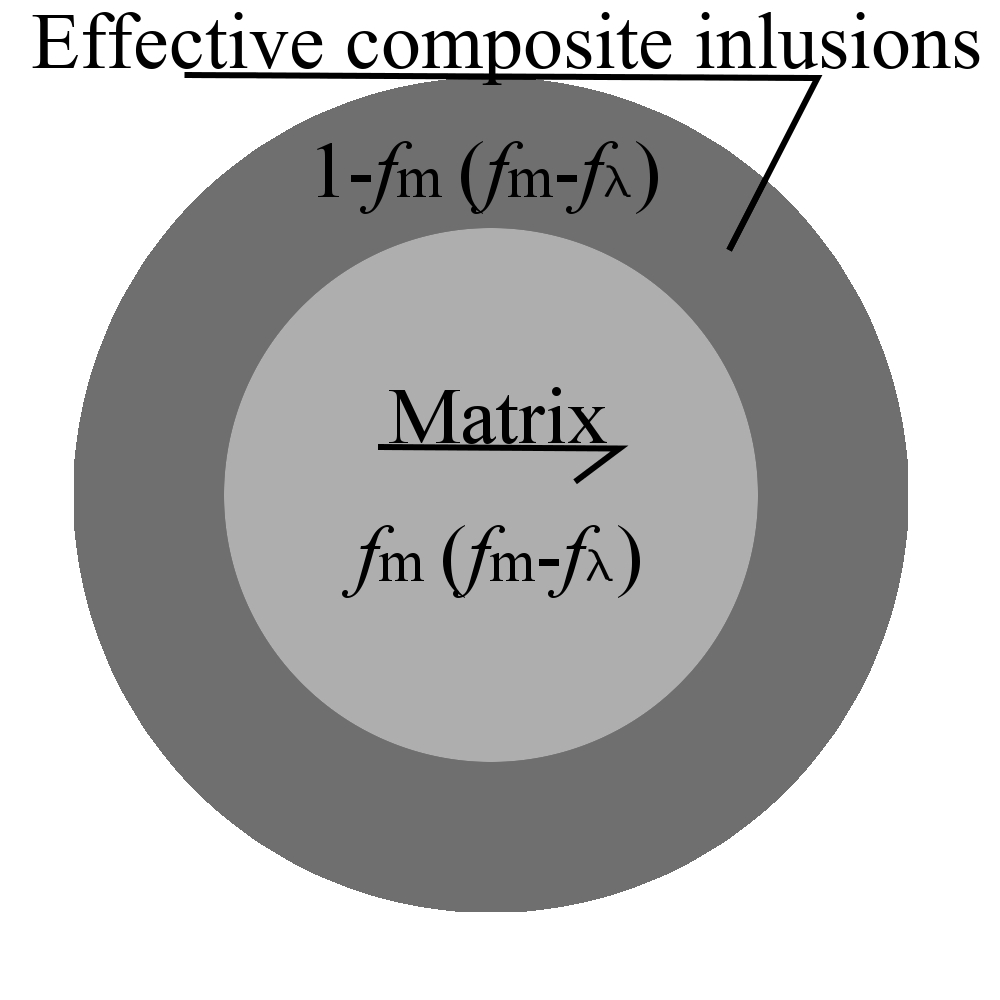}
\end{tabular}
\caption{
(a) Statistical Volume Element (SVE) of a Metal Matrix Composite (MMC) reinforced by ten periodically placed ceramic balls. (b) to (d) composite inclusion patterns, where each pattern represents one inclusion , and (e) the MT* pattern which represents the matrix outside the composites inclusion regions. 
\label{Fig:MRP_10_spheres}}
\end{figure}

In this paper, we explore the replace Mori-Tanaka method configuration of the MRP model \citep{Majewski2022}.
The model implementation in a step-by-step manner is described in \ref{Ap:Model}.
In the present section, two configurations of the MRP approach are presented: with two patterns $M=2$ (see Fig.\ref{Fig:Morph}), and with multiple patterns (see Fig.\ref{Fig:MRP_10_spheres}, the number of patterns $M$ equals the number of inhomogeneities $N_{\textup{i}}$ plus one to model each inhomogeneity $M=N_{\textup{i}}+1$).

The MRP two-pattern approach (Fig.\ref{Fig:Morph}) assumes that all inhomogeneities have similar characteristics.
The pattern called "effective composite inclusion" (Fig.\ref{Fig:Morph}.b) represents the effective properties of all inclusions, i.e., their geometry (packing, shape, orientation) and mechanical properties.
For such a basic two-pattern MRP approach, as in Fig.\ref{Fig:Morph}, the effective stiffness $\overline{\mathbb{L}}$ of the composite is specified as
\begin{equation}\label{Eq:EfStiff2}
\overline{\mathbb{L}}=
f_{\textup{i}} \mathbb{L}_{\textup{i}}^{\textup{}} \overline{\mathbb{A}}_{\textup{i}}^{\textup{RMTM}}+
f_{\lambda} \mathbb{L}_{\textup{m}}^{\textup{}} \overline{\mathbb{A}}_{m}^{\textup{RMTM,1st coating}}+
f_{\textup{i}} \left( f_{\textup{m}}-f_{\lambda} \right) \mathbb{L}_{\textup{m}}^{\textup{}} \overline{\mathbb{A}}_{m}^{\textup{RMTM,2nd coating}}+
f_{\textup{m}} \left( f_{\textup{m}}-f_{\lambda} \right) \mathbb{L}_{\textup{m}}^{\textup{}} \overline{\mathbb{A}}_{\textup{m}}^{\textup{MT*}}
\,.
\end{equation}
and the volume fractions of patterns: the RMTM-type and the MT*-type, respectively, are calculated as
\begin{equation}\label{Eq:c}
c^{\textup{RMTM}}=f_{\textup{i}}+f_{\lambda}+f_{\textup{i}} \left(f_{\textup{m}}-f_{\lambda}\right)\,,\quad c^{\textup{MT}}=1-c^{\textup{RMTM}}\,,
\end{equation}  
where $f_{\textup{i}}$ is the volume fraction of inclusions, $f_{\textup{m}}=1-f_{\textup{i}}$ the volume fraction of the matrix for a two-phase material, and $f_{\lambda}$ is the volume fraction of matrix coatings around inclusions. 

%In general, the MRP patterns may represent each inclusion individually ("composite inclusion" in Fig.\ref{Fig:MRP_10_spheres}.b-d).
We introduce the "composite inclusion" term for the configuration of an inclusion surrounded by coatings: inner $\lambda$, which describes the packing of particles, and outer, which models the influence of the inhomogeneity on the nearest region.
The classical mean-field models, e.g., MT or SC, reduce the homogenization problem to the so-called one-particle problem thus they miss the packing effect.
The presented MRP approach takes into account the packing of the inclusions, especially when each inclusion is treated separately (Fig.\ref{Fig:MRP_10_spheres}), although does not consider their spatial arrangements like the cluster model \citep{Kowalczyk2021}.

The "effective composite inclusion" term means the effective material parameters of a composite whose geometry is a sum of all composites inclusions (Fig.\ref{Fig:Morph}.b or Fig.\ref{Fig:MRP_10_spheres}.b-d).
The "effective composite inclusion" are used in the last Mori-Tanaka (MT*) pattern (Fig.\ref{Fig:Morph}.c or Fig.\ref{Fig:MRP_10_spheres}.e).
This describes the effective influence of the inclusions in the RVE on the rest of the matrix, thus modelling the interaction between particles.

The MRP considers \textbf{the particle shape} using the replace Mori-Tanaka method \citep{Klusemann2012} in the composite inclusion patterns (Fig.\ref{Fig:MRP_10_spheres}.b-d).
The diluted local concentration tensor in the RMTM is calculated numerically (FEM) using the robust procedure described in \citep{Majewski2022}.
Since the numerical concentration tensor is known for an assumed shape of a particle, the MRP model additionally accounts for \textbf{the orientation of the inclusions}.
The next section \ref{SubSec:Shape} describes the RMTM procedure.
To account for \textbf{the packing effect}, the thickness of the first coating $\lambda$ is specified by the minimum distance between nearest-neighbour particles ($2\lambda_k$ for particle $k$ in Fig.\ref{Fig:Morph}).
Thus in the RMTM pattern the volume fraction of the matrix coating $ f_{\lambda}^\alpha$ (Fig.\ref{Fig:MRP_10_spheres}.b-d) depends on the distances between inhomogeneities.
We will further use the \textit{matrix packing ratio} parameter $f_{\lambda}/f_{\textup{m}}$, which takes values from 0 to 1 and satisfies  $f_{\lambda}=\Sigma_{\alpha=1}^{M-1} f_{\lambda}^\alpha$.
The volume fraction of the second coating, $f_{\textup{i}}\left(f_{\textup{m}}-f_{\lambda}\right)$ in Fig.\ref{Fig:Morph} or $f_{\textup{i}}^\alpha \left(f_{\textup{m}}-f_{\lambda}\right)$ in Fig.\ref{Fig:MRP_10_spheres}.b-d, is the fitting parameter of the MRP model.
The authors conducted many tests of various composite materials, and the volume fraction of the second coating $f_{\textup{i}}^\alpha \left(f_{\textup{m}}-f_{\lambda}\right)$ gives the best agreement between the numerical (FEM) and  analytical (MRP) approaches (see supplementary material).
We interpreted the second coating as \textit{the influence of the inclusion $f_{\textup{i}}^\alpha$ on the surrounding matrix} beyond the $f_{\lambda}$ region.
The strain concentration tensor for the second coating is calculated as a modification of RMTM with the double inclusion framework of \citep{Hori1993}.
In \citep{Hu2000} the authors pointed out the limitation of the double inclusion framework. %and the double inclusion approach remains problematic for a multilayer particle
The last pattern, denoted by MT* (Fig.\ref{Fig:Morph}.c or Fig.\ref{Fig:MRP_10_spheres}.e) represents the remaining matrix material which is beyond composite inclusions.
Composite inclusions can form continuous domains of the volume fraction $1-f_{\textup{m}}(f_{\textup{m}}-f_{\lambda})$ and, in particular, surround the remaining matrix $f_{\textup{m}}(f_{\textup{m}}-f_{\lambda})$ that is not part of composite inclusions.
The remaining matrix material is assumed as a spherical inhomogeneity, i.e., the medium of effective composite inclusion properties surrounds the remaining matrix in an isotropic manner.
MT* is a Mori-Tanaka-type problem and relates to \textit{the interaction between particles}.
Let us underline that the current MRP approach divides the matrix phase into three subdomains, and the classical MRP model \citep{Marcadon2007} considers two matrix phase subdomains.

It is worth highlighting that without the second coating $f_{\textup{i}} \left( f_{\textup{m}}-f_{\lambda} \right)$ in the RMTM pattern for $f_{\lambda}=0$ the MRP equals \textit{the inverse MT}.
The inverse MT assumes the Mori-Tanaka model with the matrix embedded in the inhomogeneity phase, i.e., reverses the meaning of material phases: the matrix is inside while the inclusion outside for the MT representation.
As it was mentioned, for $f_{\lambda}=f_{\textup{m}}$ the MRP equals \textit{the MT model}.
Since the MT model and its inverse solution are within the Hashin-Shtrikman boundaries, the MRP model is also within the HS boundaries.
Note that the employment of the MT to multi-family composites (e.g., different shapes or phases) is sometimes questionable - the diagonal symmetry of the estimated stiffness tensor is violated \citep{Kanaun2008}.
However, this is not an issue in our studies of two-phase composite (MMC) with identical particles.
Moreover, we use the isotropisation procedure at each strain increment during the non-linear response of the composite.

\begin{figure}[H]
\centering
\begin{tabular}{cc}\\
(a)&(b)\\
\raisebox{.65\height}{\includegraphics[angle=0,height=3.5cm]{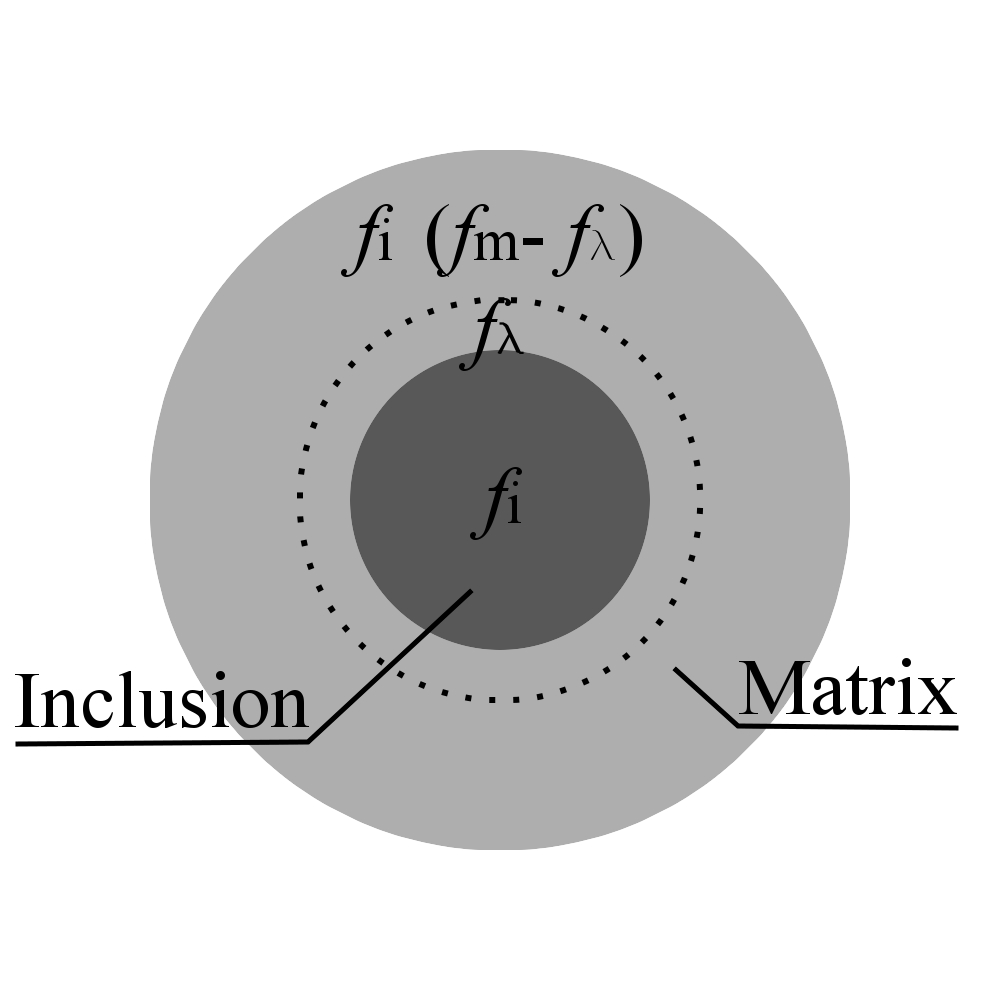}}
\raisebox{1.5\height}{\includegraphics[angle=0,height=2cm]{plus.jpg}}
\raisebox{.65\height}{\includegraphics[angle=0,height=3.5cm]{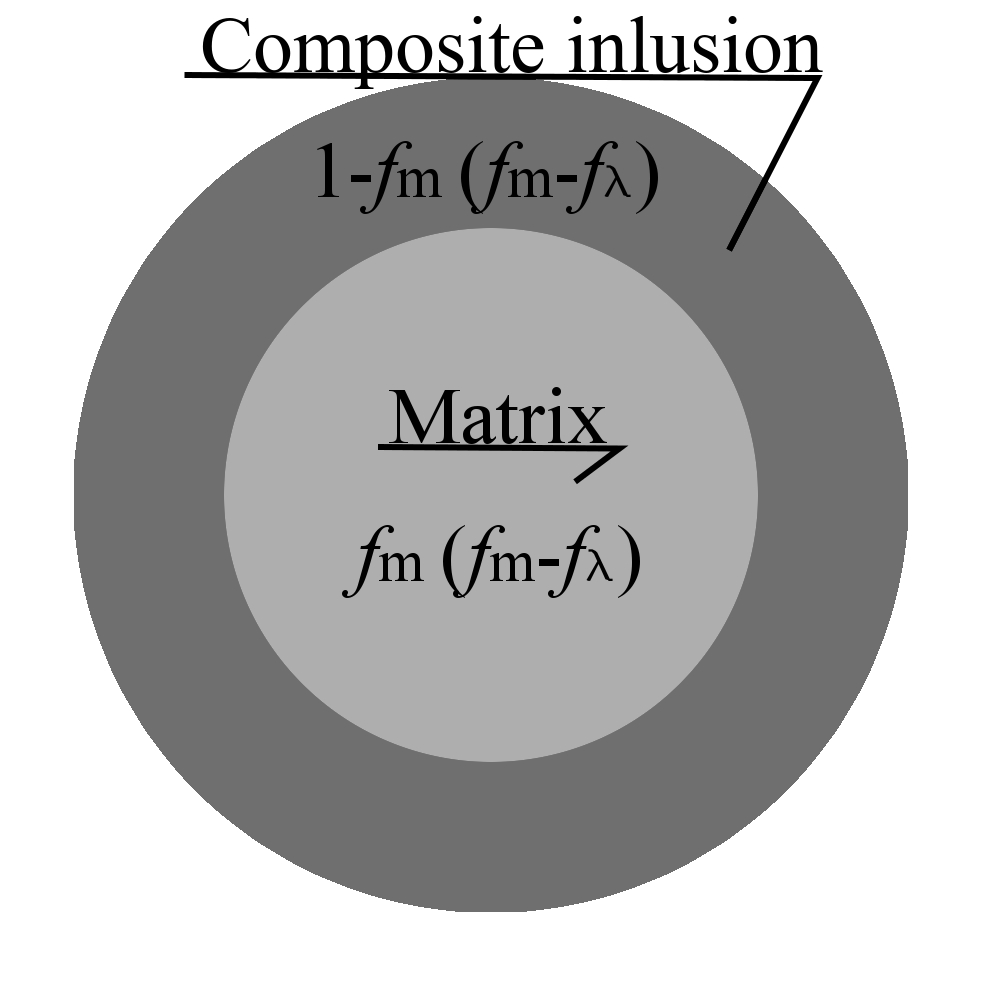}}
&
\includegraphics[angle=0,height=9cm]{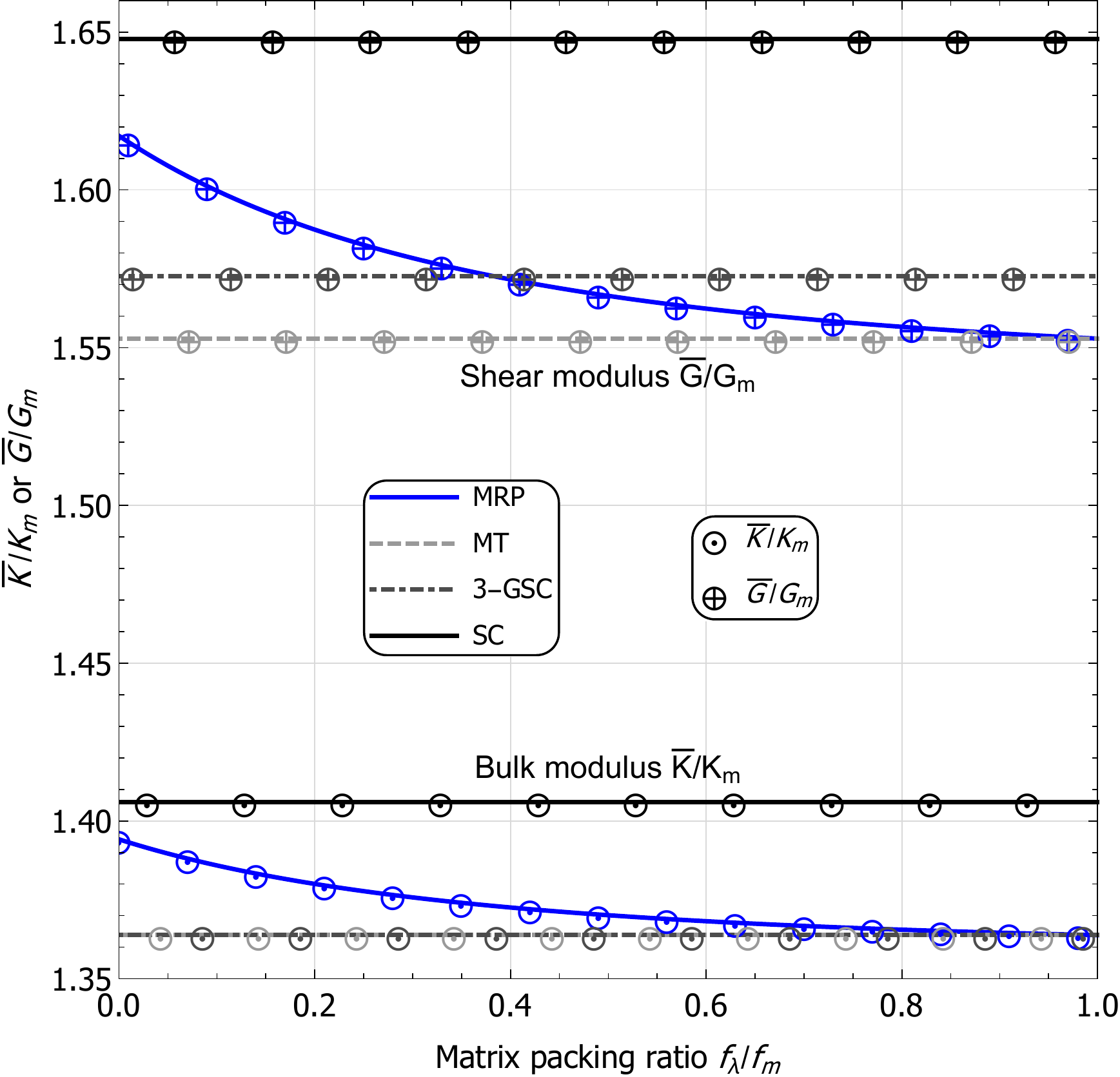}
\end{tabular}
\caption{
(a) Two-pattern MRP representation of the MMC (Tab.\ref{Tab:MMC}) reinforced by spherical inhomogenieties.
(b) The effective shear modulus $\overline{G}$ or bulk modulus $\overline{K}$ vs. the matrix packing ratio $f_{\lambda} / f_{\textup{m}}$.
The inclusions volume fraction is 30\%.
Notation:
$\left(K_{\textup{m}},G_{\textup{m}}\right)$ - bulk and shear modulus of the matrix phase,
MRP - the present variant of the MRP model based on two MT-type patterns,
MT - Mori-Tanaka method,
3-GSC - Generalized Self-Consistent scheme,
SC - Self-Consistent scheme.
\label{Fig:Keff_Geff_elastic_packing}}
\end{figure}

Fig.\ref{Fig:Keff_Geff_elastic_packing}.b illustrates the impact of the matrix packing ratio in the proposed MRP model on the effective properties of the Metal-Matrix Composite (MMC) (Tab.\ref{Tab:MMC}) reinforced with ceramic balls of 30\% volume fraction.
Fig.\ref{Fig:Keff_Geff_elastic_packing}.a illustrates the MRP approach as a two-pattern variant of spherical inhomogeneities. 
In Fig.\ref{Fig:Keff_Geff_elastic_packing}.b the effective shear modulus $\overline{G}$ and bulk modulus $\overline{K}$ are plotted as a function of the matrix packing ratio $f_{\lambda} / f_{\textup{m}}$.
The classical mean-field models results, e.g., Mori-Tanaka (MT), Self-Consistent (SC), 3-phase Generalized Self Consistent (3-GSC), do not exhibit packing effects (horizontal lines in Fig.\ref{Fig:Keff_Geff_elastic_packing}.b).
Two limit values of $f_{\lambda} / f_{\textup{m}}$: 1 and 0, respectively, are identified.
The matrix packing ratio $f_{\lambda}/f_{\textup{m}}$ is 0 when each particle touches its nearest neighbour ($\lambda_k=0 \Rightarrow f_{\lambda}=0$).
When $f_{\lambda}/f_{\textup{m}}=0$, the MRP effective material parameters are below the SC estimate for the assumed heterogeneous material in Tab.\ref{Tab:MMC}.
The matrix packing ratio $f_{\lambda}/f_{\textup{m}}$ is 1 when the entire matrix phase is used to form the coating of the inclusion in the RMTM pattern $f_{\lambda}=f_{\textup{m}}$.
Therefore when $f_{\lambda}/f_{\textup{m}}=1$, the volume fraction of the MT* pattern is zero, so the MRP solution reduces to the RMTM estimate, specifically to MT for spherical particles.

\begin{table}[H]
\centering
\begin{tabular}{lcc}
\hline
Phase & Ceramic & Metal
\\
\hline
\textup{Young's modulus,} $E$ [GPa] & 400 & 75
\\
\textup{Poisson's ratio,} $\nu$ & 0.2 & 0.3
\\
\textup{Initial yield stress,} $Y_0$ [MPa] & - & 75
\\
\textup{Plastic modulus,} $h$ [MPa] & - & 416
\\
\textup{Exponent}, $n$ & - & 0.3895
\\
\hline
\end{tabular}
\caption{Material parameters of the Metal Matrix Composite (MMC) reinforced by ceramic particles \citep{Suquet1997}. \label{Tab:MMC}}
\end{table}
\subsection{Replace Mori-Tanaka method}\label{SubSec:Shape}
Based on the paper \citep{Duschlbauer2004}, Nogales and B{\"o}hm (\citeyear{Nogales2008}) introduced a scheme for developing the classical Mori-Tanaka model, focusing mostly on thermal problems, to make it handle particles of non-ellipsoidal shapes.
This approach can be applied analogously in the case of elasticity \citep{Klusemann2012} and in the MRP model was described in \citep{Majewski2022}.
In the MRP approach, the last pattern is of the MT type (e.g. Fig.\ref{Fig:MRP_10_spheres}) and the others are modelled by the replace Mori-Tanaka method because of its applicability when the shape of inclusion is considered.
For clarity, the framework of RMTM is shortly repeated.

The RMTM is based on the MT method, whose main assumption can be written as
\begin{equation}\label{Eq:ADilcentral}
\boldsymbol{\varepsilon}_{\textup{i}}=
\mathbb{A}_{\textup{i}}^{\textup{Dil}} \boldsymbol{\varepsilon}_{\textup{m}}=
\mathbb{A}_{\textup{i}}^{\textup{MT}}\mathbf{E} 
\,,
\end{equation}
where $\mathbb{A}_{\textup{i}}^{\textup{Dil}}$ are the diluted strains concentration tensor, $\boldsymbol{\varepsilon}_{\textup{i}}$ and $\boldsymbol{\varepsilon}_{\textup{m}}$ is the averaged inhomogeneity and matrix strain, respectively, $\mathbb{A}_{\textup{i}}^{\textup{MT}}$ is the matrix Mori-Tanaka strain concentration tensor, and $\mathbf{E}$ is the macroscopic strain.
For an ellipsoidal particle, $\mathbb{A}_{\textup{i}}^{\textup{Dil}}$ is formulated as
\begin{equation}\label{Eq:ADilMT}
\mathbb{A}_{\textup{i}}^{\textup{Dil}}=
\left[\mathbb{I}+
\mathbb{P}_{\textup{m}}
\left(\mathbb{L}_{\textup{i}}-\mathbb{L}_{\textup{m}}\right)\right]^{-1}
\,,
\end{equation}
where $\mathbb{I}$ is the fourth rank identity tensors, $\mathbb{P}_{\textup{m}}$ is the polarisation tensor \citep{Eshelby1957,Hill1965} and $\mathbb{L}_{\textup{i}}$ and $\mathbb{L}_{\textup{m}}$ are the elasticity tensors of the inhomogeneity and matrix phase, respectively.
$\mathbb{A}_{\textup{i}}^{\textup{MT}}$ is the inhomogeneity Mori-Tanaka strain concentration tensor specified as
%eq.45 \citep{Klusemann12}
\begin{equation}\label{Eq:AiMT}
\mathbb{A}_{\textup{i}}^{\textup{MT}}=
\left[
f_{\textup{i}}
\mathbb{I}
+
f_{\textup{m}}
\left(
\mathbb{A}_{\textup{i}}^{\textup{Dil}}
\right)^{-1}
\right]^{-1} \,.
\end{equation}
The effective material stiffness $\overline{\mathbb{L}}_{\textup{MT}}$ can be obtained following the standard procedure of the mean-field approach \citep{Mori1973}.

The diluted strain concentration tensor of the inhomogeneity $\mathbb{A}_{\textup{i}}^{\textup{Dil}}$ (Eq.\ref{Eq:ADilMT}) is unknown in an analytical manner if the shape of inhomogeneities is non-ellipsoidal. 
Therefore, following \citep{Klusemann2012}, a phase-averaged diluted "replacement" is introduced.
Because the diluted strain concentration tensor of inhomogeneity $\mathbb{A}_{\textup{i}}^{\textup{Dil}}$ realizes the linear relation between the averaged inhomogeneity $\boldsymbol{\varepsilon}_{\textup{i}}$ and matrix $\boldsymbol{\varepsilon}_{\textup{m}}$ strains (Eq.\ref{Eq:ADilcentral}),
$\mathbb{A}_{\textup{i}}^{\textup{Dil}}$ can be obtained numerically, e.g. using finite element method, and the outcome of this numerical approach is denoted by $\mathbb{A}_{\textup{i}}^{\textup{NDil}}$.
Let us reformulate Eq.\ref{Eq:ADilcentral} and use 'RMTM' notation for clarity:
\begin{equation}\label{Eq:ANDilcentral}
\boldsymbol{\varepsilon}_{\textup{i}}=
\mathbb{A}_{\textup{i}}^{\textup{NDil}} \boldsymbol{\varepsilon}_{\textup{m}}=
\mathbb{A}_{\textup{i}}^{\textup{RMTM}}\mathbf{E} 
\,.
\end{equation}
The procedure using FEM is explained in the next section (Sec.\ref{Sec:Numer}).
Finally, the strain concentration tensor $\mathbb{A}_{\textup{i}}^{\textup{RMTM}}$, which is equivalent to $\mathbb{A}_{\textup{i}}^{\textup{MT}}$ in Eq.\ref{Eq:ADilcentral}, is given by
\begin{equation}\label{Eq:AiRMTM}
\mathbb{A}_{\textup{i}}^{\textup{RMTM}}=
\left[
f_{\textup{i}}
\mathbb{I}
+
f_{\textup{m}}
\left(
\mathbb{A}_{\textup{i}}^{\textup{NDil}}
\right)^{-1}
\right]^{-1} \,.
\end{equation}
In the MRP framework (Sec.\ref{Sec:MRP}, Eq.\ref{Eq:MRPCentral}--\ref{Eq:MRPCentral2}), the strain concentration tensor of RMTM $\mathbb{A}_{\textup{i}}^{\textup{RMTM}}$ (Eq.\ref{Eq:AiRMTM}) is used for the composite inclusion patterns $\alpha=1,..,M-1$, and $\mathbb{A}_{\textup{i}}^{\textup{MT}}$  (Eq.\ref{Eq:AiMT}) is taken in the last MT* pattern $\alpha=M$ for the remaining matrix.

\begin{figure}[H]
\centering
\begin{tabular}{ccccc}
&(a)&(b)&(c)&(d)\\
\includegraphics[angle=0,height=3.5cm]{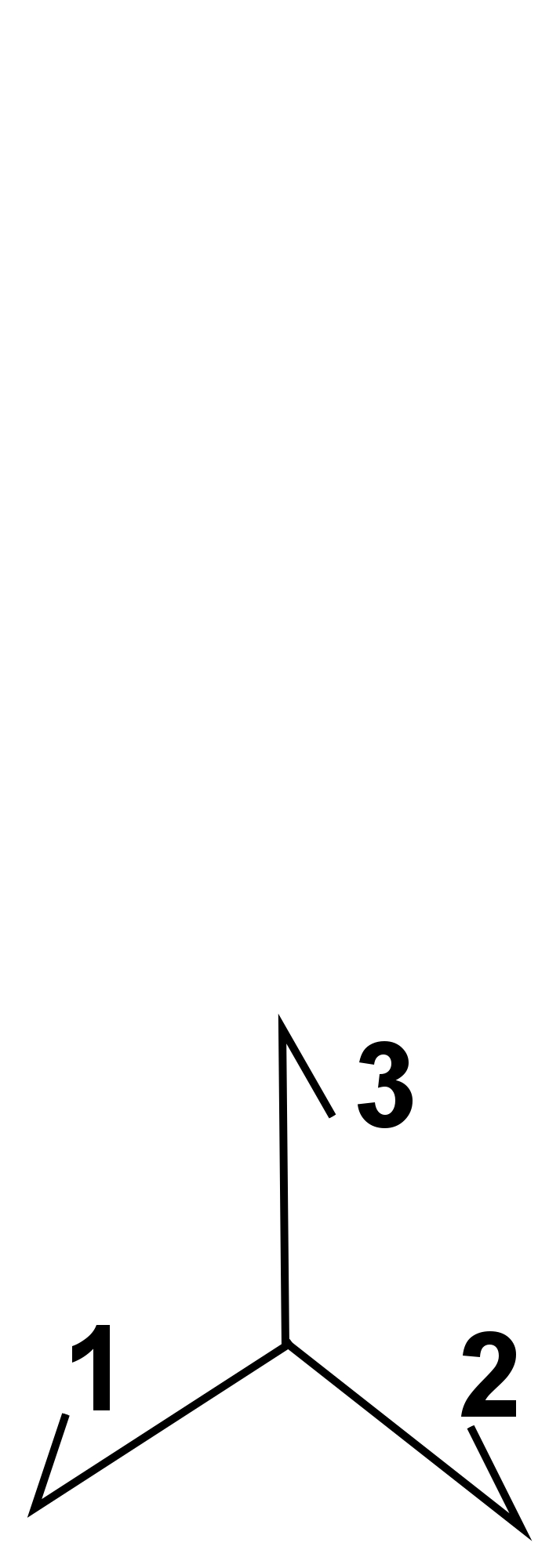}&
\includegraphics[angle=0,height=3.5cm]{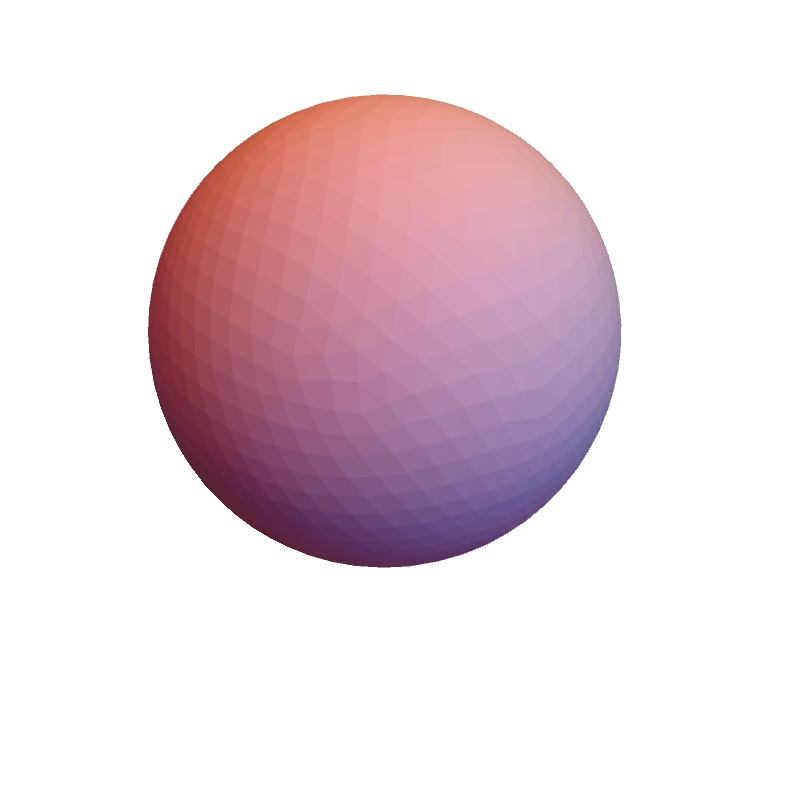}&
\includegraphics[angle=0,height=3.5cm]{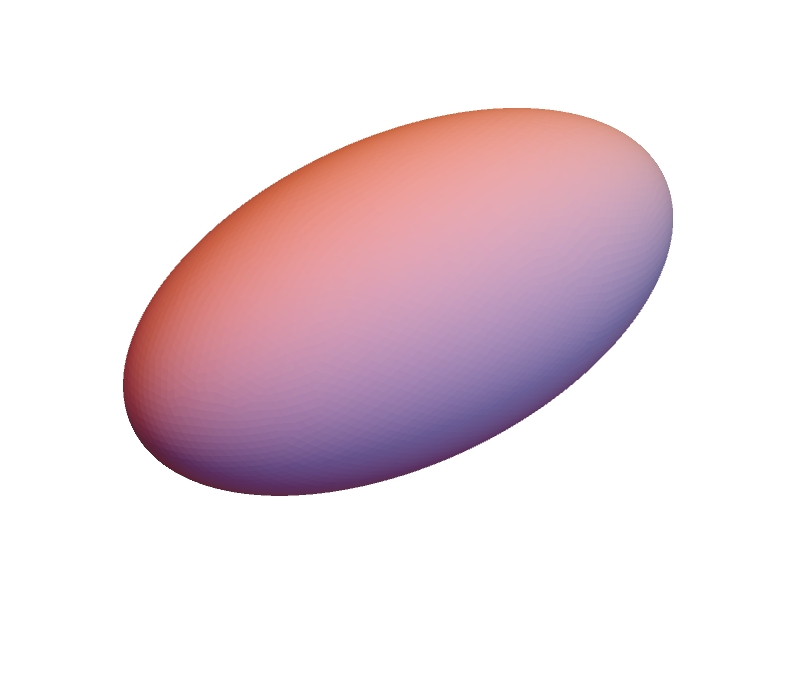}&
\includegraphics[angle=0,height=3.5cm]{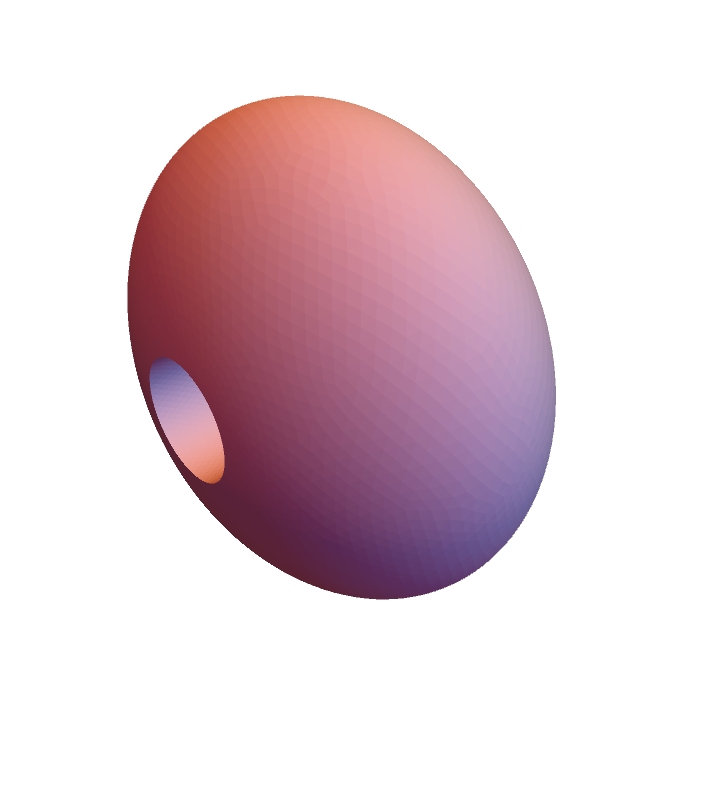}&
\includegraphics[angle=0,height=3.5cm]{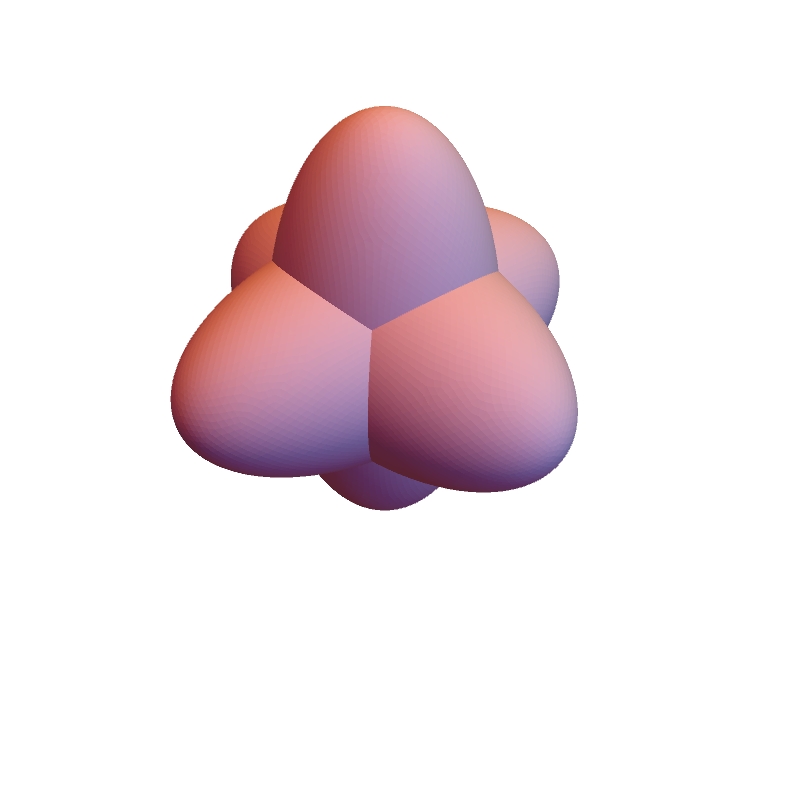}
\end{tabular}
\caption{
Selected particles shapes:
(a) ball (sphere), (b) prolate spheroid $(2a,a,a)$, (c) oblate spheroid $(2/3a,a,a)$ with a drilled cylinder $a/d_{\textup{Cyl}}=4$, and
(d) three spheroids: $(2a,a,a)$, $(a,2a,a)$, and $(a,a,2a)$ crossing at right angles.
\label{Fig:Shapes}}
\end{figure}

Four selected shapes of particles are studied in the rest of the paper, namely: 
\begin{itemize}
\item a sphere (Fig.\ref{Fig:Shapes}.a),
\item a prolate spheroid with semi-axes $(2a,a,a)$, (Fig.\ref{Fig:Shapes}.b),
\item an oblate spheroid with semi-axes $(2/3a,a,a)$, with a cylindrical cavity in the 'X' principal axis direction (drilled oblate spheroid); the cylinder's diameter is assumed to satisfy $a/d_{\textup{Cyl}}=4$ (Fig.\ref{Fig:Shapes}.c); the drilled cavity is filled by the matrix phase in the material composite (e.g. Fig.\ref{Fig:FEM_RVE_Shapes}.c),
\item three prolate spheroids with semi-axes $(2a,a,a)$ crossing at right angles (Fig.\ref{Fig:Shapes}.d).
\end{itemize}
The ellipsoidal shape (Fig.\ref{Fig:Shapes}.a and b) was analysed as a verification of the algorithm for $\mathbb{A}_{\textup{i}}^{\textup{NDil}}$ in \citep{Majewski2022}.
Let us emphasize, that if the inhomogeneity is drilled and filled with a matrix phase (e.g., Fig.\ref{Fig:Shapes}.c the drilled spheroid), the matrix in the inhomogeneity is part of the volume fraction of the matrix phase $f_{\textup{m}}$, not $f_{\textup{i}}$.
%The volume ratio of the matrix coating around the inclusion $f_{\lambda}$ to the matrix phase in the RVE $f_{\textup{m}}$ is the matrix packing ratio of the inhomogeneity $f_{\lambda} / f_{\textup{m}}$.
%The remaining matrix in the MT* pattern includes the matrix in the inhomogeneity.
%When inclusions touch each other, $\lambda=0 \Rightarrow f_{\lambda}=0$, the matrix packing ratio $f_{\lambda} / f_{\textup{m}}$ is 0.

The numerical strain concentration tensor $\mathbb{A}_{\textup{i}}^{\textup{NDil}}$ is driven solely by the symmetry of the inhomogeneity.
The numerical strain concentration tensor $\mathbb{A}_{\textup{i}}^{\textup{NDil}}$ of a spherical inhomogeneity is isotropic,
for the crossed spheroids (Fig.\ref{Fig:Shapes}.d) $\mathbb{A}_{\textup{i}}^{\textup{NDil}}$ has cubic symmetry, viz.
\begin{equation}\label{Eq:ACube}
\mathbb{A}^{\textup{Cub}}=
A^{\textup{Cub}}_0\mathbb{I}^{\rm{P}}+
A^{\textup{Cub}}_1(\mathbb{K}-\mathbb{I}^{\textup{P}})+
A^{\textup{Cub}}_2(\mathbb{I}-\mathbb{K})\,,
\end{equation}
and if the inhomogeneities are the prolate or drilled oblate spheroids (Fig.\ref{Fig:Shapes}.b and c), the symmetry group is that for transverse isotropy:
\begin{equation}\label{Eq:ATrans}
\mathbb{A}^{\textup{Trans}}=
A^{\textup{Trans}}_0 \mathbb{I}^{\textup{P}}+
A^{\textup{Trans}}_1 \mathbb{P}_{1}+
A^{\textup{Trans}}_2 \mathbb{P}_{2}+
A^{\textup{Trans}}_3 \mathbb{P}_{3}+
%A^{\textup{Trans}}_4 \mathbb{P}_{4}+
\frac{1}{\sqrt{6}}\left(A^{\textup{Trans}}_{12} \mathbf{d}\otimes\mathbf{I} +A^{\textup{Trans}}_{21}  \mathbf{I}\otimes\mathbf{d} \right)\,.
\end{equation}
The projectors $\mathbb{P}_K$ and the remaining tensorial quantities, e.g. $\mathbb{I}^{\textup{P}}$, are listed in  \ref{Ap:AppendixProjektory}.
The components $A_K$ depends on the material parameters of phases and in further work (Sec.\ref{SubSec:Reg}) are expressed as functions $A_K \left[ K_{\textup{i}},G_{\textup{i}},K_{\textup{m}},G_{\textup{m}}\right]$ (Eq.\ref{Eq:funANDilapprox}, \ref{Eq:hydrostaticdeviatoricparts}, \ref{Eq:ANDilIso}) of the bulk and shear moduli of the two-phase composite.
In general, when the inclusion has an irregular shape, its concentration tensor is usually anisotropic. The numerical simulations should be carried out in such a way as to identify all independent components of the concentration tensor \citep{Kowalczyk2009}.
\subsection{Extension to the elastic-plastic response}\label{SubSec:Lin}
The incremental linearization proposed by Hill (\citeyear{Hill1965}) is adopted to apply the MRP approach to elastic-plastic materials, like it was described in \citep{Majewski2020}.
For clarity, the framework of the linearization procedure is recapitulated below.
Two forms of the linearization scheme: secant \citep{Tandon1988} and tangent \citep{Hill1965} are considered depending on the definition of the current stiffness tensor.
The linearized constitutive relations in the phases for the secant (superscript s) and tangent (superscript t) procedure at each strain increment are:
\begin{equation}\label{Eq:LinStSiec}
\boldsymbol{\dot{\sigma}} =\mathbb{L}^{\textup{t}} \cdot \boldsymbol{\dot{\varepsilon}}
\quad , \qquad
{\boldsymbol{\sigma} =\mathbb{L}^{\textup{s}} \cdot \boldsymbol{\varepsilon}}\,,
\end{equation}
respectively, where the current secant ($\mathbb{L}^{\rm{s}}$) or tangent ($\mathbb{L}^{\rm{t}}$) elastic-plastic stiffness tensor is applied.

The effective elastic-plastic response of a Metal-Matrix Composite (MMC, Tab.\ref{Tab:MMC}) is analysed, as in \citep{Kursa2014}.
The MMC has an elastic-plastic metal matrix and elastic ceramic inclusions.
%Table \ref{Tab:MMC} stores the material parameters of the MMC.
The elastic-plastic metal matrix material is considered as a ductile material governed by linear elasticity and the Huber-von Mises yield function $f \left(  \boldsymbol{\sigma}_{\textup{m}}  \right)$ with the associated flow rule:
\begin{equation}\label{Eq:LinHMa}
\boldsymbol{\dot{\varepsilon}}_{\textup{m}}=\boldsymbol{\dot{\varepsilon}}^{\textup{e}}_{\textup{m}}+\boldsymbol{\dot{\varepsilon}}^{\textup{p}}_{\textup{m}}\,,\quad
f \left( \boldsymbol{\sigma}_{\textup{m}} \right) =
\sqrt{\frac{3}{2} \boldsymbol{\sigma^{'}}_{\textup{m}} \cdot \boldsymbol{\sigma^{'}}_{\textup{m}} }
- Y \left( \varepsilon^{\textup{p}}_{\textup{eq}} \right)
\leq 0
\, , \quad
\boldsymbol{\dot{\varepsilon}}^{\textup{p}}_{\textup{m}} = \lambda \frac{3 \boldsymbol{\sigma^{'}}_{\textup{m}}  }{2 Y}
\, ,
\end{equation}
where: $(.)^{e}$ and $(.)^{p}$ are elastic and plastic parts, respectively, the apostrophe $(.)^{'}$ denotes the deviatoric part of the tensor, the yield stress $Y \left(\varepsilon^{\textup{p}}_{\textup{eq}} \right)$ is a function of the equivalent plastic strain $\varepsilon^{\textup{p}}_{\textup{eq}}$, and $\lambda\geq0$ is a plastic multiplier.
The yield stress $Y \left(\varepsilon^{\textup{p}}_{\textup{eq}} \right)$ is assumed in the form of isotropic hardening with a power law:
\begin{equation}\label{Eq:LinY}
Y \left(\varepsilon^{\textup{p}}_{\textup{eq}} \right) =
Y_0 +
h \, \left(\varepsilon^{\textup{p}}_{\textup{eq}} \right)^n
\, , \quad
\dot{\varepsilon}^{\textup{p}}_{\textup{eq}}=\sqrt{\frac{2}{3} \boldsymbol{\dot{\varepsilon}}^{\textup{p}}_{\textup{m}} \cdot \boldsymbol{\dot{\varepsilon}}^{\textup{p}}_{\textup{m}}} = \lambda
\, .
\end{equation}
The yield stress has been adopted in the form of isotropic hardening with a power law due to the good compatibility of the experimental and numerical results, e.g., \citep{Kursa2018}.

Following \citep{Kursa2018}, the secant elastic-plastic stiffness tensor $\mathbb{L}^{\textup{s}}_{\textup{m}}$ in Eq.\ref{Eq:LinStSiec} can be expressed, for a proportional loading path, in the isotropic form:
\begin{equation}\label{Eq:LinSec}
\mathbb{L}^{\textup{s(iso)}}_{\textup{m}} = 3 K_{\textup{m}} \mathbb{I}^{\textup{P}}+
2G^{\textup{s}}_{\textup{m}} \left(\boldsymbol{\varepsilon}^{\textup{p}}_{\textup{eq}} \right) \mathbb{I}^{\textup{D}} \quad ,
\end{equation}
where
\begin{equation}\label{Eq:LinUpr2a}
2G^{\textup{s}}_{\textup{m}} \left(\boldsymbol{\varepsilon}^{\textup{p}}_{\textup{eq}} \right) = \frac{|| \boldsymbol{\sigma}^{'}_{\textup{m}} ||}{|| \boldsymbol{\varepsilon}^{'}_{\textup{m}} || } =
\frac{\sqrt{\boldsymbol{\sigma}^{'}_{\textup{m}}\cdot \boldsymbol{\sigma}^{'}_{\textup{m}}}}{\sqrt{\boldsymbol{\varepsilon}^{'}_{\textup{m}} \cdot \boldsymbol{\varepsilon}^{'}_{\textup{m}}}} \quad .
\end{equation}
In the case of tangent linearization of the constitutive law, the current elastic-plastic stiffness tensor of the matrix $\mathbb{L}^{\textup{t}}_{\textup{m}}$ in Eq.\ref{Eq:LinStSiec} is defined as:
\begin{equation}\label{Eq:LinStyOsno}
\mathbb{L}^{\textup{t}}_{\textup{m}} =
3K_{\textup{m}} \mathbb{I}^{\textup{P}} \, + \,
2G^{\textup{t}}_{\textup{m}} \left(\boldsymbol{\varepsilon}^{\textup{p}}_{\textup{eq}} \right) \,
\boldsymbol{N} \otimes \boldsymbol{N}  \, + \,
2G_{\textup{m}} \left(\mathbb{I}^{\textup{D}} -
\boldsymbol{N} \otimes \boldsymbol{N} \right) \, ,
\end{equation}
in which:
\begin{equation}\label{Eq:LinStyOsno2}
G^{\textup{t}} _{\textup{m}}  \left(\varepsilon^{\textup{p}}_{\textup{eq}} \right)
=
G_{\textup{m}}  \frac{nh \left(\varepsilon^{\textup{p}}_{\textup{eq}} \right)^{n-1}}{nh \left(\varepsilon^{\textup{p}}_{\textup{eq}} \right)^{n-1} +3G_{\textup{m}} }\,, \quad
\boldsymbol{N}=\frac{\boldsymbol{\sigma^{'} }_{\textup{m}}}{\sqrt{\boldsymbol{\sigma^{'} }_{\textup{m}} \cdot \boldsymbol{\sigma^{'}}_{\textup{m}}}}
\, ,
\end{equation}
and $\left(K_{\textup{m}},G_{\textup{m}}\right)$ are the elastic bulk and shear modulus, respectively, and $\boldsymbol{N}$ is a unit vector of the deviatoric part of the stress tensor $\boldsymbol{\sigma}^{'}_{\textup{m}}$ in stress space.

The current tangent stiffness tensor of the matrix phase $\mathbb{L}^{\textup{t}}_{\textup{m}}$ is anisotropic despite the matrix phase has been assumed as isotropic.
Therefore, the current tangent stiffness tensor of the matrix is isotropized following \citep{Chaboche2005,Kursa2018} for two reasons.
The first one is to use the formula for concentration tensors of the RMTM and MT model, which are available only for isotropic materials \citep{Mori1973,Klusemann2012}.
The second one is to avoid excessive stiffness of the elastic-plastic response of the composite \citep{Chaboche2005}.
Thus, the tangent incremental variant is subjected to the isotropization procedure,   namely:
\begin{equation}\label{Eq:LinTan}
\mathbb{L}^{\textup{t(iso)}}_{\textup{m}}  = 3 K_{\textup{m}}  \mathbb{I}^{\textup{P}}+
2G^{\textup{t}}_{\textup{m}}  \left(\varepsilon^{\textup{p}}_{\textup{eq}} \right) \mathbb{I}^{\textup{D}}\,.
\end{equation}

Kursa et al. (\citeyear{Kursa2018}) presented the framework of the algorithm with details provided in Box 1 of their paper.
The tangent $G^{\textup{t}} _{\textup{m}}$ (Eq.\ref{Eq:LinStyOsno2}) and secant shear modulus $G^{\textup{s}} _{\textup{m}}$ (Eq.\ref{Eq:LinUpr2a}) are employed in the algorithm as a function of the deviatoric parts of the stress and strain fields.
The non-linear response of the MMC is obtained using the mean-field MRP approach in an incremental iterative procedure.
The current stiffness tensor, defined by Eq.(\ref{Eq:LinSec}) and Eq.(\ref{Eq:LinTan}) for the secant and tangent procedure, respectively, is updated with the current accumulated plastic strain $\varepsilon^{\textup{p}}_{\textup{eq}}$.
In the secant approach, the current stress and strain tensors are sufficient to derive $\mathbb{L}^{\textup{s}}_{\textup{m}}$, while in the tangent variant, the algorithm uses the incremental stress and strain.
For the calculation of MRP in the elastic-plastic regime, a modification has been made to calculate each pattern's concentration factors for the composite phases $\bar{\mathbb{A}}_k^{\alpha}$ (\ref{Ap:Model}) at each incremental time step $ t +\, \bigtriangleup t $.
The local strain measures $\varepsilon_k^{\alpha}$ is calculated by $\bar{\mathbb{A}}_k^{\alpha}$ (Eq.\ref{Eq:MRPCentral2}) and relies on the volume averages.
The plastic propagation is taken into account in the model by considering different zones within the metal-matrix phase, but it could be improved (Fig.\ref{Fig:nGSC} comparison between the MRP and n-GSC schemes).

\begin{figure}[H]
\centering
\begin{tabular}{cc}\\
(a)&(c)\\
\raisebox{.2\height}{\includegraphics[angle=0,height=3.5cm]{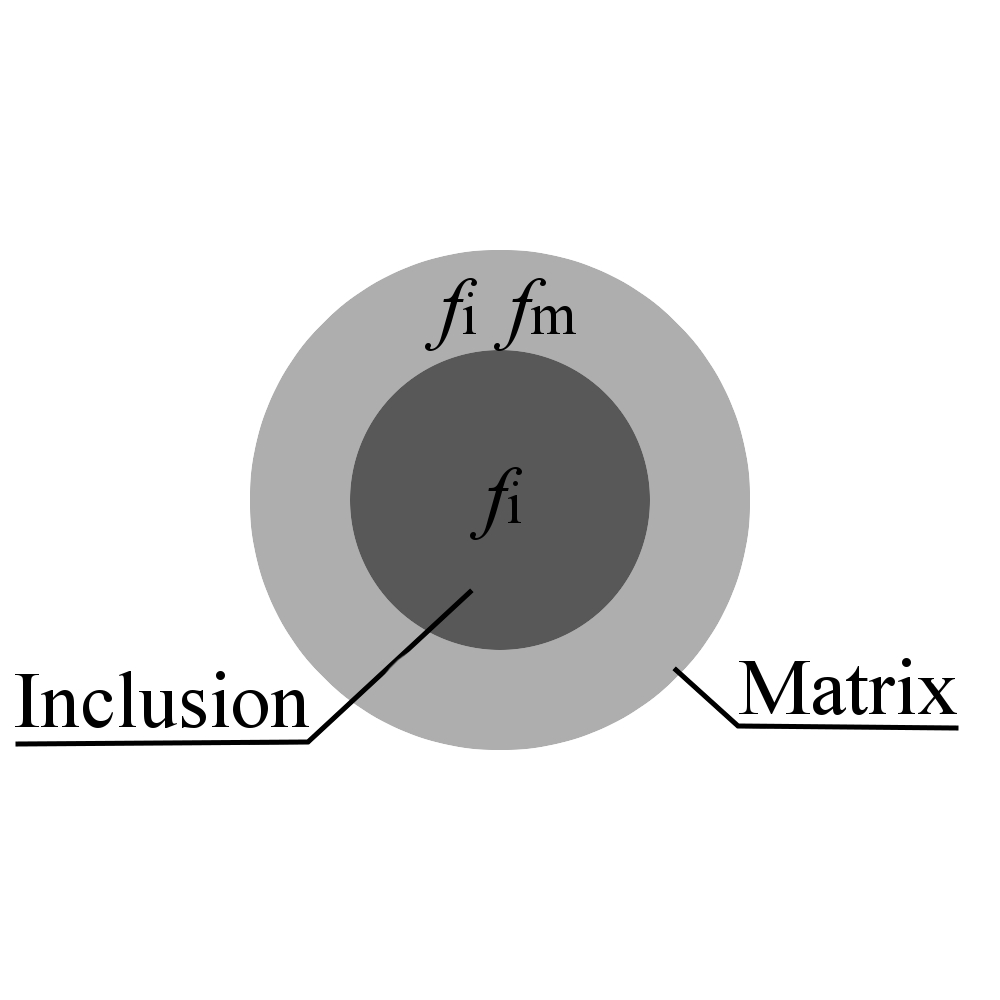}}
\raisebox{0.8\height}{\includegraphics[angle=0,height=2cm]{plus.jpg}}
\raisebox{.2\height}{\includegraphics[angle=0,height=3.5cm]{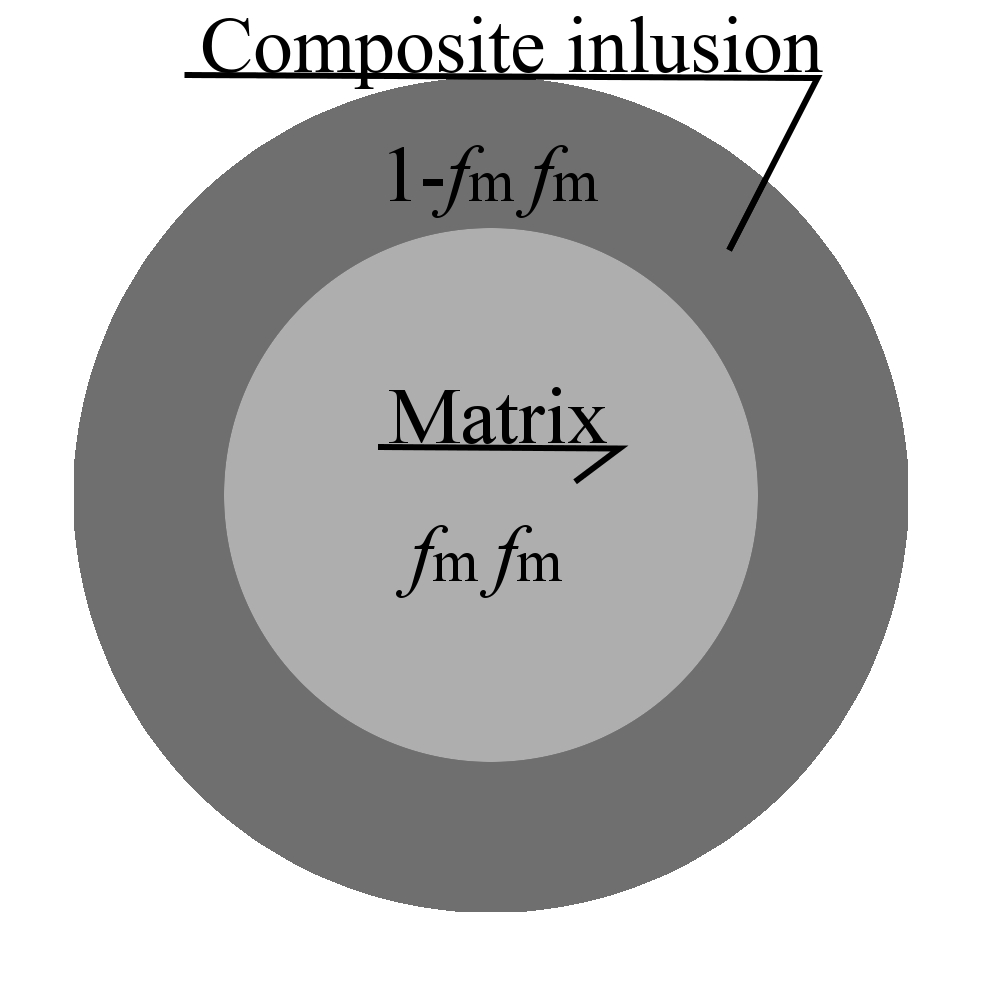}}&
\multirow{3}{*}[3.5cm]{
\includegraphics[angle=0,height=9cm]{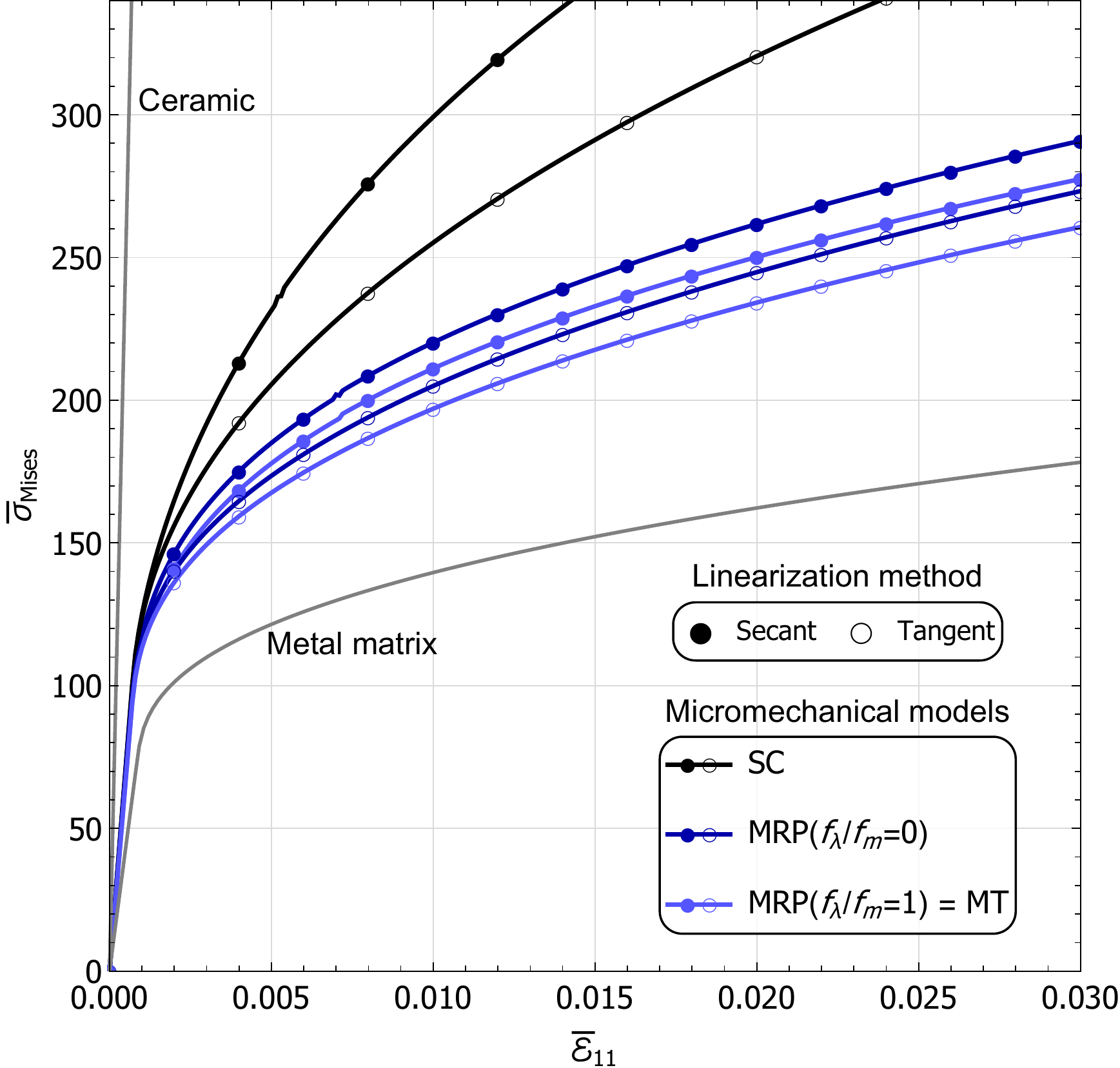}
}\\
(b)&\\
\raisebox{.1\height}{\includegraphics[angle=0,height=3.5cm]{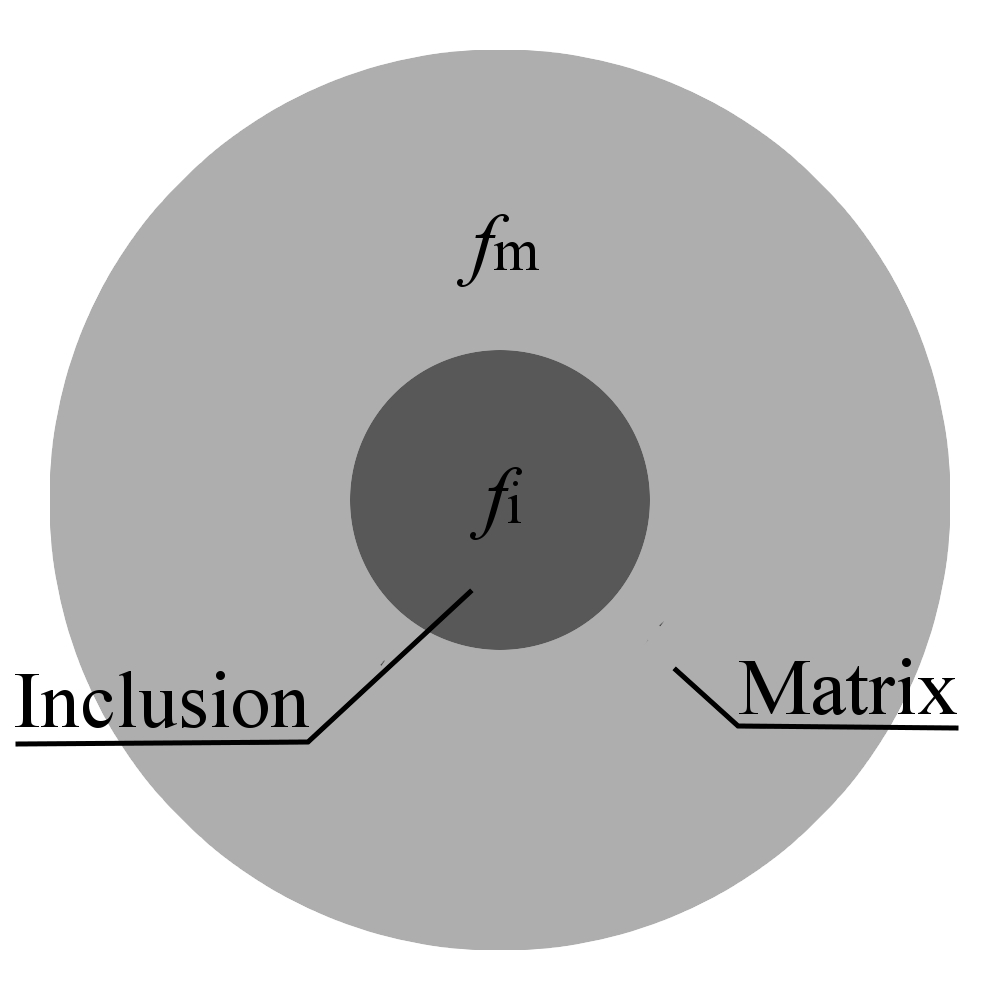}}&
\end{tabular}
\caption{
(a) and (b) the MRP representation of the MMC (Tab.\ref{Tab:MMC}) reinforced by ceramic balls with the packing ratio $f_{\lambda} / f_{\textup{m}}$ equal to: 0 or 1, respectively.
(c) The mean-field estimates of the elastic-plastic response of the MMC with $f_{\textup{i}}=0.30$, in the isochoric tension test : Huber-von Mises equivalent stress $\overline{\sigma}_{\textup{Mises}}$ vs. effective strain component $\overline{\varepsilon}_{11}$ in the direction of elongation.
SC - the Self-Consistent scheme,
MRP $f_{\lambda} / f_{\textup{m}}=0$ and
MRP $f_{\lambda} / f_{\textup{m}}=1$ - the Morphologically Representative Pattern-based approach for the packing ratio equal to: 0 (particles are in contact), or 1 (equivalent to Mori-Tanaka solution).
Two linearization schemes: secant (s) and tangent (t) are shown.
\label{Fig:Mises_tan_sec_0_1_spheres}}
\end{figure}

\begin{figure}[H]
\centering
\begin{tabular}{cc}\\
(a)&(b)\\
\raisebox{.7\height}{\includegraphics[angle=0,height=3.5cm]{pattern_inc_FIG3_6.jpg}}
\raisebox{1.6\height}{\includegraphics[angle=0,height=2cm]{plus.jpg}}
\raisebox{.7\height}{\includegraphics[angle=0,height=3.5cm]{pattern_free_FIG3_6.jpg}}
&
\includegraphics[angle=0,height=9cm]{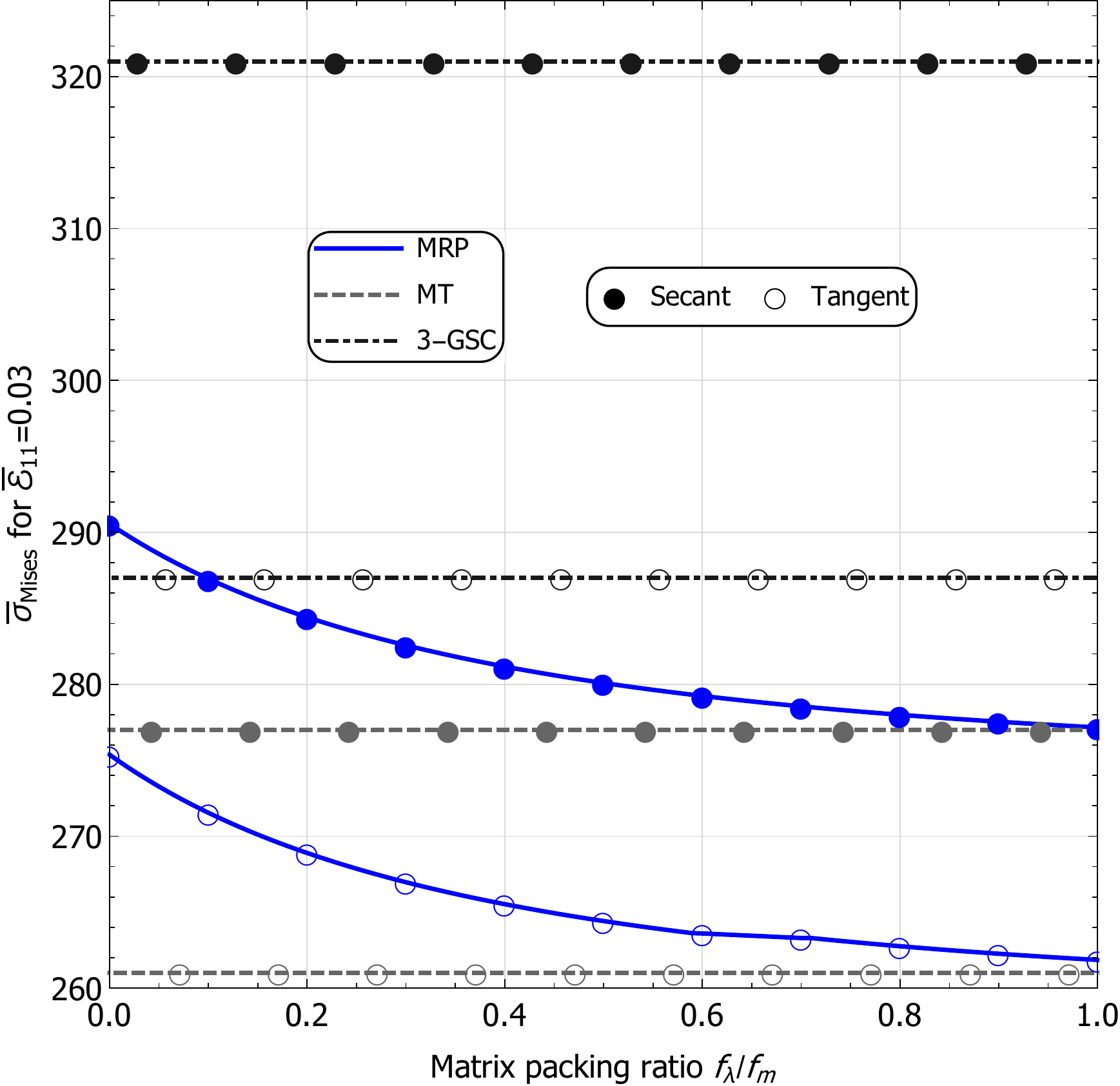}
\end{tabular}
\caption{
(a) two-pattern MRP representation of the MMC (Tab.\ref{Tab:MMC}) reinforced by spherical inhomogenieties.
(b) Huber-von Mises equivalent stress $\overline{\sigma}_{\textup{Mises}}$ for the effective strain component $\overline{\varepsilon}_{11}=0.03$ vs. the matrix packing ratio $f_{\lambda} / f_{\textup{m}}$.
Notation:
MRP - the MRP model based on two MT-type patterns,
MT - the Mori-Tanaka method,
3-GSC - the Generalized Self-Consistent scheme.
Two linearization schemes: secant and tangent, are presented.
\label{Fig:Smises_plastic_packing}}
\end{figure}

The essential aspects of the two linearization methods: secant and tangent, when combined with the MRP averaging scheme, are presented in Fig.\ref{Fig:Mises_tan_sec_0_1_spheres}.c for MMC reinforced by ceramic balls with the volume fraction of 30\%.
Fig.\ref{Fig:Mises_tan_sec_0_1_spheres}.c shows MRP results for two limit cases of the matrix packing ratio: 0 (MRP patterns in Fig.\ref{Fig:Mises_tan_sec_0_1_spheres}.a) or 1 (MRP representation in Fig.\ref{Fig:Mises_tan_sec_0_1_spheres}.b).
The self-consistent scheme estimates are presented for comparison with the MRP.
Fig.\ref{Fig:Mises_tan_sec_0_1_spheres}.c presents the effective Huber-von Mises equivalent stress as a function of the `11` component of the strain in the isochoric tension test.
The composite response obtained using secant linearization $\bullet$ is stiffer than the one from tangent linearization $\circ$.
When the matrix packing ratio goes to 0, the effective response of the assumed composite is stiffer.
The MRP estimates are below the SC results.
Fig.\ref{Fig:Smises_plastic_packing}.b presents the effective Huber-von Mises equivalent stress $\overline{\sigma}_{\textup{Mises}}$ for the effective strain component $\overline{\varepsilon}_{11}=0.03$ as a function of the matrix packing ratio $f_{\lambda} / f_{\textup{m}}$.
Fig.\ref{Fig:Smises_plastic_packing}.a shows representation of the basic two-pattern MRP.
Like previously, the MMC is reinforced by 30\% volume of ceramic balls.
The classical mean-field models: Mori-Tanaka and 3-phase Generalized Self Consistent do not consider packing effects.
The MT and 3-GSC results are horizontal lines in Fig.\ref{Fig:Smises_plastic_packing}, just as previously in Fig.\ref{Fig:Keff_Geff_elastic_packing}.b for elastic composite properties.
Results of two linearization schemes: secant and tangent, are presented.
Similar trends are observed for $\overline{\sigma}_{\textup{Mises}}$ vs. $f_{\lambda} / f_{\textup{m}}$ (Fig.\ref{Fig:Smises_plastic_packing}.b) and $\left(\overline{G},\overline{K}\right)$ vs. $f_{\lambda} / f_{\textup{m}}$ (Fig.\ref{Fig:Keff_Geff_elastic_packing}.b) .
In the MRP approach based on SC and n-GSC \citep{Majewski2020} the linearization procedure affected the relation between $\overline{\sigma}_{\textup{Mises}}$ and $f_{\lambda} / f_{\textup{m}}$ in comparison to the elastic properties of the composite \citep{Majewski2017}.
The secant linearization leads to much too stiff estimates of the effective behaviour. As a consequence, the tangent method was employed for the rest of the publication.

\section{Numerical calculations \label{Sec:Numer}}
The computer code was implemented in the Wolfram Mathematica (www.wolfram.com) environment.
FE meshes for the SVEs were generated using NetGen \citep{Schoberl1997}.
It is worth emphasizing that the reliability of the FEM simulations depends on the quality of the mesh.
The FE analyses were conducted in the AceFEM environment \citep{Korelc2002}.

\subsection{Procedure for generation of cubic volume elements with randomly-placed inclusions with periodic structure}
\begin{figure}[H]
\centering
\begin{tabular}{cccc}
(a)&(b)&(c)&(d)\\
\includegraphics[angle=0,height=4cm]{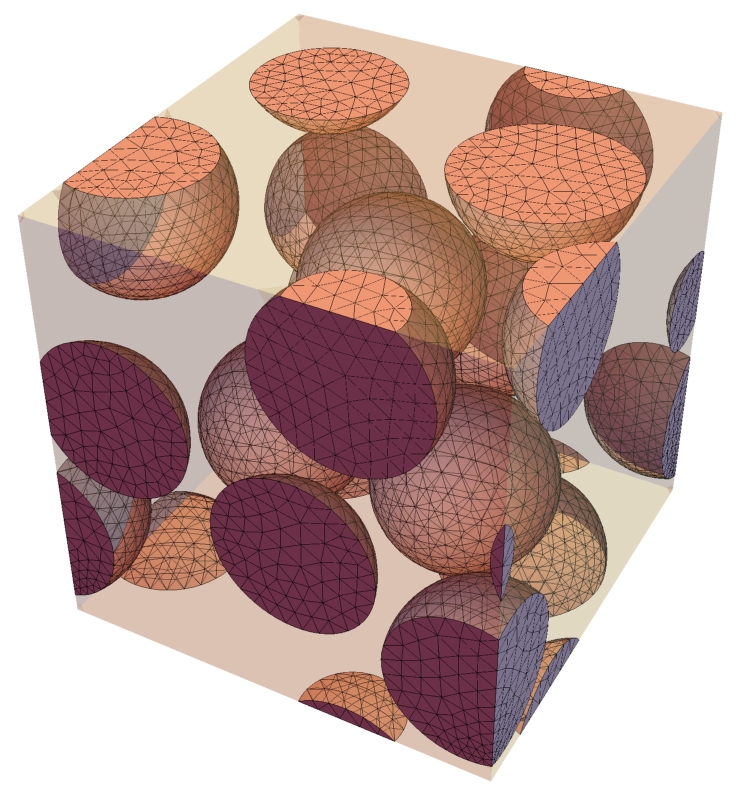}&
\includegraphics[angle=0,height=4cm]{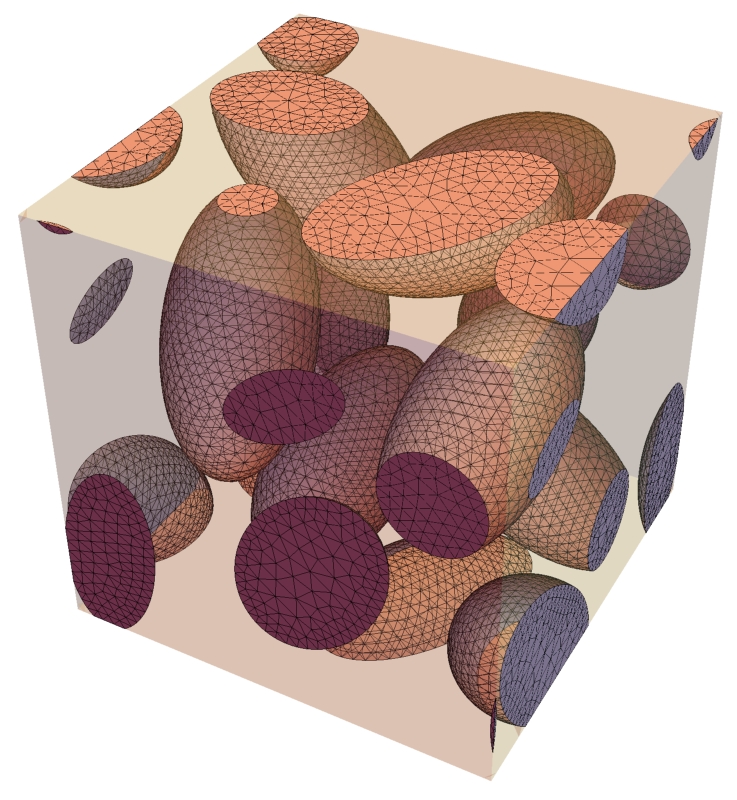}&
\includegraphics[angle=0,height=4cm]{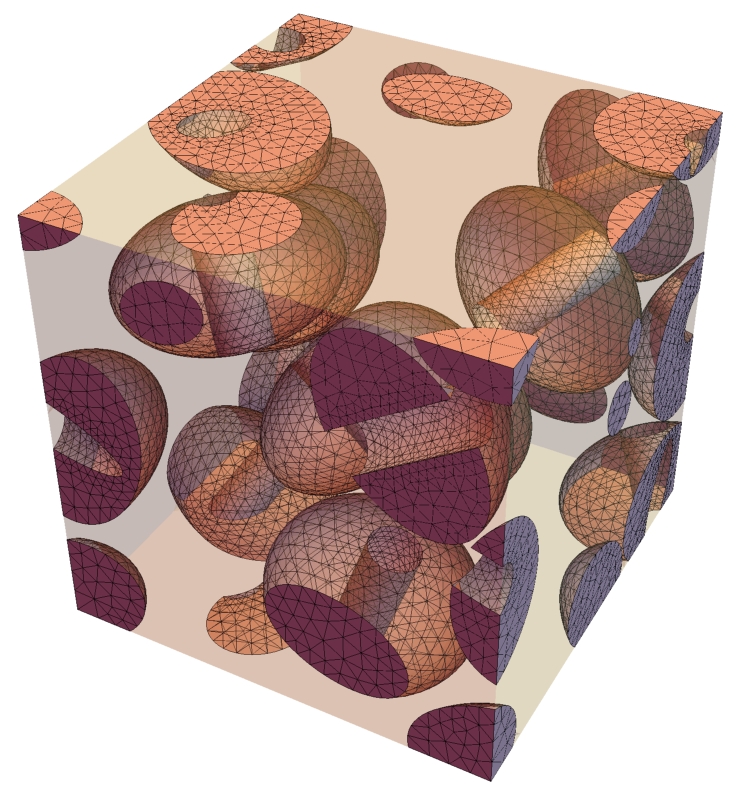}&
\includegraphics[angle=0,height=4cm]{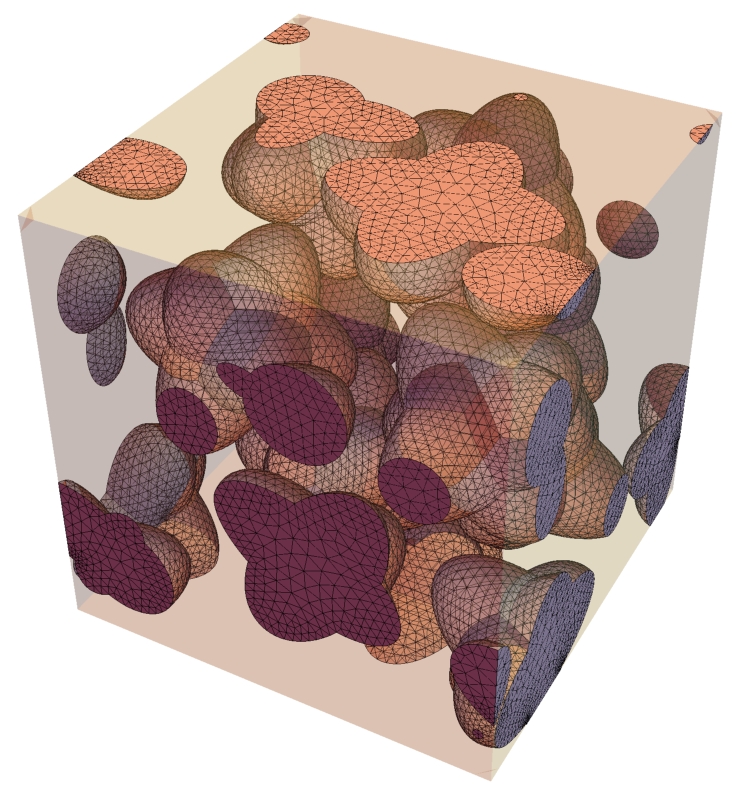}
\end{tabular}
\caption{
Statistical Volume Elements with periodic boundaries and randomly generated structures. The volume fraction of ten inclusions is equal to 0.3\,.
The shapes of inclusions are:
(a) spheres (Fig.\ref{Fig:Shapes}.a),
(b) prolate spheroids (Fig.\ref{Fig:Shapes}.b),
(c) drilled oblate spheroids (Fig.\ref{Fig:Shapes}.c),
and (d) crossed spheroids (Fig.\ref{Fig:Shapes}.d).
Meshes are shown for particles only.
\label{Fig:FEM_RVE_Shapes}}
\end{figure}

%The finite element method simulations were performed on samples with both regular and random arrangements of inclusions  \citep{Majewski2017, Majewski2020, Majewski2022}.
In this publication, the MMC with random arrangements of different shapes of particles (Fig.\ref{Fig:FEM_RVE_Shapes}) is examined to highlight the MRP's ability to represent each particle by one pattern allowing one to study differences in response between them.
The results for regular arrangements of inhomogeneities were published in papers:
\citep{Majewski2017} dealing with elastic properties of composites with spherical inclusions and \citep{Majewski2022} with various shapes of particles.

The statistical volume element (e.g. Fig.\ref{Fig:FEM_RVE_Shapes}) has a random microstructure, which is produced using the discrete element method system Yade \citep{Yade}.
SVE occupies a $1\times 1\times 1$ periodic cube.
Majewski et al. (\citeyear{Majewski2017}) described the methodology for generating spherical particles.
In short, the procedure runs as follows.
First, $n$ non-overlapping elastic and frictionless spheres of radius $R_{\textup{i}}$ are placed  in an enlarged cube of dimensions $2\times 2\times 2$.
The radius $R_{\textup{i}}$ corresponds to the prescribed volume fraction $f_{\textup{i}}$ of the target unit cube, i.e. $n\frac{4}{3}\pi R_{\textup{i}}^3=f_{\textup{i}}$.
The sparse preliminary setting of spheres is obtained using a built-in random sequential addition \citep{Torquato2002} functionality of Yade.
Second, the periodic cell is compressed uniformly to its target dimensions $1\times 1\times 1$, forcing the spheres closer together.
During the shrinking, the spheres mix by undergoing elastic collisions and crossing the periodic walls.
If at the end of the process the sphere centres are separated by more than $2 R_{\textup{i}}$, then the locations of the spheres are accepted as the positions of inhomogeneities in the particulate composite with spherical inclusions (Fig.\ref{Fig:FEM_RVE_Shapes}.a). 
For non-spherical shapes of inhomogeneities (Fig.\ref{Fig:FEM_RVE_Shapes}.b-d), subsequent particles with random orientations were added in such a manner that they did not intersect the already placed inhomogeneities.
The smallest distance between the surfaces of the inclusions $\lambda$ (see Fig.\ref{Fig:Morph}) was obtained numerically in order to determine the matrix packing ratio.
The procedure ignores the cavities in drilled spheroids (Fig.\ref{Fig:Shapes}.c), i.e., treats all inhomogeneities as filled.
If the smallest distance between the surfaces of the inclusions was shorter than 0.0005, the random structure was generated once again for the sake of  efficiency of FEM simulations.

%%%tekst napisany przezemnie, moze sie przydac
%The Representative Volume Element (RVE) (e.g. Fig.\ref{Fig:FEM_RVE_Shapes}) is the cubic volume element $1\times 1\times 1$ and has random arragement of particles, which is produced using the Discrete Element Method (DEM) system Yade \citep{Yade} based on dynamic procedure.
%The methodology has been described in detail in the previous paper \citep{Majewski2017}, when applied to spherical particles.
%The most crucial steps are repeted for clarity.
%First, $n$ non-overlapping elastic frictionless balls of radius $R_{\textup{i}}$ are located in tow tmies bigger cube $2\times 2\times 2$.
%Initial random placement of spheres is obtained with the aid of a built-in random sequential addition \citep{Widom1966,Torquato2002} functionality of Yade.
%Particels occupy a determined volume fraction $f_{\textup{i}}$ of $1\times 1\times 1$ cube, i.e. $n\frac{4}{3}\pi R_{\textup{i}}^3=f_{\textup{i}}$,
%plus the radius are artificially increased by  $0.001$ to ensure that there remains a gap between the actual particles of radius $R_{\textup{i}}$.
%Second, the periodic cell is compressed uniformly to its target dimensions $1\times 1\times 1$, forcing the spheres to be closer together - the spheres mix by undergoing elastic collisions and crossing the periodic walls.
\subsection{Numerical concentration tensors \label{SubSec:NumConcTens}}
\begin{figure}[H]
\centering
\begin{tabular}{cc}
\includegraphics[angle=0,height=4cm]{UklWsp.jpg}&
\includegraphics[angle=0,height=6cm]{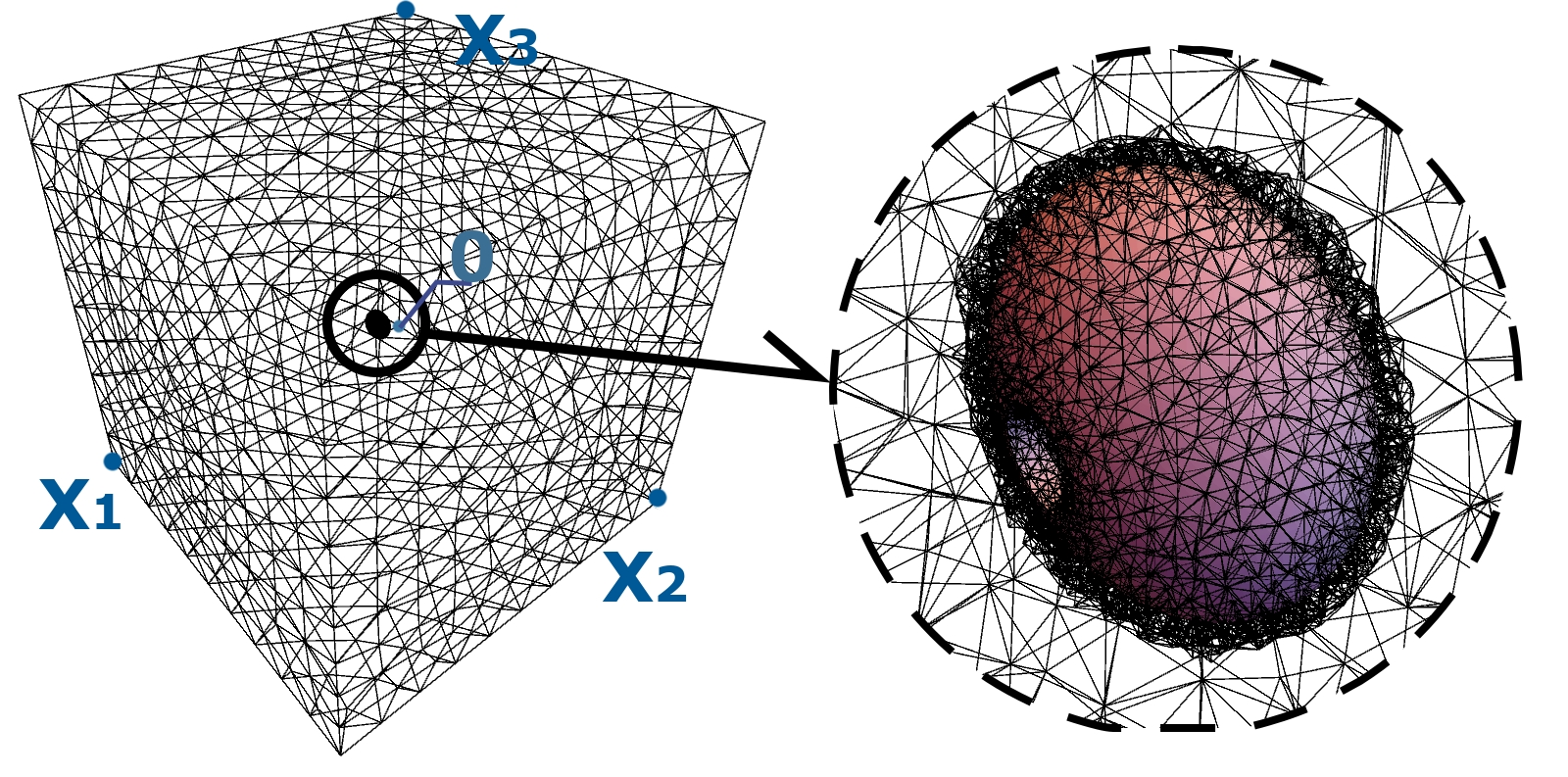}
\end{tabular}
\caption{
As an example, a drilled oblate spheroid is embedded into a large, but finite matrix region, to calculate the numerical diluted strain concentration tensor $\mathbb{A}_{\textup{i}}^{\textup{NDil}}$ using FEM.
\label{Fig:UVRMTM}}
\end{figure}

In this paper the diluted strain concentration tensor of a non-ellipsoidal inhomogeneity, $\mathbb{A}_{\textup{i}}^{\textup{Dil}}$, is obtained numerically, $\mathbb{A}_{\textup{i}}^{\textup{NDil}}$, using the finite element method.
To complete this task, three-dimensional numerical models of a large but finite matrix with a single heterogeneity having assumed properties and shape, and occupying a volume fraction $f_{\textup{i}}=10^{-4}$, as it is showed in Fig.\ref{Fig:UVRMTM}, are generated and simulated by FEM.
The mesh size was selected so that a subsequent mesh refinement does not affect the calculated concentration tensors noticeably, i.e., the difference is less than 0.1\%.
As an alternative, one can use the open-access software AMAT \citep{Markov2020}.
The advantage of the AMAT software is the formulation in terms of volume integral equations, which limits the calculation of the fields to the region occupied by the inclusion only.
In this work, we used our FEM codes due to our future plans for code modification to model propagation of cracks.

The macroscopic response of a composite (Fig.\ref{Fig:UVRMTM}) - a diluted inhomogeneity in a finite matrix - is driven by the symmetry of the particle.
Consequently, the fourth-order tensors $\mathbb{A}_{\textup{i}}^{\textup{NDil}}$ is:
\begin{itemize}
\item isotropic for the spherical inhomogeneity (Fig.\ref{Fig:Shapes}.a),
\item of cubic symmetry (Eq.\ref{Eq:ACube}) for the crossed spheroids (Fig.\ref{Fig:Shapes}.d),
\item transversely isotropic (Eq.\ref{Eq:ATrans}) for the prolate and oblate spheroids (Fig.\ref{Fig:Shapes}.b and c).
\end{itemize}
To derive $\mathbb{A}_{\textup{i}}^{\textup{NDil}}$ (Eq.\ref{Eq:ACube}-\ref{Eq:ATrans}) sets of periodic displacement boundary conditions were imposed \ref{Ap:AppendixPeriodic} (for details see \cite{Majewski2022}).
The displacement boundary conditions are based on the eigensubspaces of $\mathbb{A}_{\textup{i}}^{\textup{NDil}}$.
The relevant formulae for the projectors stemming from the spectral decomposition of the fourth-order tensors of cubic symmetry and transverse isotropy are given by Eq.\ref{Eq:ACube}-\ref{Eq:ATrans}, and Eq.\ref{Eq:PCube}-\ref{Eq:PTransSShear2}.
The representations of the imposed strains in the basis $\{\mathbf{m}_i\}$ aligned with the main symmetry axes of the cell (e.g. Fig.\ref{Fig:UVRMTM}) are:
\begin{equation}\label{Eq:BCTransStrainsText}
\begin{gathered}
E_{ij}^{\left(0\right)}=
\left(\begin{array}{ccc} d & 0& 0 \\ 0 & d & 0 \\ 0 & 0 & d \end{array}\right)
\,,\quad
E_{ij}^{\left(1\right)}=
\left(\begin{array}{ccc} d & 0& 0 \\ 0 & -d/2 & 0 \\ 0 & 0 & -d/2 \end{array}\right)
\,,\quad
E_{ij}^{\left(2\right)}=
\left(\begin{array}{ccc} 0 & 0& 0 \\ 0 & 0 & d \\ 0 & d & 0 \end{array}\right)
\,,\quad
E_{ij}^{\left(3\right)}=\left(\begin{array}{ccc} 0 & d& 0 \\ d & 0 & 0 \\ 0 & 0 & 0 \end{array}\right)\,,
\end{gathered}
\end{equation}
where $d$ specifies the strain magnitude.
For cubic anisotropy (Eq.\ref{Eq:ACube}) the analysis is performed for $\mathbf{E}^{\left(0\right)}$ through $\mathbf{E}^{\left(2\right)}$, and for transverse isotropy (Eq.\ref{Eq:ATrans}) for $\mathbf{E}^{\left(0\right)}$ through $\mathbf{E}^{\left(3\right)}$. 

For the displacement boundary conditions $\mathbf{E}^{(n)}$ from Eq.(\ref{Eq:BCTransStrainsText}) the components of $A^{\textup{Cub}}$ (Eq.\ref{Eq:ACube}) and $A^{\textup{Trans}}$ (Eq.\ref{Eq:ATrans}) are:
\begin{equation}\label{Eq:Acomponents}
\begin{gathered}
A^{\textup{Trans}}_0=\frac{\langle\boldsymbol{\varepsilon}\rangle^{\left(0\right)}_{11}+2\langle\boldsymbol{\varepsilon}\rangle^{\left(0\right)}_{22}}{3d}\,,\quad
A^{\textup{Trans}}_{12}\approx A^{\textup{Trans}}_{21}=\frac{\langle\boldsymbol{\varepsilon}\rangle^{\left(0\right)}_{11}-\langle\boldsymbol{\varepsilon}\rangle^{\left(0\right)}_{22}+\langle\boldsymbol{\varepsilon}\rangle^{\left(1\right)}_{11}+2\langle\boldsymbol{\varepsilon}\rangle^{\left(1\right)}_{22}}{3d}\,,\quad
A^{\textup{Trans}}_1=\frac{2\left(\langle\boldsymbol{\varepsilon}\rangle^{\left(1\right)}_{11}-\langle\boldsymbol{\varepsilon}\rangle^{\left(1\right)}_{22}\right)}{3d}\,,\quad
 \\ 
A^{\textup{Trans}}_2=\frac{\langle\boldsymbol{\varepsilon}\rangle^{\left(2\right)}_{23}}{d}\,,\quad
A^{\textup{Trans}}_3=\frac{\langle\boldsymbol{\varepsilon}\rangle^{\left(3\right)}_{12}}{d}\,,\quad
%A^{\textup{Trans}}_4=\frac{\langle\boldsymbol{\varepsilon}\rangle^{\left(4\right)}_{22}}{d}\,,
\\
A^{\textup{Cub}}_0=\frac{\langle\boldsymbol{\varepsilon}\rangle^{\left(0\right)}_{11}}{d}\,,\quad
A^{\textup{Cub}}_1=\frac{\langle\boldsymbol{\varepsilon}\rangle^{\left(1\right)}_{11}}{d}\,,\quad
A^{\textup{Cub}}_2=\frac{\langle\boldsymbol{\varepsilon}\rangle^{\left(2\right)}_{23}}{d}\,,\quad
\end{gathered}
\end{equation}
where $\boldsymbol{\varepsilon}$ is the local strain tensor in the inhomogeneity domain, $\langle \cdot \rangle$ is the volume averaging operation defined as $1/V \int_V (\cdot)dV$, and $\langle\boldsymbol{\varepsilon}\rangle^{\left(n\right)}_{ij}$ is the component $ij$ of the inhomogeneity's average strain in response to the displacement BC given by $\mathbf{E}^{\left(n\right)}$.
The simplification $A^{\textup{Trans}}_{12}\approx A^{\textup{Trans}}_{21}$ was postulated because of negligible $A^{\textup{Trans}}_{12}$ and $A^{\textup{Trans}}_{21}$ values in comparison to $A^{\textup{Trans}}_{0}$, $A^{\textup{Trans}}_{1}$, $A^{\textup{Trans}}_{2}$, and $A^{\textup{Trans}}_{3}$ \citep{Majewski2022}.
\section{Results\label{Sec:Results}}
\subsection{Concentration tensors\label{SubSec:Reg}}
\begin{figure}[H]
\centering
\begin{tabular}{ccc}
(a)&(b)\\
\includegraphics[angle=0,height=8cm]{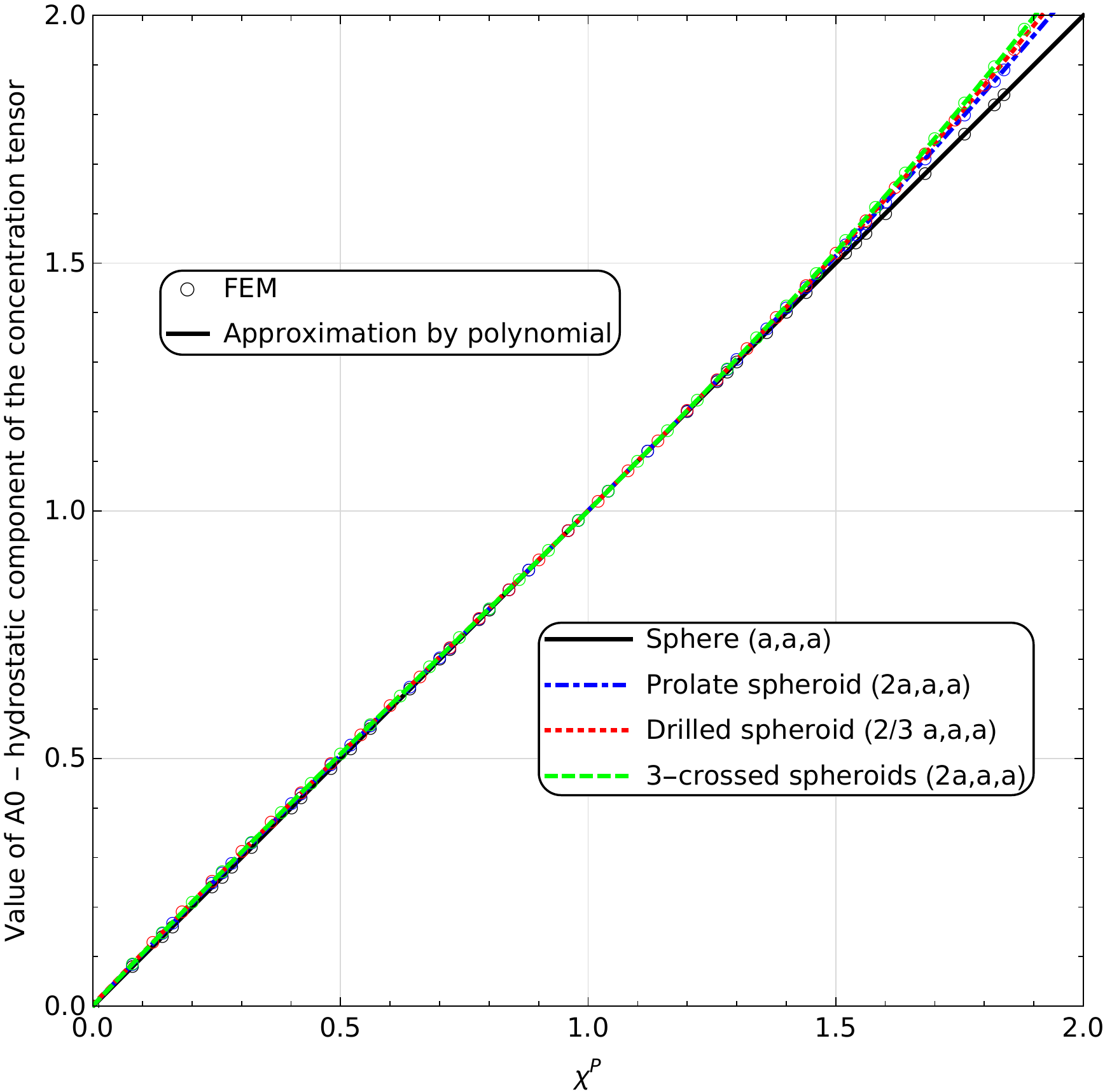}&
\includegraphics[angle=0,height=8cm]{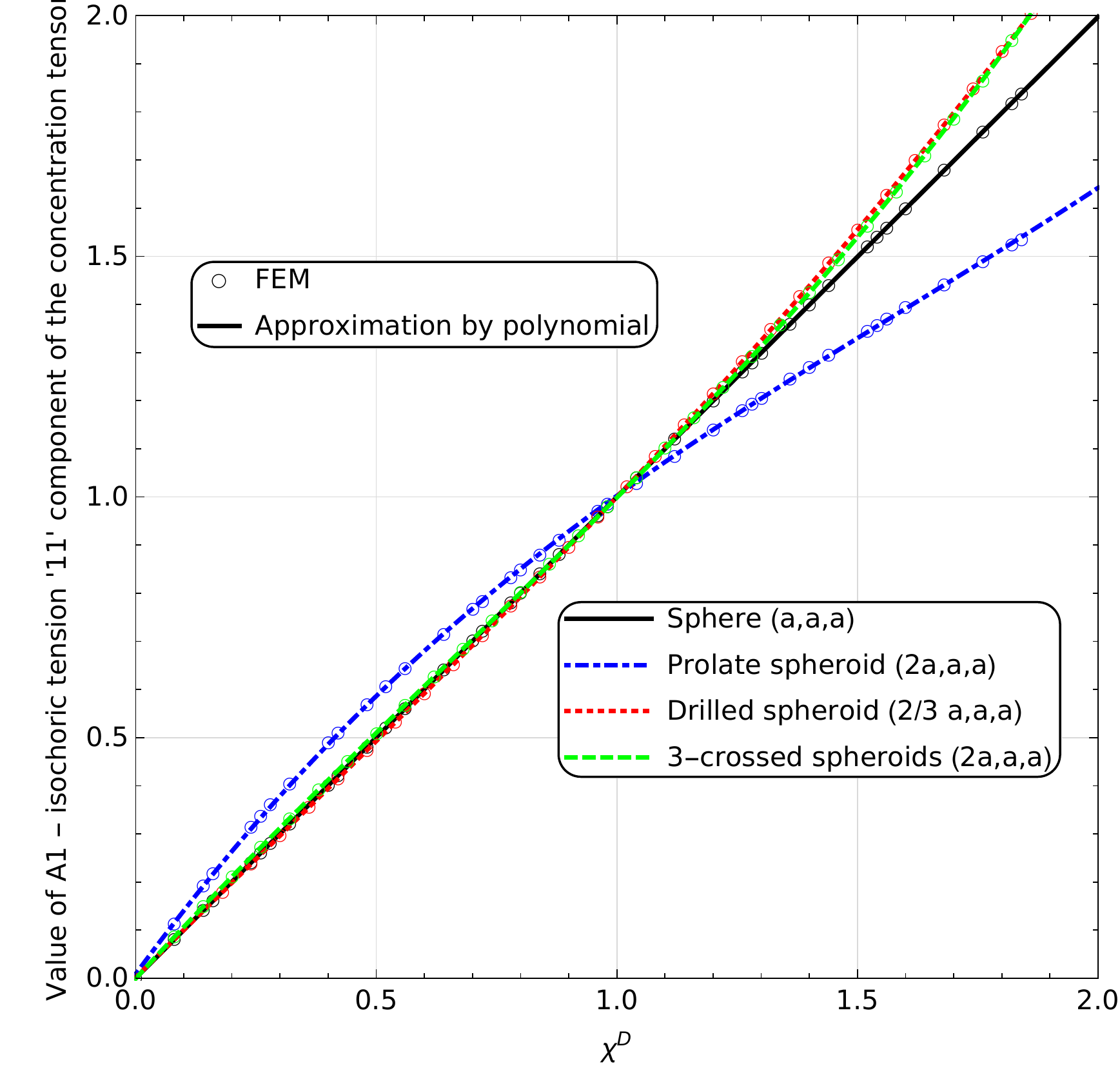}\\
(c)\\
\includegraphics[angle=0,height=8cm]{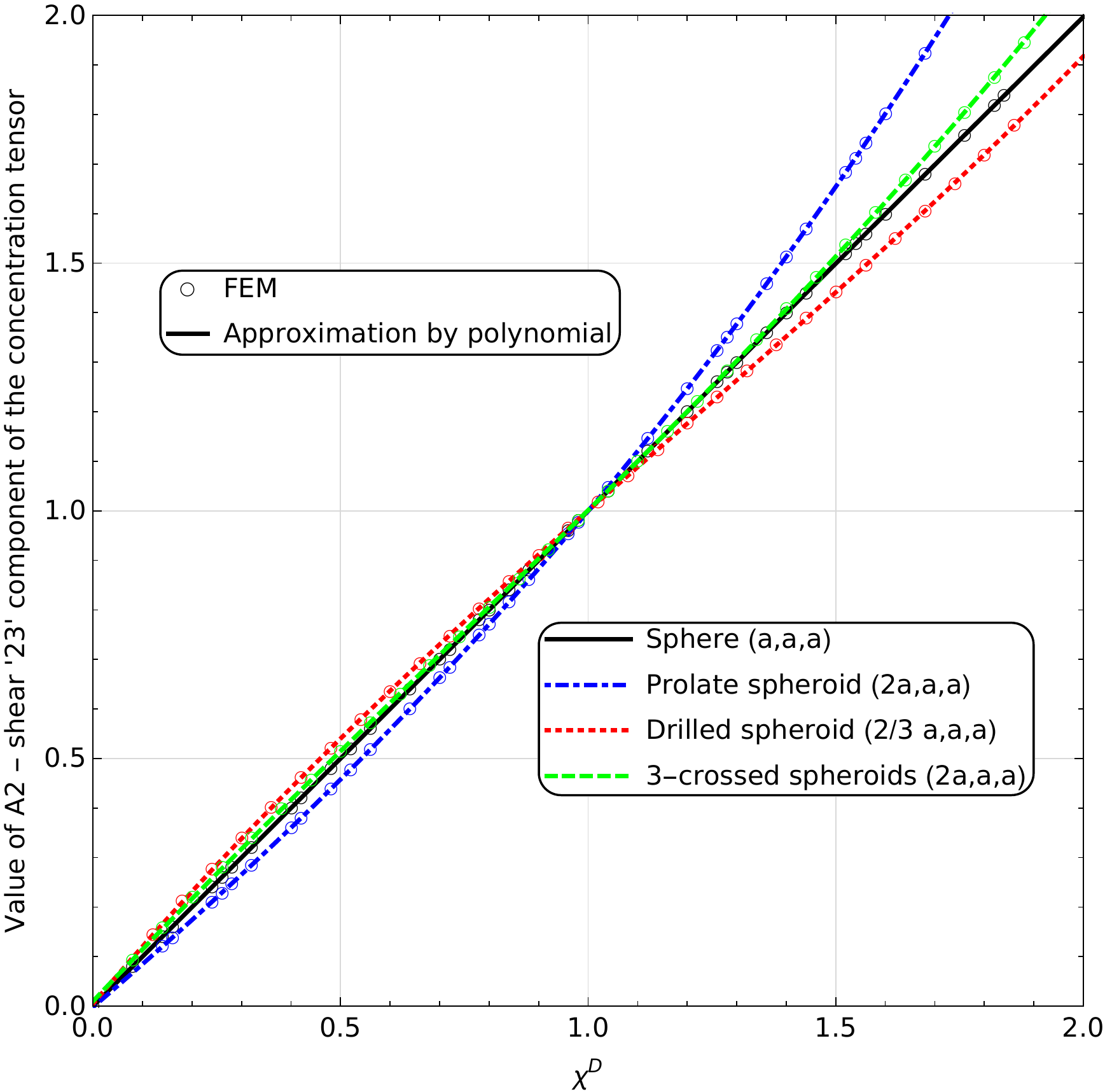}
\end{tabular}
\caption{
Components of the numerical diluted concentration tensor $\mathbb{A}_{\textup{i}}^{\textup{NDil}}$: (a) $A_0$ hydrostatic, (b) $A_1$ isochoric tension '11', and (c) $A_2$ shear '23', vs. MT's local concentration tensor of a spherical inhomogeneity: (a) $\chi^{\rm{P}}$ bulk , and (b)-(c) $\chi^{\rm{D}}$ deviatoric part (Eq.\ref{Eq:ANDilIso}).
FEM - numerical strain concentration tensor from Finite Element Method simulation,
solid lines - polynominal approximation of numerical strain tensors for different material parameters.
For the shapes of particles see Fig.\ref{Fig:Shapes}.
\label{Fig:Results_cons_tens_tests}}
\end{figure}

\begin{figure}[H]
\centering
\begin{tabular}{cc}
(a)&(b)\\
\includegraphics[angle=0,height=8cm]{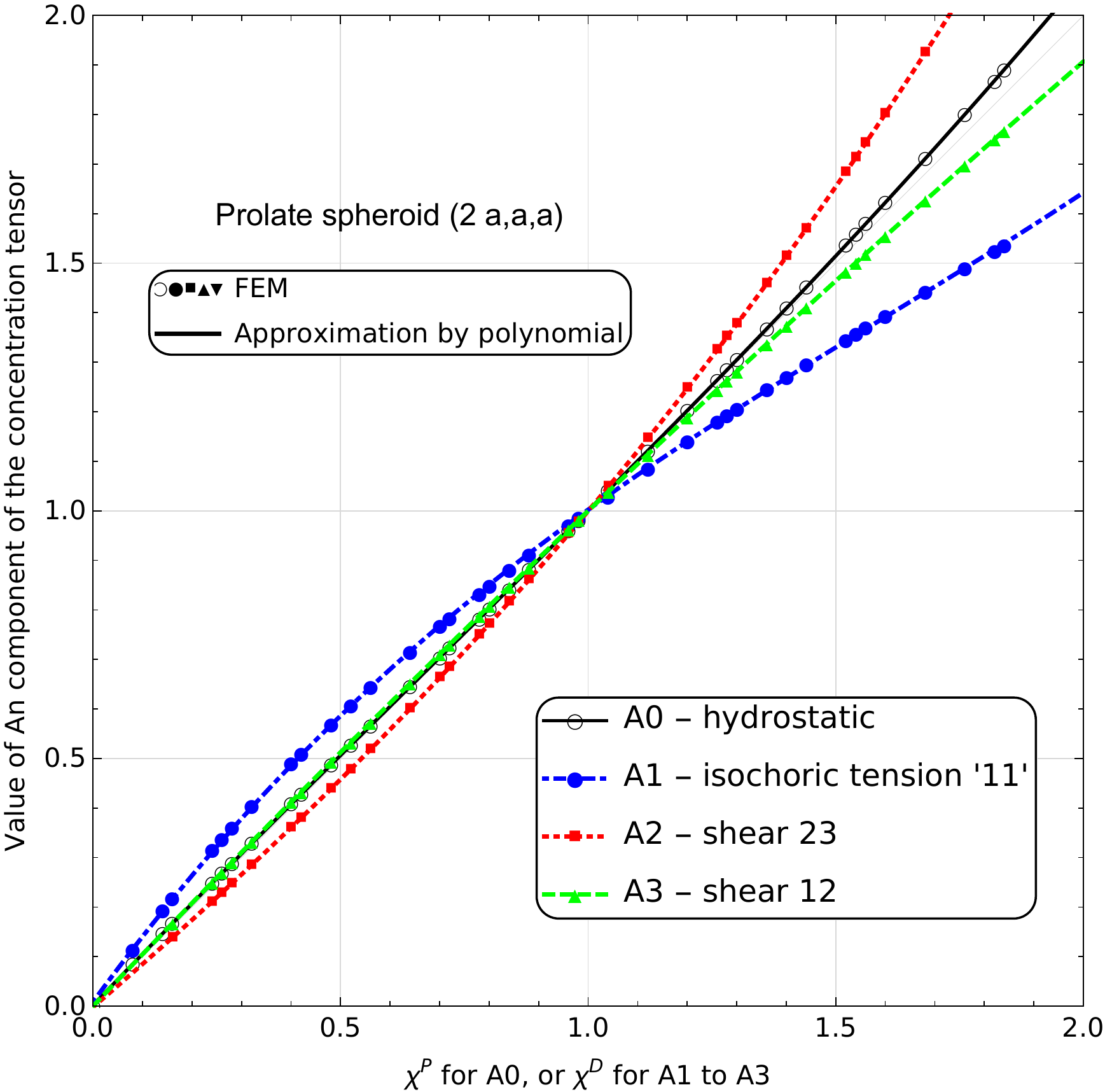}&
\includegraphics[angle=0,height=8cm]{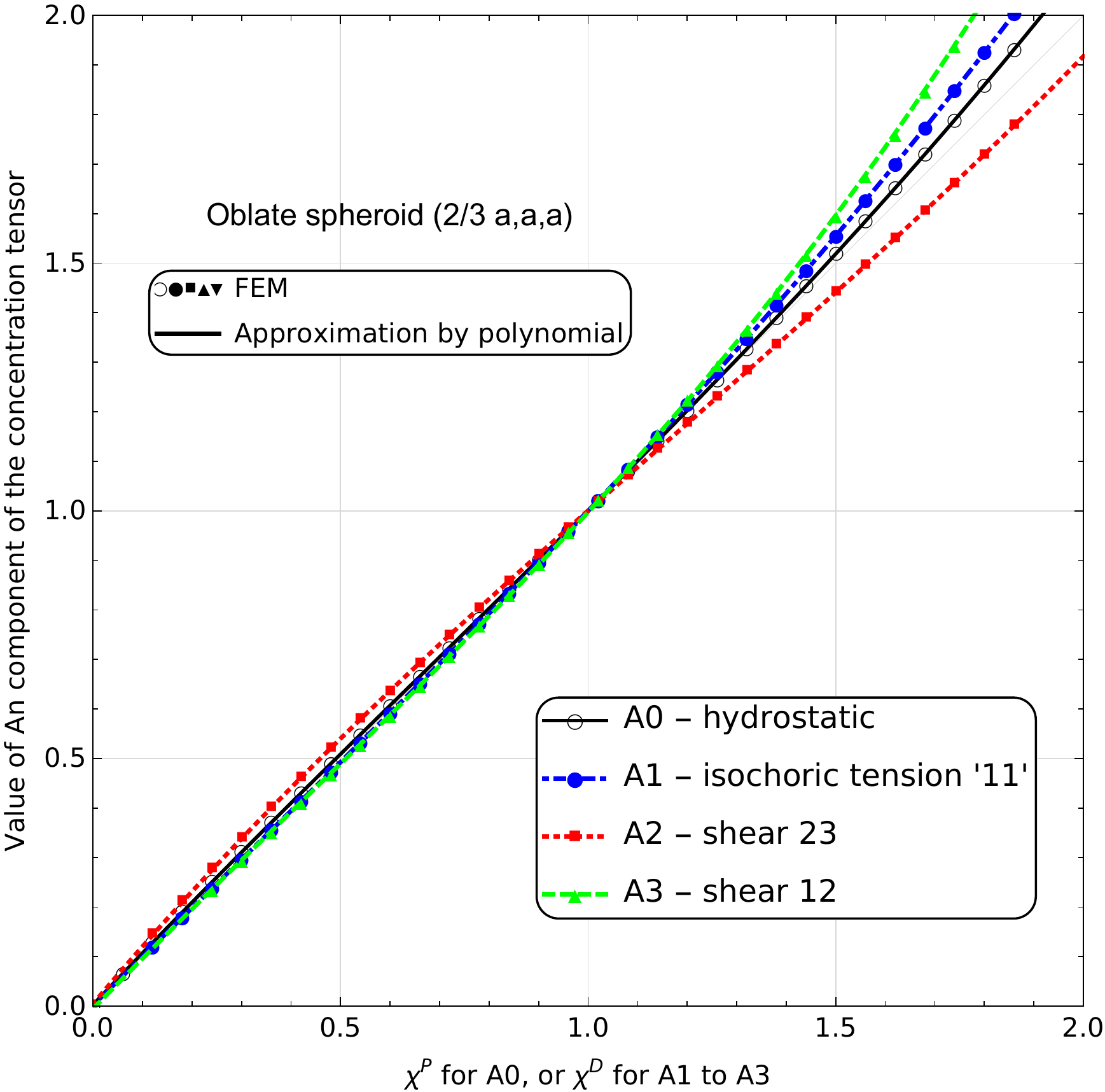}
\end{tabular}
\caption{
Components of the numerical diluted concentration tensor $\mathbb{A}_{\textup{i}}^{\textup{NDil}}$ ($A_0$ to $A_3$), vs. MT's local concentration tensor of a spherical inhomogeneity ($\chi^{\rm{P}}$ and $\chi^{\rm{D}}$ from Eq.\ref{Eq:ANDilIso}).
(a) prolate spheroid, (b) drilled oblate spheroid.
FEM - numerical strain concentration tensor from Finite Element Method simulation,
solid lines - polynomial approximation of numerical strain concentration tensors for different material parameters (Eq.\ref{Eq:funANDilapprox}).
\label{Fig:Results_cons_tens_shapes}}
\end{figure}

Majewski et al. (\citeyear{Majewski2022}) verified the procedure of finding the numerical strain concentration tensor in the elastic regime.
The metal matrix starts to flow plastically following equations \ref{Eq:LinHMa}-\ref{Eq:LinY} during the process of elastic-plastic deformation of the MMC. 
When plastic deformation occurs in the metal matrix, the bulk modulus of metal matrix remains unchanged, while the shear modulus decreases (Eq.\ref{Eq:LinSec} or \ref{Eq:LinStyOsno}) and changes the concentration tensor.
For this reason, the components $A_n$, $n=0,..,3$, of the numerical concentration tensor $\mathbb{A}_{\textup{i}}^{\textup{NDil}}$ have been determined as a third degree polynomial $A_n\left[\chi\right]$
\begin{equation}\label{Eq:funANDilapprox}
A_n\left[\chi\right] \overset{\mathrm{approx.}}{=} a_n+b_n \chi +c_n \chi^2 +d_n \chi^3
\, ,
\end{equation}
in which the variable $\chi$ depends on the material parameters
$\left( K_{\textup{i}},G_{\textup{i}},K_{\textup{m}},G_{\textup{m}}  \right)$ of the composite.
Due to the trend of numerical results (Fig.\ref{Fig:Results_cons_tens_tests} and \ref{Fig:Results_cons_tens_shapes}), we decided to approximate the concentration tensor with a polynomial.
The third-order polynomial approximates the results accurately enough.
The components $A_{12}$ and $A_{21}$ of $\mathbb{A}^{\textup{Trans}}$ (Eq.\ref{Eq:ATrans}) are near 0 for the presented results \citep{Majewski2022}, thus in the present analysis $A_{12}=A_{21}=0$ was assumed in all cases.
After the analysis of the influence of material parameters on the components of the numerical diluted concentration tensor, it turned out that the best variables are the ratios of material parameters:
\begin{equation}\label{Eq:hydrostaticdeviatoricparts}
\chi^{\rm{P}}=\frac{4 G_{\textup{m}}+3 K_{\textup{m}}}{4 G_{\textup{m}}+3 K_{\textup{i}}} \quad\mathrm{for}\, n=0\,,\quad\mathrm{and}\qquad
\chi^{\rm{D}}=\frac{5 G_{\textup{m}}\left(4 G_{\textup{m}}+3 K_{\textup{m}}\right)}{6 G_{\textup{i}}\left(2 G_{\textup{m}}+K_{\textup{m}}\right)+G_{\textup{m}}\left(8 G_{\textup{m}}+9 K_{\textup{m}}\right)}\quad\mathrm{for}\,n=1,2,3\,,
\end{equation}
where $\chi$ is expressed as the hydrostatic ($n=0$) or deviatoric ($n=1,2,3$) part of the Mori-Tanaka local concentration tensor determined for the spherical inclusion $\left(K_{\textup{i}},G_{\textup{i}} \right)$ placed in the matrix $\left(K_{\textup{m}},G_{\textup{m}} \right)$.
Thus, the numerical diluted strain concentration tensor of the spherical inhomogeneity is calculated as 
\begin{equation}\label{Eq:ANDilIso}
\mathbb{A}^{\textup{NDil,Iso}}=
A^{\textup{Iso}}_0\left[\chi^{\rm{P}}\right] \mathbb{I}^{\rm{P}}+
A^{\textup{Iso}}_1\left[\chi^{\rm{D}}\right] \mathbb{I}^{\rm{D}}
=
\frac{4 G_{\textup{m}}+3 K_{\textup{m}}}{4 G_{\textup{m}}+3 K_{\textup{i}}} \mathbb{I}^{\rm{P}}+
\frac{5 G_{\textup{m}}\left(4 G_{\textup{m}}+3 K_{\textup{m}}\right)}{6 G_{\textup{i}}\left(2 G_{\textup{m}}+K_{\textup{m}}\right)+G_{\textup{m}}\left(8 G_{\textup{m}}+9 K_{\textup{m}}\right)} \mathbb{I}^{\rm{D}}=
\chi^{\rm{P}} \mathbb{I}^{\rm{P}}+
\chi^{\rm{D}} \mathbb{I}^{\rm{D}}
\,,
\end{equation}
where $a_n=c_n=d_n=0$ and $b_n=1$ (Eq.\ref{Eq:funANDilapprox}) for $n=0$ or $1$.
The numerical form $\mathbb{A}^{\textup{NDil,Iso}}$ (Eq.\ref{Eq:ANDilIso}) equals its analytical formulation $\mathbb{A}^{\textup{Dil,Iso}}$.

Figure \ref{Fig:Results_cons_tens_tests}.a-c shows the $\mathbb{A}^{\textup{NDil}}$ components $A_0$, $A_1$, and $A_2$, respectively, as a function of the MMC material parameters $\left( K_{\textup{i}},G_{\textup{i}},K_{\textup{m}},G_{\textup{m}}  \right)$.
The numerical concentration tensor was determined for four shapes of inclusions (Fig.\ref{Fig:Shapes}).
FEM simulations (markers in Fig.\ref{Fig:Results_cons_tens_tests}) were carried out for several variants of the material parameter $(K_{\textup{i}},G_{\textup{i}},K_{\textup{m}},G_{\textup{m}})$.
The functions approximating the concentration tensor were determined based on the obtained FEM results (lines in Fig.\ref{Fig:Results_cons_tens_tests}).
Fig.\ref{Fig:Results_cons_tens_tests}.a shows the results for the hydrostatic component $A_0$ of $\mathbb{A}^{\textup{NDil}}$.
As seen in Fig.\ref{Fig:Results_cons_tens_tests}.a, for material parameters $\left( G_{\textup{i}},K_{\textup{m}},G_{\textup{m}}  \right)$ as in Tab.\ref{Tab:MMC} and $\chi^{\rm{P}} > 1$, $A_0$ of: the prolated spheroid, the drilled spheroid, and the crossed spheroids is larger than that for the spherical inclusions, i.e., the studied non-spherical inclusions with $K_{\textup{i}}<K_{\textup{m}}$ will accommodate a larger hydrostatic deformation than the spherical ones.
For $\chi^{\rm{P}} < 1$, the differences between the values of $A_0$ for the presented shapes of inclusions are small.

In further analysis, let us assume material parameters $\left( G_{\textup{i}},K_{\textup{i}},K_{\textup{m}}  \right)$ as in Tab.\ref{Tab:MMC}.
As it was assumed (Eq.\ref{Eq:LinHMa}), the metal matrix is elastic-plastic and its shear modulus $G_{\textup{m}}$ decreases during plastic deformation of the MMC.
In the isochoric tension test ($\mathbf{E}_1$ from Eq.\ref{Eq:BCTransStrainsText}), the influence of the inclusions shape on the concentration tensor function is clearly visible (Fig.\ref{Fig:Results_cons_tens_tests}.b).
When $G_{\textup{m}}<G_{\textup{i}}$ ( Fig.\ref{Fig:Results_cons_tens_tests}.b $\chi^{\rm{D}} < 1$), which is always true during plastic deformation of the matrix, the prolated spheroid undergoes a larger and larger deformation in comparison to the spheres.
The respective values for the remaining shapes are similar to those for spherical inclusions.
However, when the shear modulus of the inclusions is smaller than that of the matrix $G_{\textup{m}}>G_{\textup{i}}$ ( Fig.\ref{Fig:Results_cons_tens_tests}.b $\chi^{\rm{D}} > 1$), elongated spheroids undergo smaller deformations than the others, which may be applicable in the case of damaged areas, so when instead of elastic-plastic model, the elastic damage model is used.
The last graph, Fig.\ref{Fig:Results_cons_tens_tests}.c, shows the results of the shear test in the 2-3 plane ($\mathbf{E}_2$ from Eq.\ref{Eq:BCTransStrainsText}).
Let us postulate $G_{\textup{m}}$ as before and the remaining material parameters $\left( G_{\textup{i}},K_{\textup{i}},K_{\textup{m}}\right)$ as in the MMC composite (Tab.\ref{Tab:MMC}).
Along with the weakening of the matrix shear modulus $G_{\textup{m}}$, the highest value of $A_2$ was found for drilled spheroids, the lowest for prolated spheroids,
while $A_2$ for crossed spheroids was greater than for spherical inclusions when $G_{\textup{m}}\neq G_{\textup{i}}$. 
Next, Fig.\ref{Fig:Results_cons_tens_shapes} presents the numerical concentration tensor of: (a) a prolate spheroid, and (b) an oblate spheroid.
Both are tranversely isotropic thus the components $A_0$ to $A_3$ are shown.
A larger difference in the components $\mathbb{A}_{\textup{i}}^{\textup{NDil}}$ (Fig.\ref{Fig:Results_cons_tens_shapes}.a) is observed for the prolate spheroid  compared to the oblate drilled spheroid (Fig.\ref{Fig:Results_cons_tens_shapes}.b).
The component $A_3$, which results from the shear test in plane 1-2 ($\mathbf{E}_3$ from Eq.\ref{Eq:BCTransStrainsText}), shows a similar trend to $A_1$ (Fig.\ref{Fig:Results_cons_tens_shapes}).
\subsection{Spherical shape of particles\label{SubSec:Spheres}}
\begin{figure}[H]
\centering
\begin{tabular}{cc}
(a)&(b)\\
pattern 1\qquad \qquad \qquad \qquad pattern 2 &
\multirow{4}{*}[0cm]{
\includegraphics[angle=0,height=9cm]{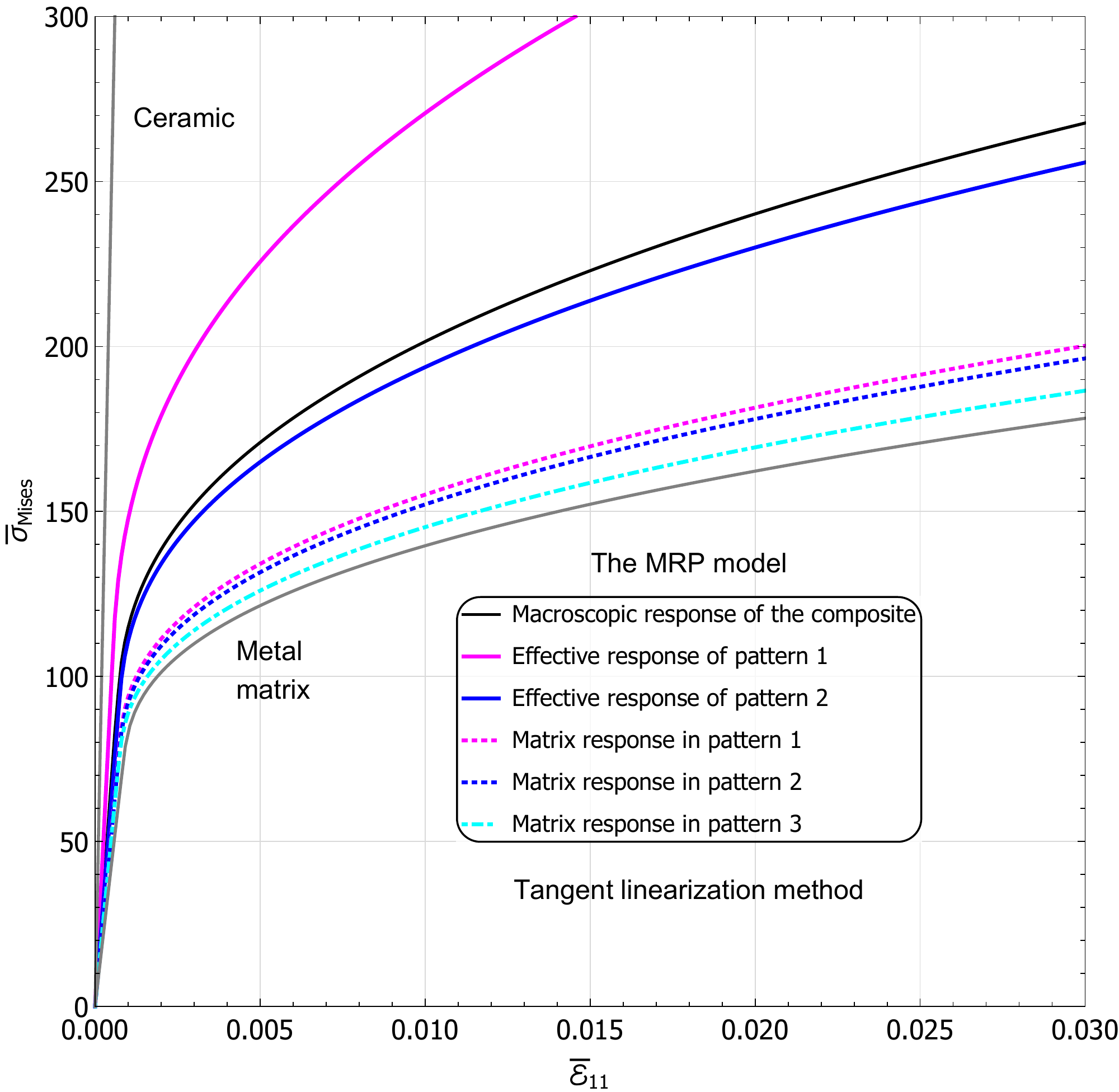}
}
\\
\raisebox{0.1\height}{\includegraphics[angle=0,height=3.5cm]{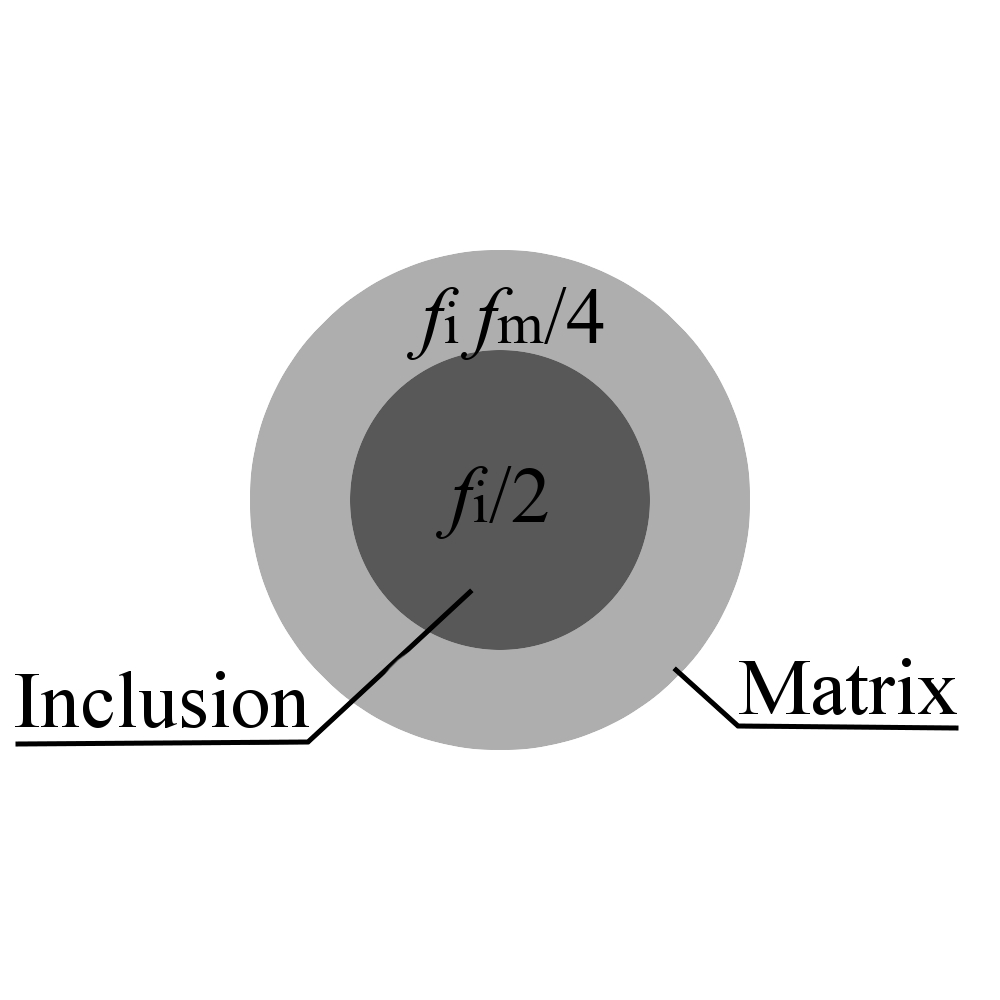}}
\raisebox{0.6\height}{\includegraphics[angle=0,height=2cm]{plus.jpg}}
\raisebox{0.1\height}{\includegraphics[angle=0,height=3.5cm]{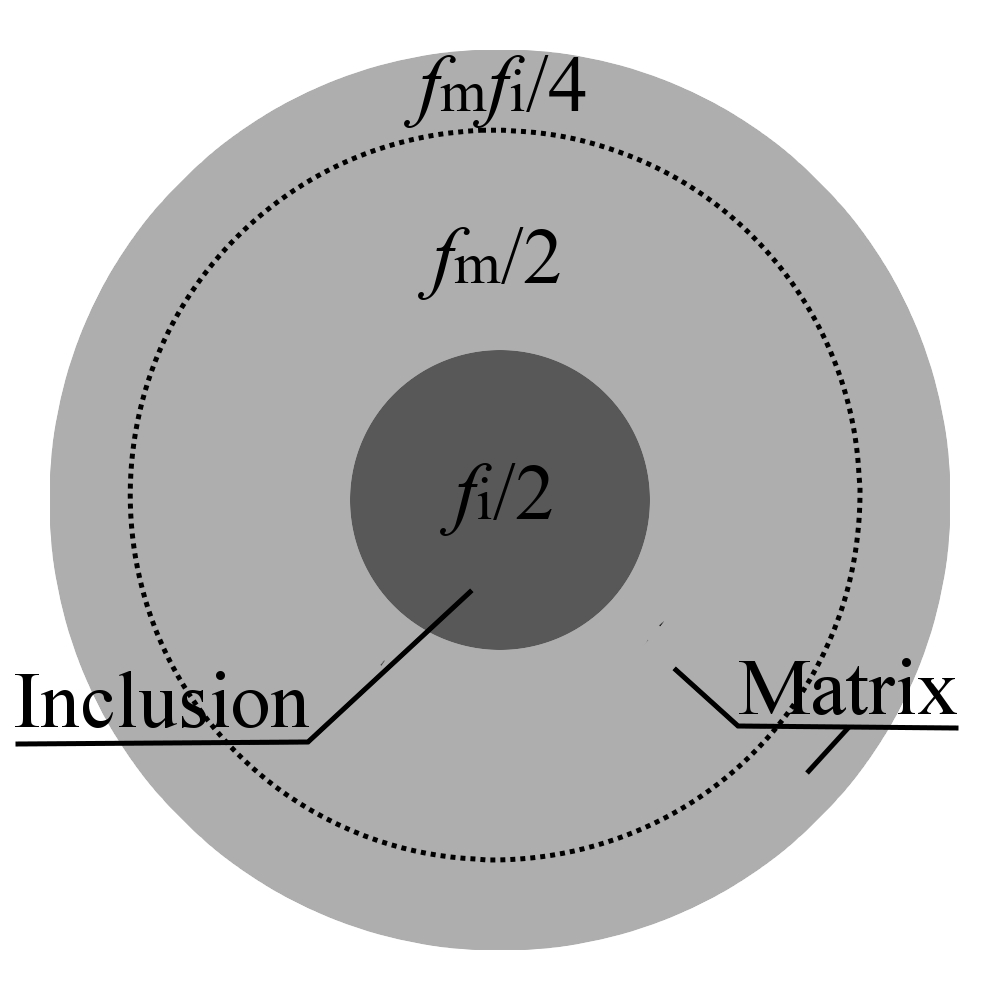}}&
\\
\quad pattern 3&
\\
\raisebox{0.6\height}{\includegraphics[angle=0,height=2cm]{plus.jpg}}
\raisebox{0.1\height}{\includegraphics[angle=0,height=3.5cm]{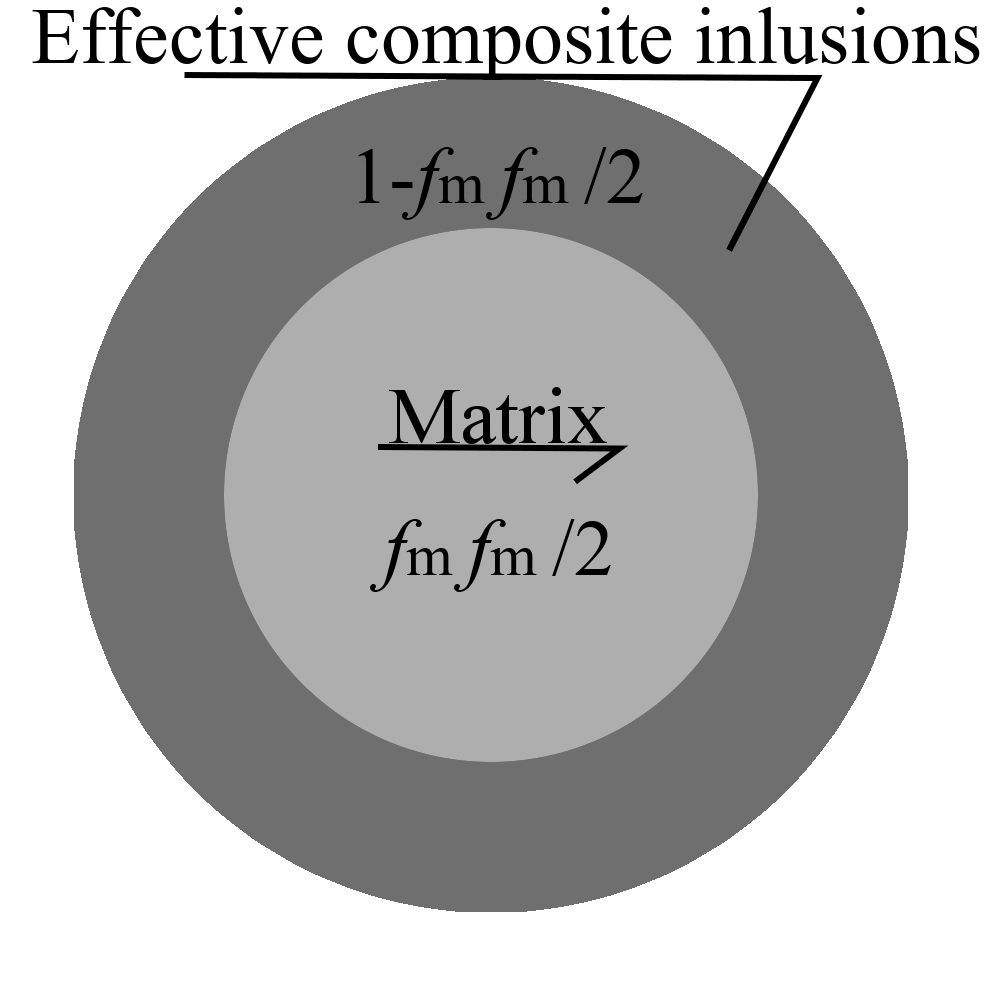}}
\qquad &
\end{tabular}
\caption{
(a) The MMC (Tab.\ref{Tab:MMC}) reinforced by the ceramic spherical inclusions represented by three patterns in the MRP model: (1) $f_{\lambda}=0$,  (2) $f_{\lambda}=f_{\textup{m}}/2$, and (3) the rest of the matrix phase.
(b) The MRP estimation of the elastic-plastic response of the MMC with $f_{\textup{i}}=0.30$ in the isochoric tension test: Huber-von Mises equivalent stress $\overline{\sigma}_{\textup{Mises}}$ vs. effective strain component $\overline{\varepsilon}_{11}$ in the direction of elongation.
The effective response of some pattern (b) means the overall behaviour of a composite made of that pattern.
Tangent linearization scheme is employed.
\label{Fig:Results_spheres}}
\end{figure}

First, we present the results for the MMC reinforced with ceramic spherical inclusions with the volume fraction $f_{\textup{i}}=0.3$.
Fig.\ref{Fig:Results_spheres} shows the MRP estimates of the equivalent Huber-von Mises stress as a function of the effective strain.
At the macro level isochoric tension process (Eq.\ref{Eq:BCTransStrainsText} $\mathbf{E}_1$) is assumed.
Let us assume that the MRP approach postulates that the composite microstructure can be described by three patterns (Fig.\ref{Fig:Results_spheres}.a).
The MRP model considers two groups of particles (with the volume fraction $f_{\textup{i}}/2$ each):
\begin{enumerate}
\item particles in a cluster, where the particles touch each other, $f_{\lambda}=0$,
\item disperse particles, which have the matrix packing ratio $f_{\lambda}/f_{\textup{m}}=1/2$,
\item the rest of the matrix phase.
\end{enumerate}
In Fig.\ref{Fig:Results_spheres}, the MRP approach distinguishes patterns responses:
pattern 1 (magenta line) has a much stiffer response than the macroscopic response of the composite (black line).
According to the MRP predictions particles in the clusters, like pattern 1 with $f_{\lambda}=0$, 
have higher stresses, which could cause composite failure in this area.
Response of the metal-matrix phase (dashed lines) depends on the pattern:
in pattern 1 the MRP predicts a stiffer response of the matrix phase than in pattern 2.
It is due to, that the volume fraction of the metal phase in pattern 1 is smaller than in pattern 2, $f_{\lambda}^1<f_{\lambda}^2$.
In other words, the ratio between metal layer to stiff inclusion is smaller $\left(f_{\textup{i}} f_{\textup{m}}/4\right)/\left(f_{\textup{i}}/2\right)$.
The MRP model evaluates the mean stress of the matrix, which is beyond the composites inclusions (pattern 3) as the lowest.

\begin{figure}[H]
\centering
\begin{tabular}{ccccc}
(a)&&(b)&(c)&(d)\\
\includegraphics[angle=0,height=4cm]{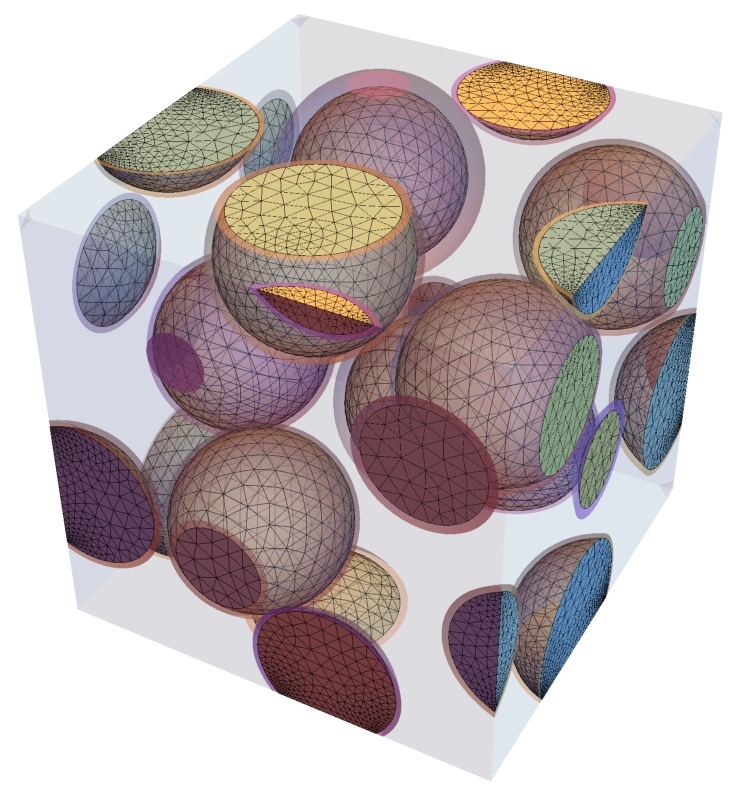}&
\includegraphics[angle=0,height=4cm]{strz.jpg}&
\includegraphics[angle=0,height=4cm]{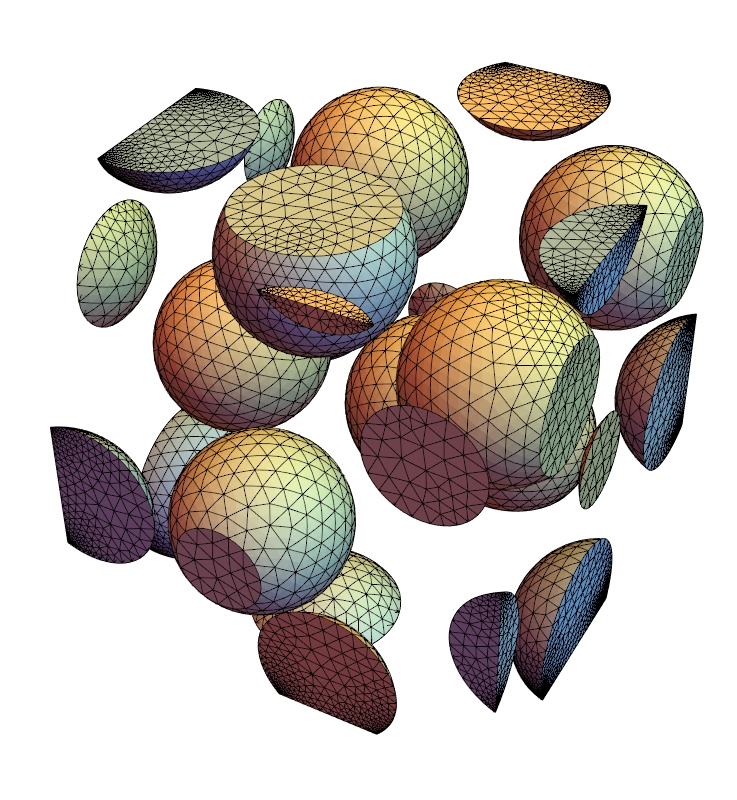}&
\includegraphics[angle=0,height=4cm]{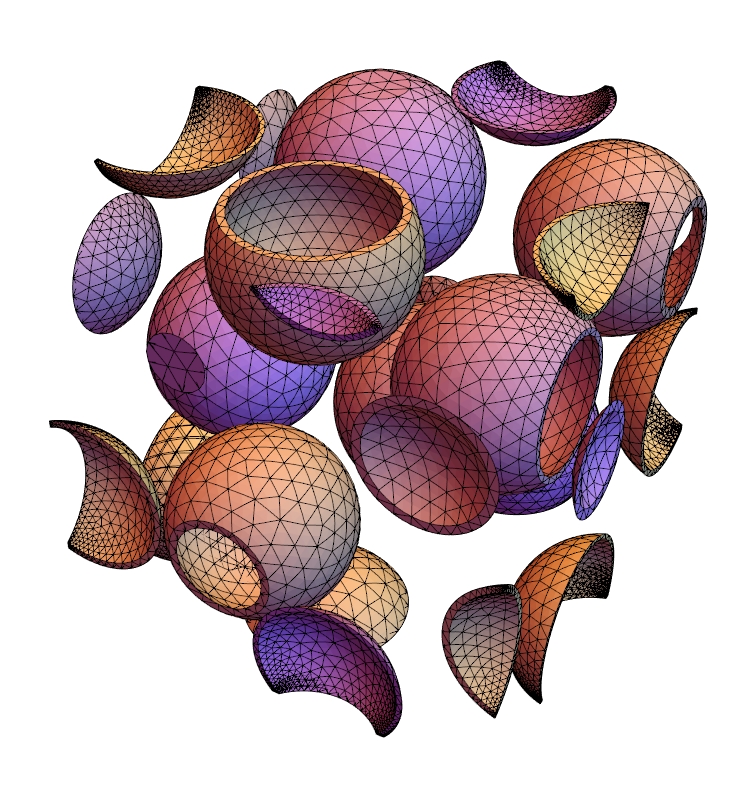}&
\includegraphics[angle=0,height=4cm]{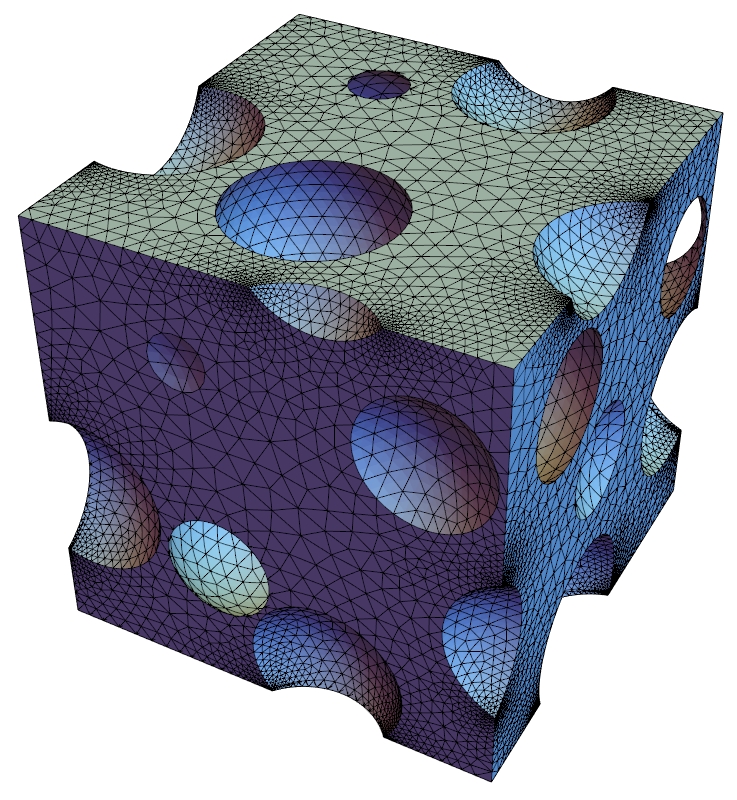}
\end{tabular}
\caption{FEM model of the SVE (a) of the MMC reinforced by ten balls. Each spherical particle (b), each coating $\lambda$ (c), and the rest of the matrix (d) are meshed separately .\label{Fig:FEM_10_spheres}}
\end{figure}

\begin{figure}[H]
\centering
\begin{tabular}{cc}
(a)&(b)\\
\includegraphics[angle=0,height=8cm]{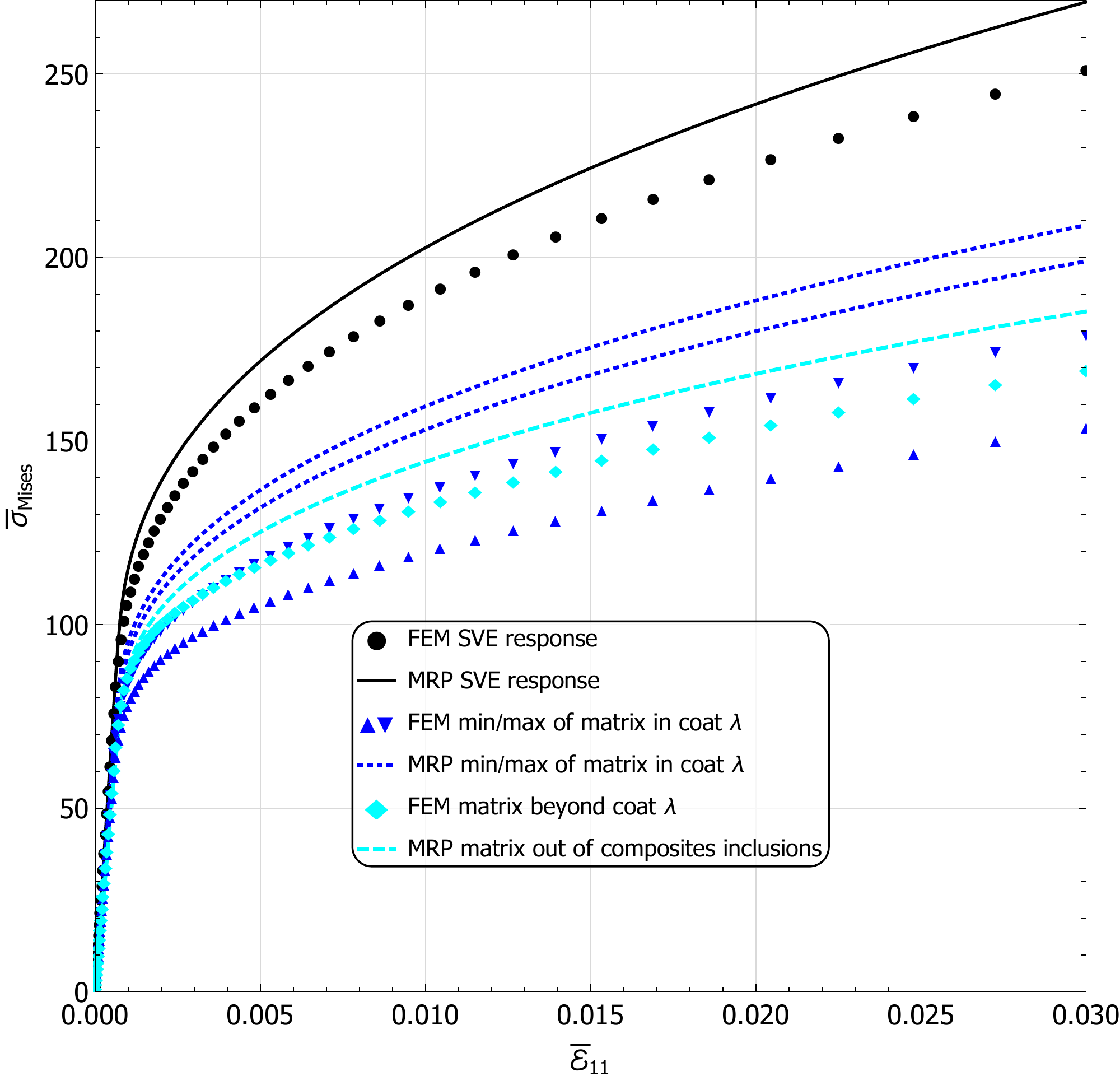}&
\includegraphics[angle=0,height=8cm]{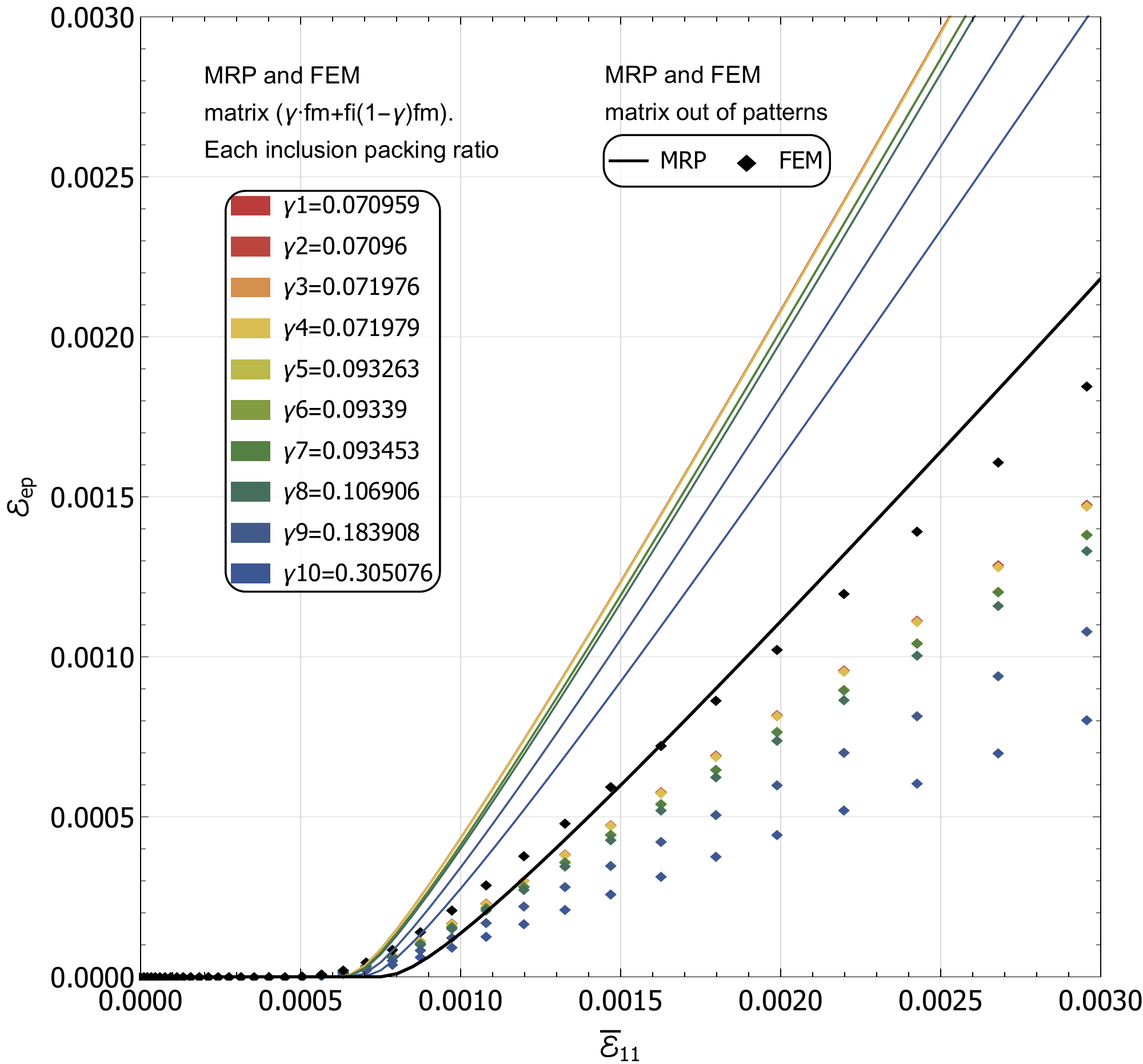}
\end{tabular}
\caption{
11-pattern MRP estimates (see schematic in Fig.\ref{Fig:MRP_10_spheres}) and numerical homogenisation (FEM SVE with 10 inclusions shown in Fig.\ref{Fig:FEM_10_spheres}) of the elastic-plastic response of the MMC with $f_{\textup{i}}=0.30$, in the isochoric tension test :
(a) overall and averaged per phase Huber-von Mises equivalent stress $\overline{\sigma}_{\textup{Mises}}$,
(b) accumulated plastic strain $\varepsilon_{\textup{ep}}$ vs. effective strain component $\overline{\varepsilon}_{11}$ in the direction of elongation.
Tangent linearization scheme is employed.
(a) FEM and MRP results of: the SVE response - black, matrix in the selected $\lambda$ coating (min/max according to FEM results) - blue, rest of the matrix - cyan.
(b) The plastic strain of the matrix of each of the composite inclusions 1 to 10. Legend colours sorted according to the packing ratio. The rest of the matrix phase - black.
\label{Fig:Results_spheres_Epl_Mis}}
\end{figure}

To better grasp the packing effect of particles, a random SVE (Fig.\ref{Fig:FEM_10_spheres}.a)  was meshed observing division into the following domains:
the spherical inclusions Fig.\ref{Fig:FEM_10_spheres}.b,
matrix coatings $\lambda$  around particles Fig.\ref{Fig:FEM_10_spheres}.c,
and the matrix beyond the coatings Fig.\ref{Fig:FEM_10_spheres}.d.
The mesh size was chosen so that the difference in the obtained results (Fig.\ref{Fig:Results_spheres_Epl_Mis}.a) is less than 3\% after a subsequent mesh refinement.
The volume fraction of the 10 inclusions in the volume is 30\%.
It should be noted, that each particle and each coating $\lambda$ is meshed and later simulated separately enabling easy assessment of its average response for future comparison with the MRP model predictions (Fig.\ref{Fig:MRP_10_spheres}).

Let us study a selected composite with the SVE shown in Fig.\ref{Fig:FEM_10_spheres}.a.
Fig.\ref{Fig:Results_spheres_Epl_Mis} presents the SVE's responses in the isochoric tension test.
In the MRP approach each of the 10 inclusions is represented by its 'pattern' (see the schematic in Fig.\ref{Fig:MRP_10_spheres}).
The MRP tangent linearization estimates the equivalent Huber-von Mises stress $\overline{\sigma}_{\textup{Mises}}$ closer to the FE computational results than the secant linearization.
Likewise, the tangent linearization gave a better agreement between the computational homogenisation and the micromechanical model in \citep{Majewski2020}.
Thus only the MRP results, obtained by the tangent linearization, are introduced in Fig.\ref{Fig:Results_spheres_Epl_Mis}.
Fig.\ref{Fig:Results_spheres_Epl_Mis}.a shows the equivalent Mises stress of two selected matrix coatings $\lambda$, which have the highest and lowest values of $\sigma_{\textup{Mises}}$ among all coatings $\lambda$.
The MRP outcomes are shown for selected matrix coatings $\lambda$, marked by 'min' and 'max' in Fig.\ref{Fig:Results_spheres_Epl_Mis}.a.
The MRP scheme estimates a lower difference between 'min' and 'max' $\lambda$ coatings than the numerical simulation.
The MRP approach estimates the response of the matrix beyond coatings $\lambda$ (Fig.\ref{Fig:MRP_10_spheres}.e) as the most compliant within the MMC (Fig.\ref{Fig:Results_spheres}.b and Fig.\ref{Fig:Results_spheres_Epl_Mis}.a).
However, the $\sigma_{\textup{Mises}}$ of the matrix beyond coatings $\lambda$ obtained by the FE simulation (Fig.\ref{Fig:FEM_10_spheres}.d) is between 'min' and 'max' for all ten coatings $\lambda$ (Fig.\ref{Fig:Results_spheres_Epl_Mis}.a light blue markers).

Fig.\ref{Fig:Results_spheres_Epl_Mis}.b presents the plastic strain $\varepsilon_{\textup{ep}}$ in each of the matrix coatings assigned to the 10 inclusions in the studied SVE (Fig.\ref{Fig:FEM_10_spheres}.a).
Each pattern is shown separately and sorted by the increasing matrix packing ratio $\gamma^{\alpha}= f_{\lambda}^{\alpha}/f_{\textup{m}}$ for $\alpha=1,..,10$ (legend in Fig.\ref{Fig:Results_spheres_Epl_Mis}.b).
%The linearization method only slightly affects the plastic strain $\varepsilon_{\textup{ep}}$.
%The reason may be the precision of the MRP algorithm calculation and the small range of deformation in Fig.\ref{Fig:Results_spheres_Epl_Mis}.b.
Predictions of FEM and MRP are in qualitative agreement.
In the MRP and FE simulations, the smaller the value of the matrix packing ratio $\gamma^{\alpha}= f_{\lambda}^{\alpha}/f_{\textup{m}}$ for the coating is, the higher the accumulated plastic strain becomes (Fig.\ref{Fig:Results_spheres_Epl_Mis}.b).
A plausible explanation is the higher stress level in the particles with a smaller matrix packing ratio.
Fig.\ref{Fig:Results_spheres} shows a similar trend for the limit cases $ f_{\lambda}/f_{\textup{m}}=0$ and $f_{\lambda}/f_{\textup{m}}=1/2$.
Although the plastic strains of the inclusions patterns differ between the MRP and computational homogenisation, the responses of the remaining matrix phase are comparable (Fig.\ref{Fig:Results_spheres_Epl_Mis}.b black lines and points).
 
\begin{figure}[H]
\centering
\begin{tabular}{cc}
\\
\includegraphics[angle=0,height=3cm]{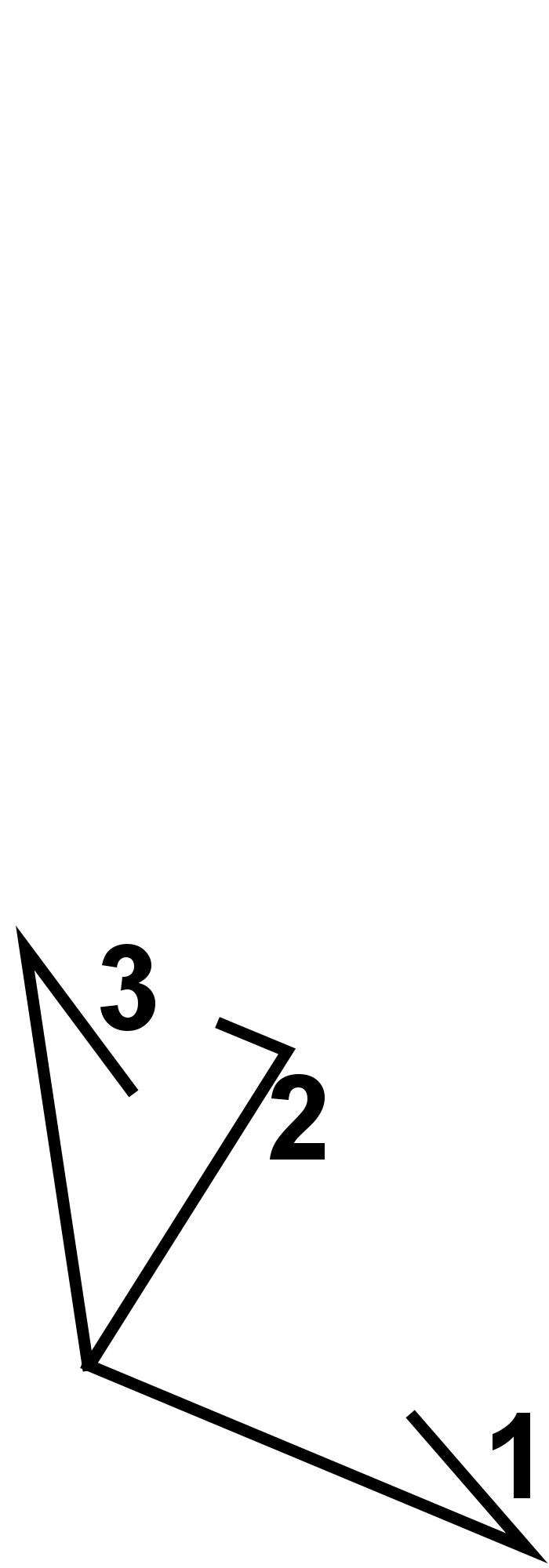}&
\includegraphics[angle=0,height=8cm]{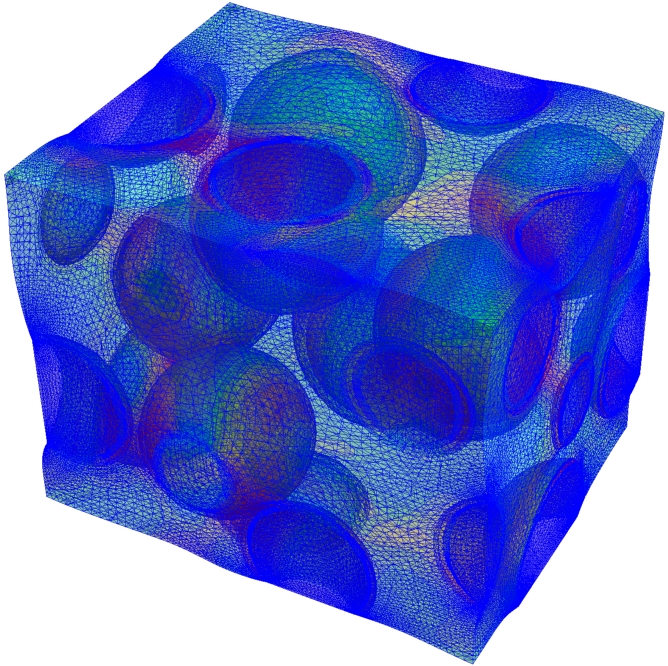}
\includegraphics[angle=0,height=4cm]{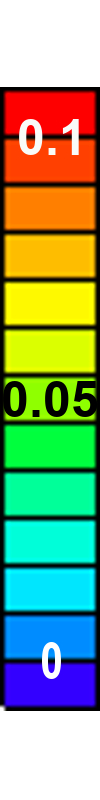}
\end{tabular}
\caption{
Accumulated plastic strain within the statistical volume element with a randomly generated structure (Fig.\ref{Fig:FEM_10_spheres}), with boundaries fulfilling periodicity.
The volume fraction of the 10 inclusions is 30\%.
\label{Fig:Results_Mises_plast}}
\end{figure}

Fig.\ref{Fig:Results_Mises_plast} demonstrates the spatial distribution of accumulated plastic strain $\varepsilon^{\textup{p}}_{\textup{eq}}$ in the studied SVE (Fig.\ref{Fig:FEM_10_spheres}).
The highest plastic deformation regions of the matrix phase with the highest $\varepsilon^{\textup{p}}_{\textup{eq}}$ occur between particles, especially in direction '1' of the isochoric tension test.
Fulfilment of the periodic BC is clearly seen in the deformed shape of the SVE.

\subsection{Various shapes of particles\label{SubSec:Shapes}}
\begin{figure}[H]
\centering
\begin{tabular}{cc}
(a)&(b)\\
\includegraphics[angle=0,height=8cm]{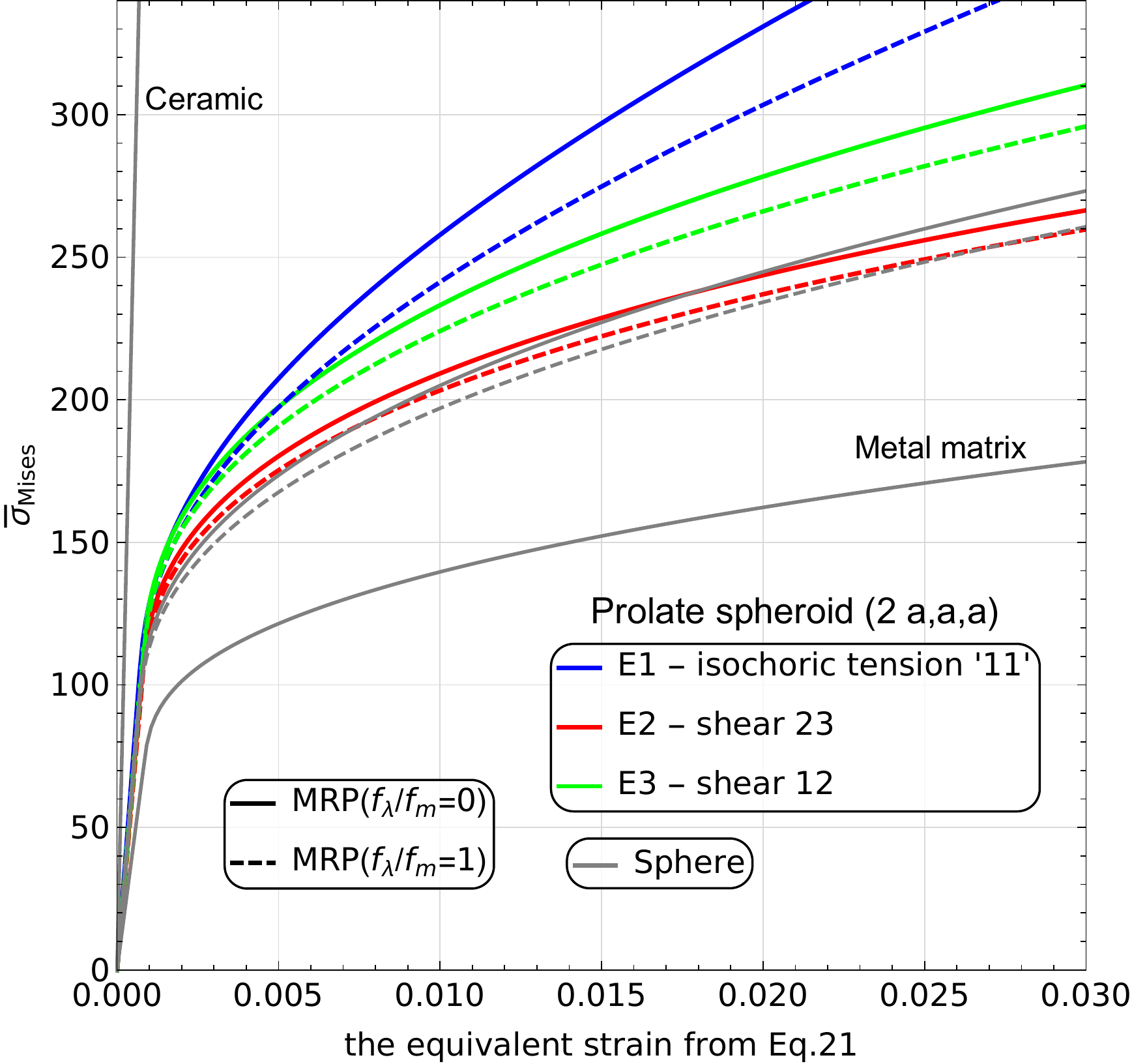}&
\includegraphics[angle=0,height=8cm]{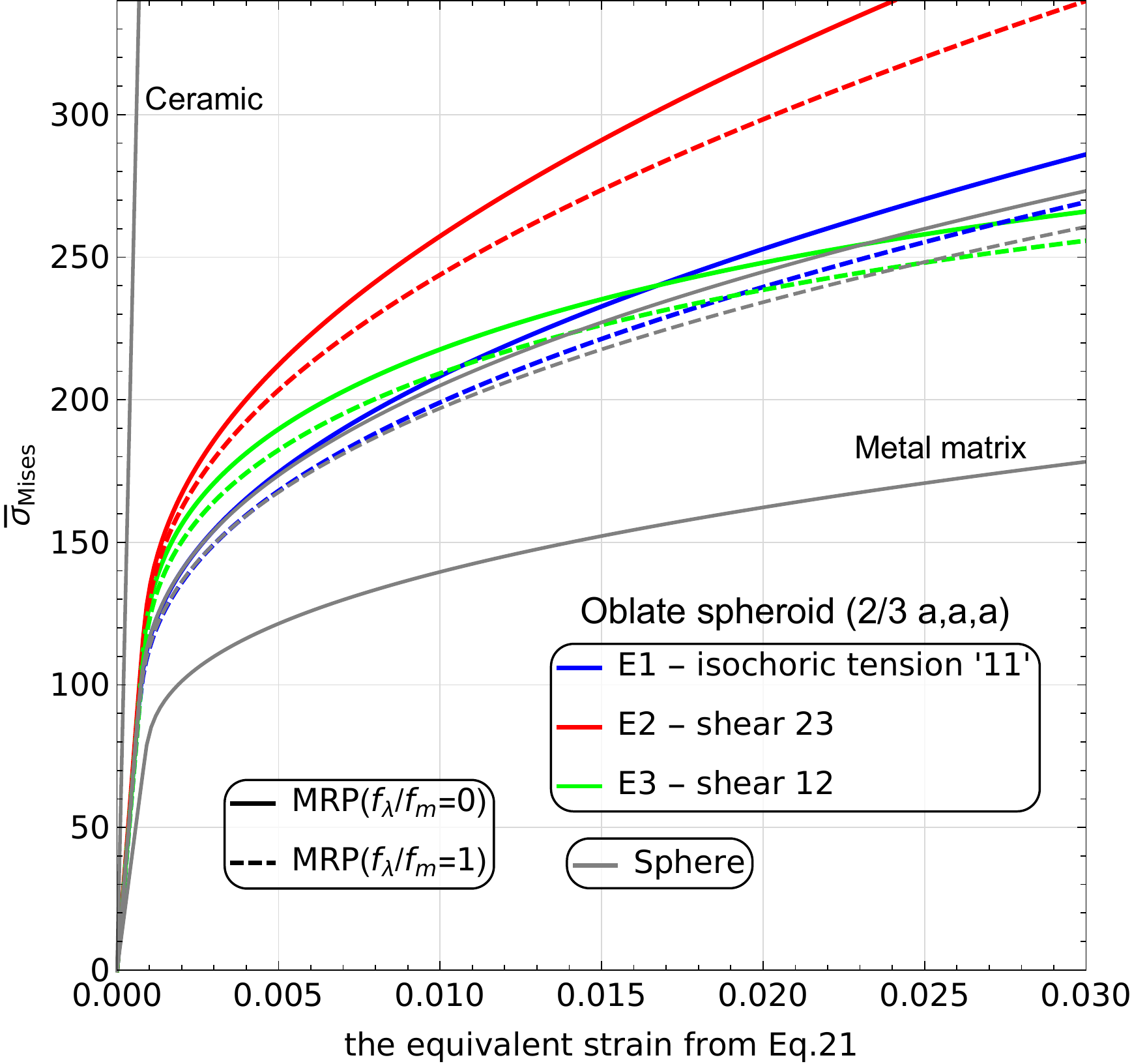}\\
(c)\\
\includegraphics[angle=0,height=8cm]{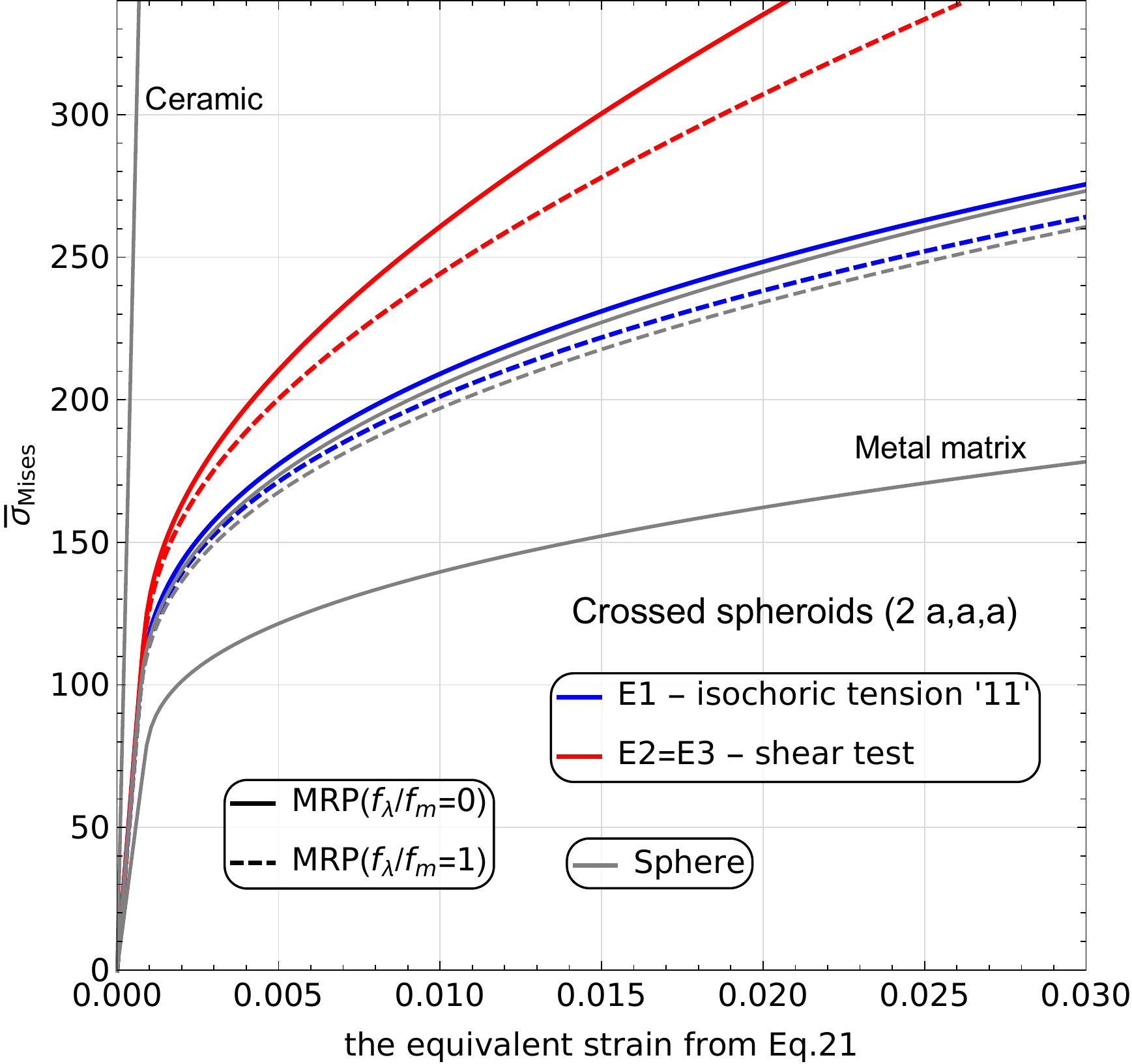}
\end{tabular}
\caption{
The MRP estimates of the MMC elastic-plastic response with $f_{\textup{i}}=0.30$: the Huber-von Mises equivalent stress $\overline{\sigma}_{\textup{Mises}}$, vs. the equivalent strain ((a) $d$, (b) and (c) $d\,2/\sqrt{3}$ from Eq.\ref{Eq:BCTransStrainsText}).
Two limit cases of the packing ratio are presented: $f_{\lambda}/f_{\textup{m}}=0$ and $f_{\lambda}/f_{\textup{m}}=1$.
Shapes of particles: (a) prolate spheroids, (b) drilled oblate spheroids, (c) crossed spheroids.
The composite response is based on the deviatoric components of the numerical concentration tensors: A1, A2, and A3 (Fig.\ref{Fig:Results_cons_tens_tests}-\ref{Fig:Results_cons_tens_shapes}), by the imposed strains in BC: E1, E2, and E3 (Eq.\ref{Eq:BCTransStrainsText}).
Only the tangent linearization scheme is presented.
\label{Fig:Results_shapes_Mis_MRP}}
\end{figure}

%Fig.\ref{Fig:Results_spheres} and Fig.\ref{Fig:Results_spheres_Epl_Mis} showed that the MRP tangent linearization gives results closer to the FE simulation outcomes than the MRP secant linearization.
%Thus the results of the MRP combined with tangent linearization are presented in further results (Fig.\ref{Fig:Results_shapes_Mis_MRP}-\ref{Fig:Results_shapes_phases_2_Mis_MRP}).
Fig.\ref{Fig:Results_shapes_Mis_MRP} demonstrates the MRP estimation of the MMC elastic-plastic response for $f_{\textup{i}}=0.30$.
The variation of the Huber-von Mises equivalent stress $\overline{\sigma}_{\textup{Mises}}$ with the equivalent strain is obtained for the isochoric tension and two pure shears tests ($\mathbf{E}_1$, $\mathbf{E}_2$, and $\mathbf{E}_3$ specified by Eq.\ref{Eq:BCTransStrainsText}).
In Fig.\ref{Fig:Results_shapes_Mis_MRP} the orientation of particles is the same as in Fig.\ref{Fig:Shapes}.
The MRP results (Fig.\ref{Fig:Results_shapes_Mis_MRP}) correspond to the symmetry of the dilute strain concentration tensors: $\mathbb{A}^{\textup{Cub}}$ (Eq.\ref{Eq:ACube}) and $\mathbb{A}^{\textup{Trans}}$ (Eq.\ref{Eq:ATrans}).
In Fig.\ref{Fig:Results_shapes_Mis_MRP}, the numerical concentration tensor of inclusions is approximated by the functions (Fig.\ref{Fig:Results_cons_tens_tests}-\ref{Fig:Results_cons_tens_shapes}).
During the elastic-plastic deformation, the tangent shear modulus of the matrix $G_{\textup{m}}$ evolves according to Eq.\ref{Eq:LinStyOsno2}.
So the variable $\chi$ in the concentration tensor function Eq.\ref{Eq:funANDilapprox} goes from the initial elastic state for which it is equal to 0.305 to the value around 0.0057-0.0046 for the strain magnitude $d=0.03$ (Eq.\ref{Eq:BCTransStrainsText}) for the studied composite microstructure.
%In the authors' opinion, the Huber-von Mises stress sequence in Fig\ref{Fig:Results_shapes_Mis_MRP} mainly bases on the initial concentration tensor for 0.305 from Fig.\ref{Fig:Results_cons_tens_tests}-\ref{Fig:Results_cons_tens_shapes}, which gives initial stiffness around $d<0.005$, and undergoes yielding during the process.
Note that the highest values of $\overline{\sigma}_{\textup{Mises}}$ for the various tests $\mathbf{E}_1$, $\mathbf{E}_2$, and $\mathbf{E}_3$ (Eq.\ref{Eq:BCTransStrainsText}) correlate with the order of the concentration tensor components for $\chi=0.305$.
As an example, the response of the prolate spheroid (Fig.\ref{Fig:Results_shapes_Mis_MRP}.a) is the stiffest for the isochoric tension test $\mathbf{E}_1$, which is related to $A_1$ (above $A_2$ and $A_3$ for $\chi=0.305$).
For the sake of comparison, the response of the MMC reinforced with ceramic balls computed using MRP with tangent linearization and shown in Fig.\ref{Fig:Mises_tan_sec_0_1_spheres}.c is repeated in Fig.\ref{Fig:Results_shapes_Mis_MRP}.
The MMC reinforced by the crossed spheroids (Fig.\ref{Fig:Shapes}.d) is always stiffer than that with spherical reinforcement (grey line in Fig.\ref{Fig:Results_shapes_Mis_MRP}.c).

\begin{figure}[H]
\centering
\begin{tabular}{cc}
(a)&(b)\\
\includegraphics[angle=0,height=8cm]{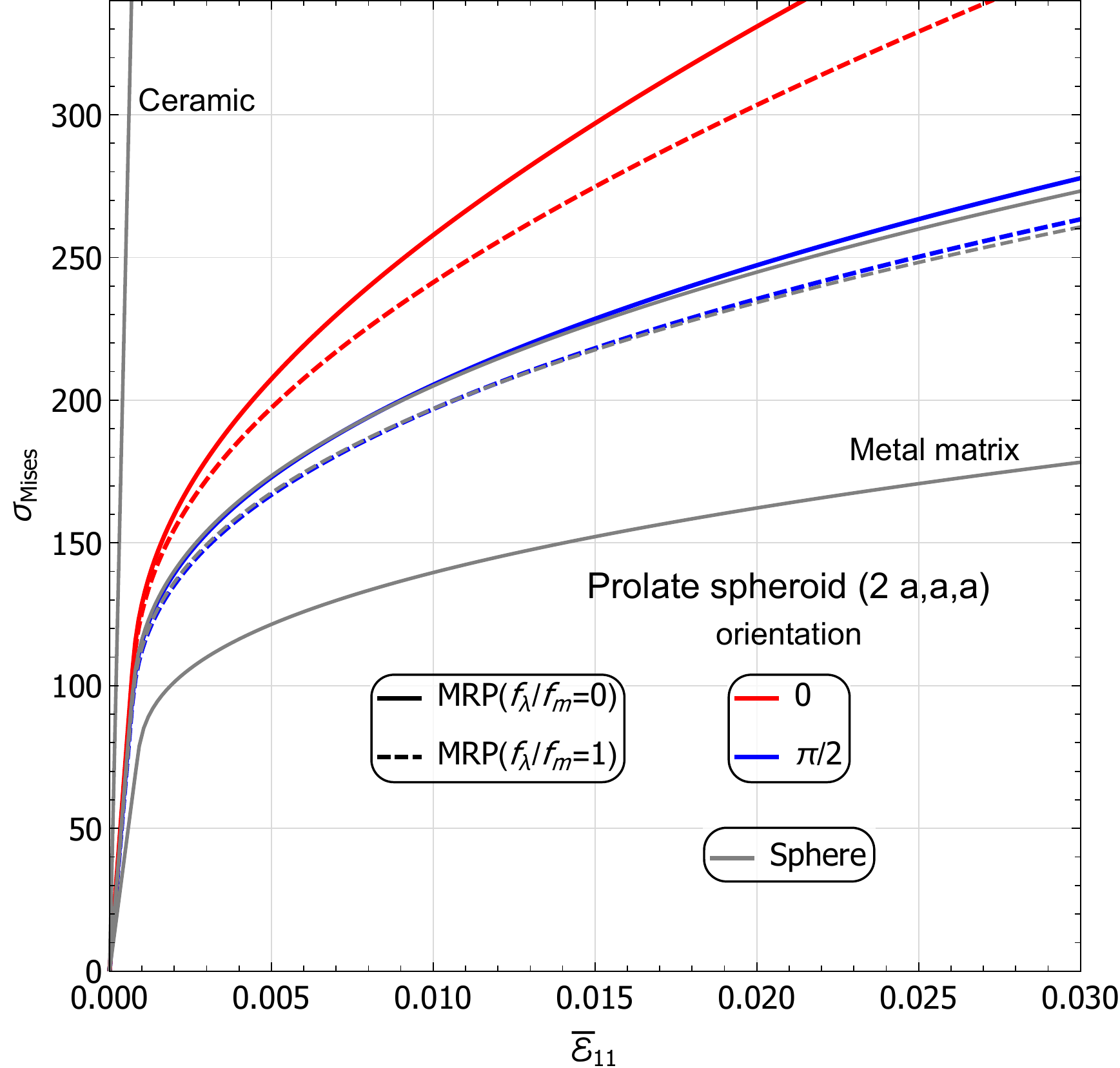}&
\includegraphics[angle=0,height=8cm]{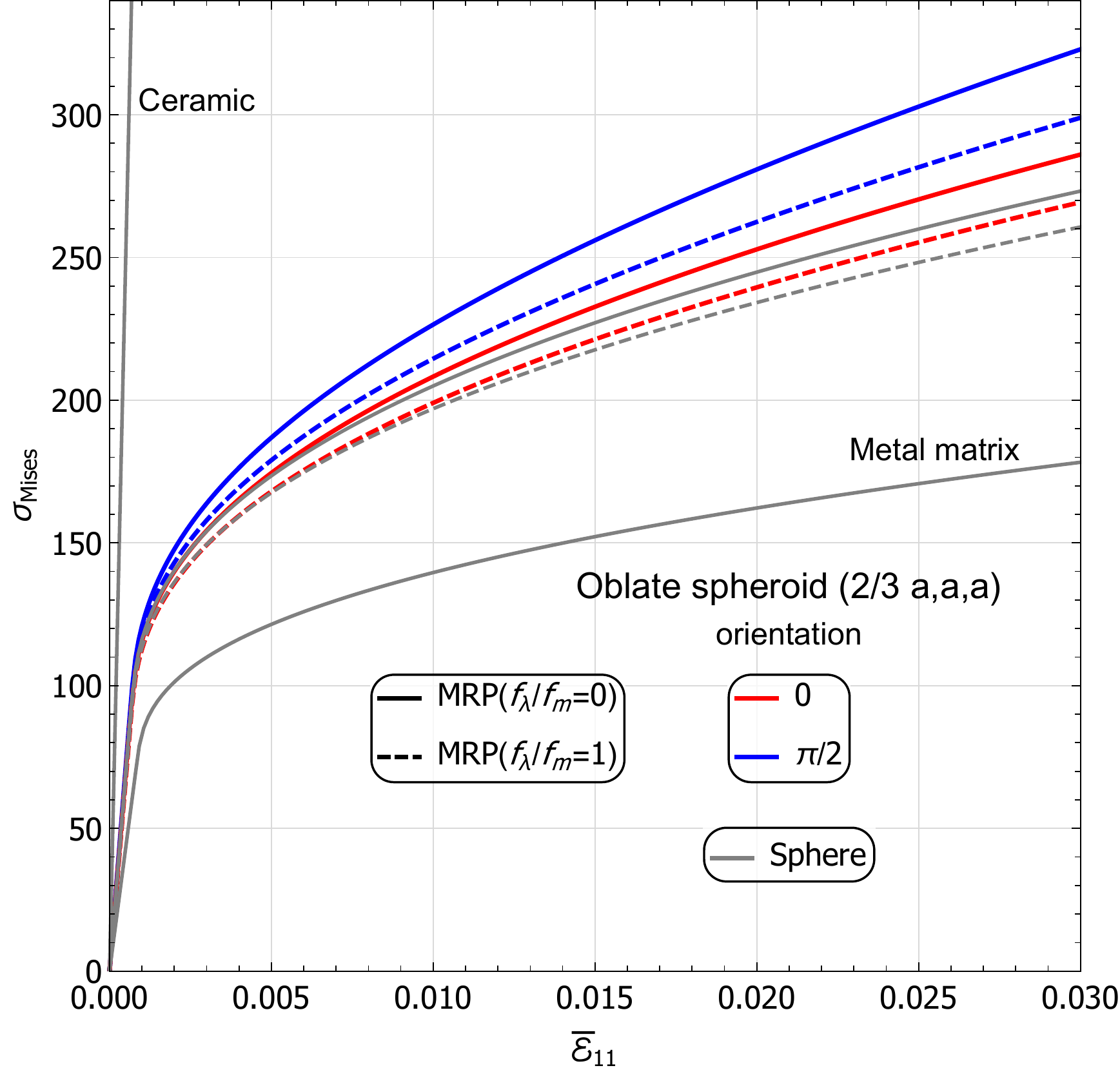}\\
(c)\\
\includegraphics[angle=0,height=8cm]{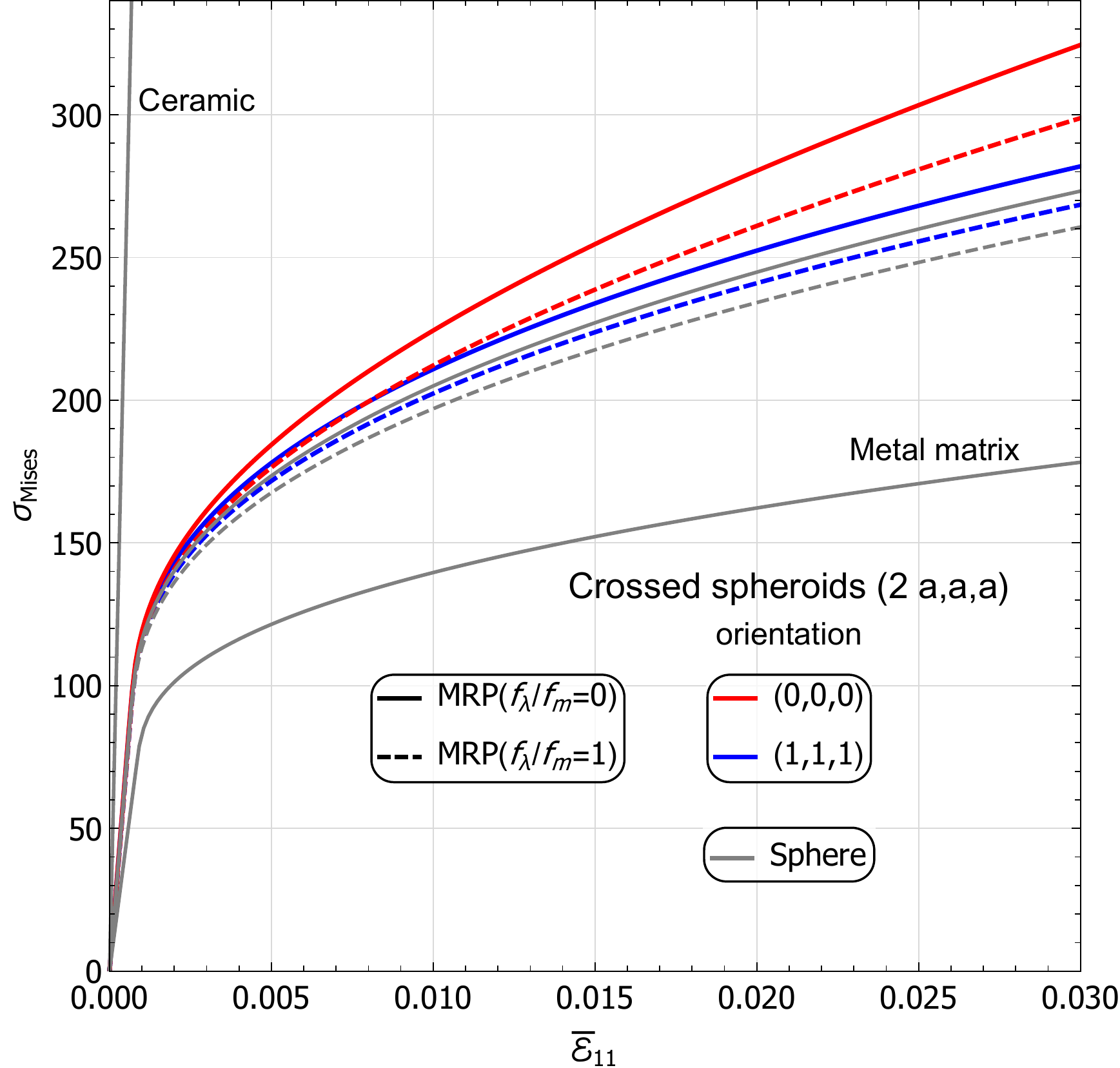}
\end{tabular}
\caption{
The Huber-von Mises equivalent stress $\overline{\sigma}_{\textup{Mises}}$, vs. the component $\overline{\varepsilon}_{11}$ of the isochoric tension test $\mathbf{E}_1$ (Eq.\ref{Eq:BCTransStrainsText}).
The MRP estimates the elastic-plastic response of the MMC (Tab.\ref{Tab:MMC}) with a fixed orientation of particles and $f_{\textup{i}}=0.30$.
Two limit cases of the packing ratio are presented: $f_{\lambda}/f_{\textup{m}}=0$ and $f_{\lambda}/f_{\textup{m}}=1$.
Particles shapes: (a) the prolate spheroids, (b) the drilled oblate spheroids, (c) the crossed spheroids, all having fixed orientations as described in the legend.
Only the tangent linearization scheme is presented.
\label{Fig:Results_shapes_Mis_MRP_2}}
\end{figure}

In Fig.\ref{Fig:FEM_RVE_Shapes}, the MMC is reinforced by randomly placed ceramic particles of fixed shape and various orientations.
The MRP scheme also takes into account the orientation of the inclusions.
Let us study the anisotropy of the MMC response when the ceramic particles are placed randomly but are aligned.
The composite is under the isochoric tension test $\mathbf{E}_1$ (Eq.\ref{Eq:BCTransStrainsText}) with the direction of tension changing with respect to the inclusions' axes.
Fig.\ref{Fig:Results_shapes_Mis_MRP_2} presents the MRP estimations for directions of isochoric tension delivering the highest and the lowest level of equivalent Huber-von Mises stresses.
The prolate spheroid has the stiffest response (Fig.\ref{Fig:Results_shapes_Mis_MRP_2}.a) for the elongation $\mathbf{E}_1$ along direction '1' (Fig.\ref{Fig:Shapes}.b), which is simultaneously the direction of the softest response of the oblate spheroid (Fig.\ref{Fig:Results_shapes_Mis_MRP_2}.b).
The three-crossed-spheroids particle (Fig.\ref{Fig:Results_shapes_Mis_MRP_2}.c) has the stiffest response along one of the main axes:1, 2 or 3 (Fig.\ref{Fig:Shapes}.d), and is the most compliant in the direction inclined at the same angle with respect to all main axes.
In all cases shown in Fig.\ref{Fig:Results_shapes_Mis_MRP_2}.a-c, the Huber-von Mises stress of the spherical particles (grey line) lies below the others.

\begin{figure}[H]
\centering
\begin{tabular}{cc}
(a)&(b)\\
\includegraphics[angle=0,height=8cm]{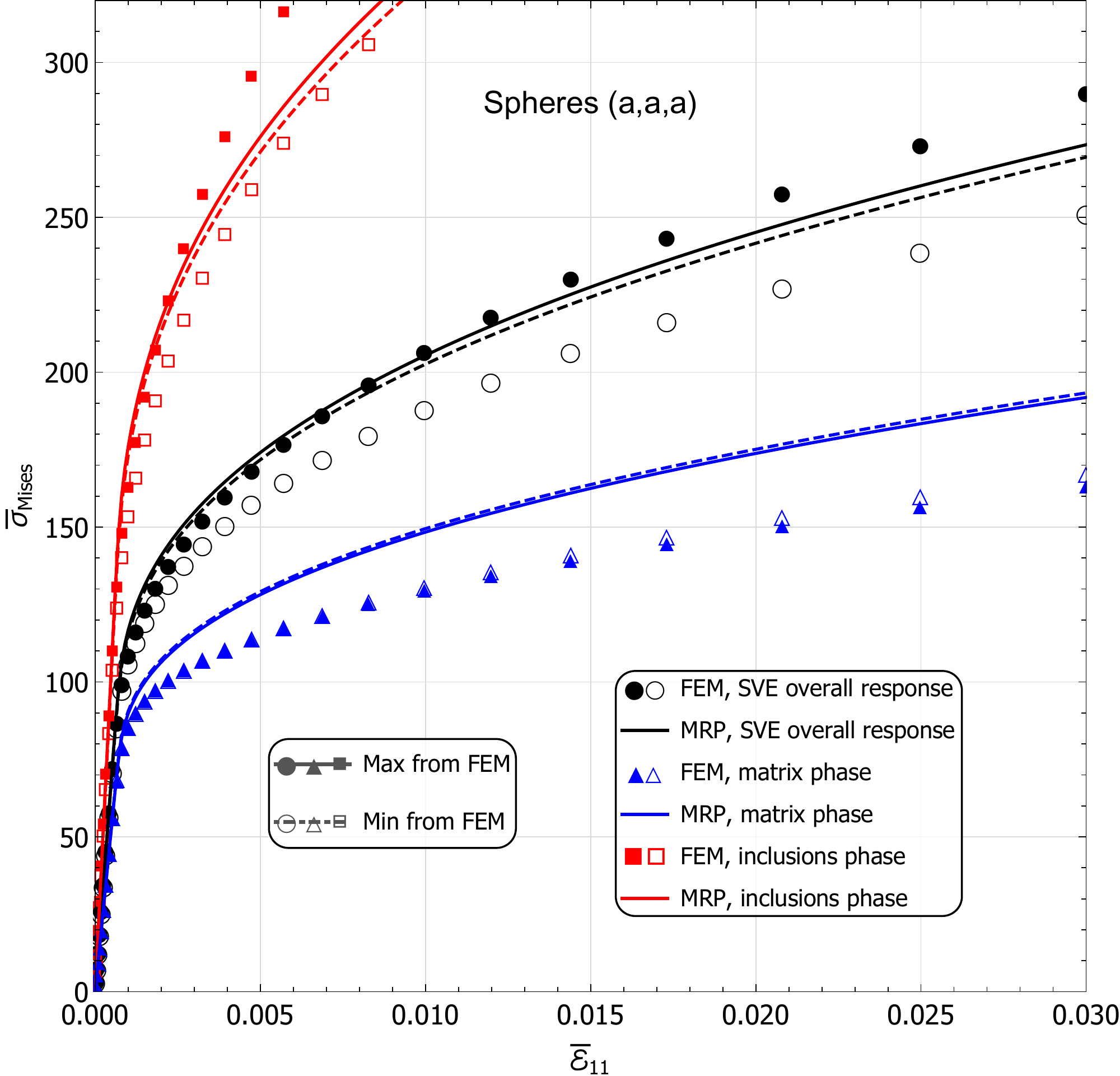}&
\includegraphics[angle=0,height=8cm]{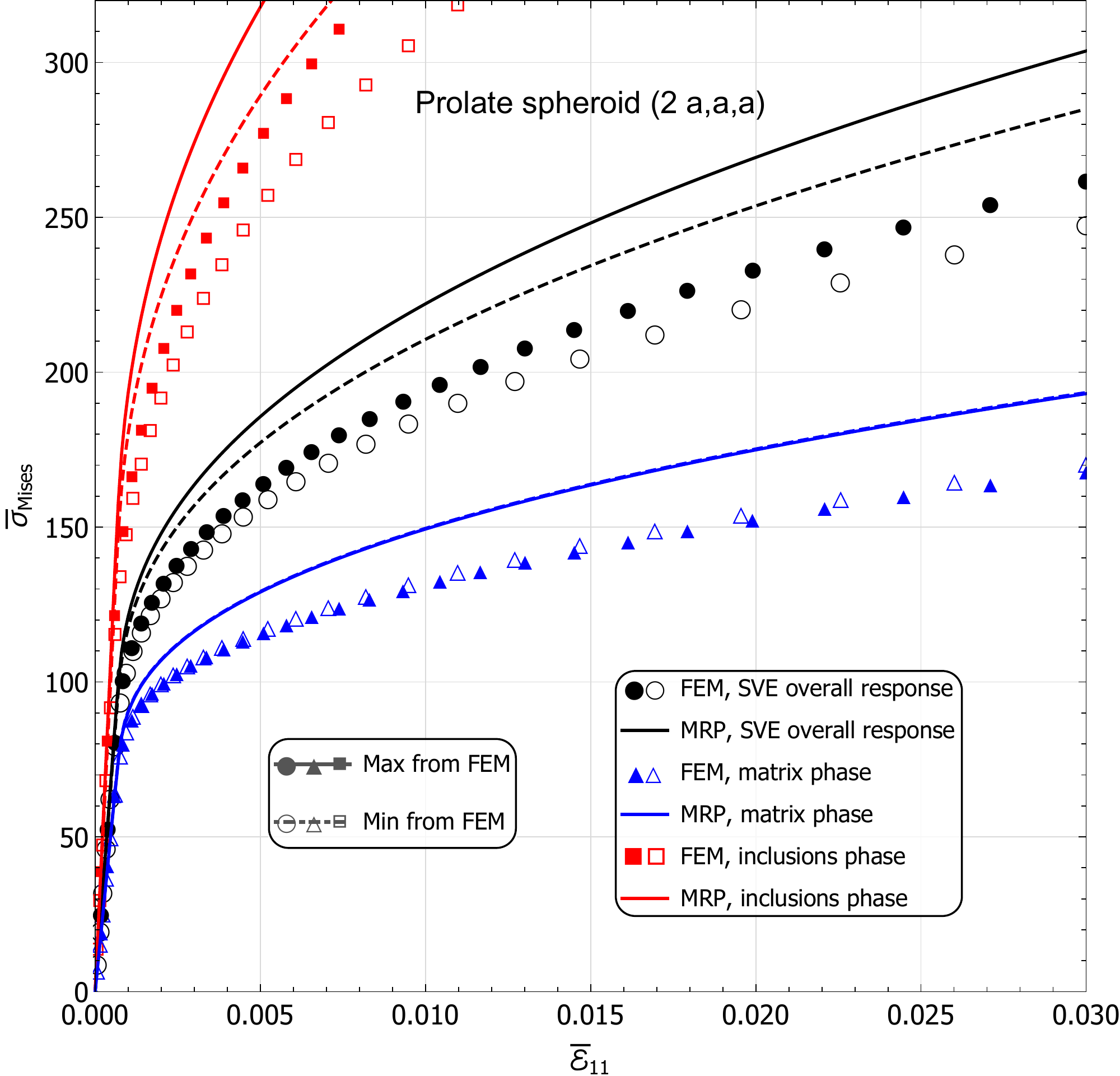}\\
(c)&(d)\\
\includegraphics[angle=0,height=8cm]{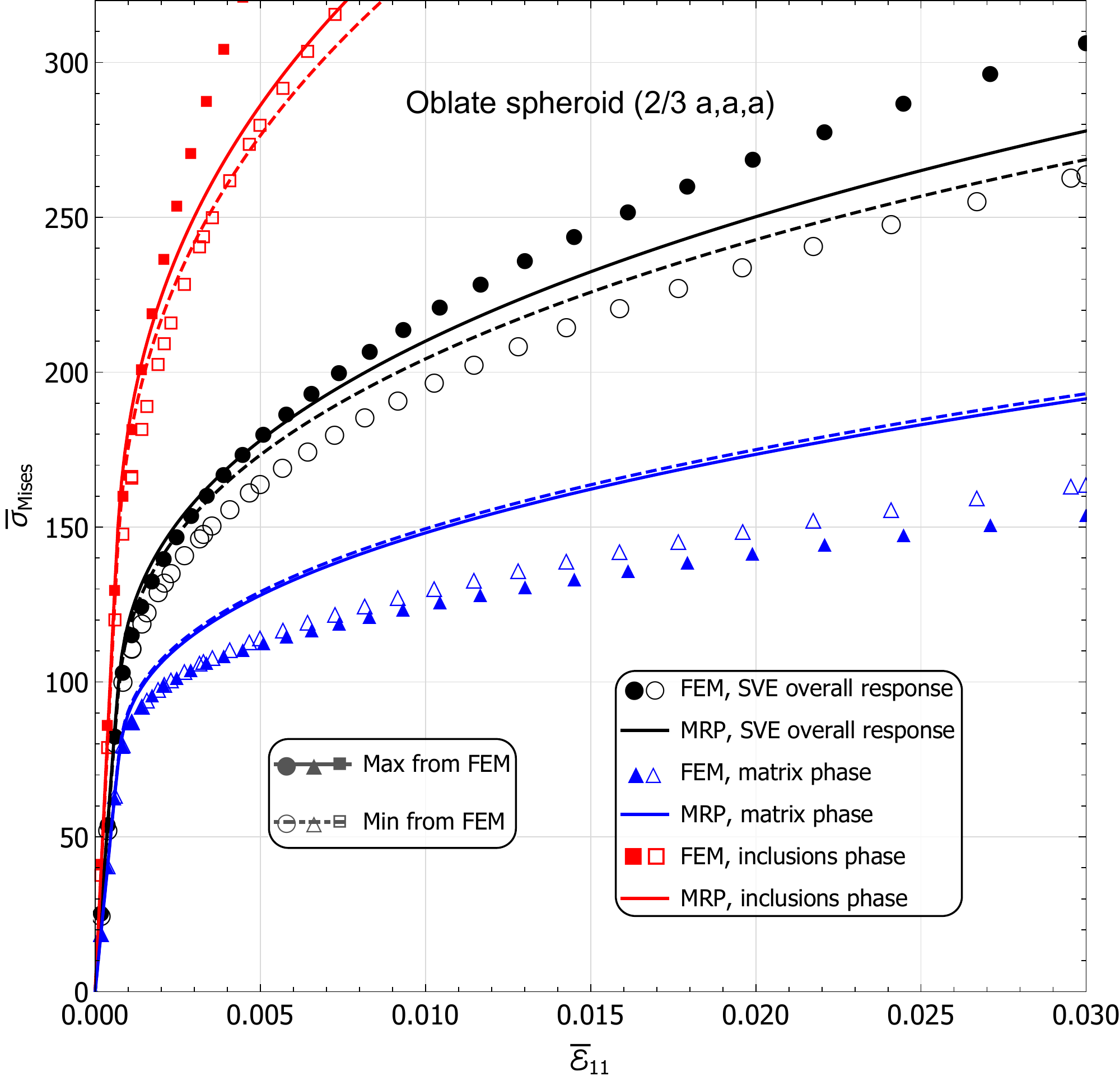}&
\includegraphics[angle=0,height=8cm]{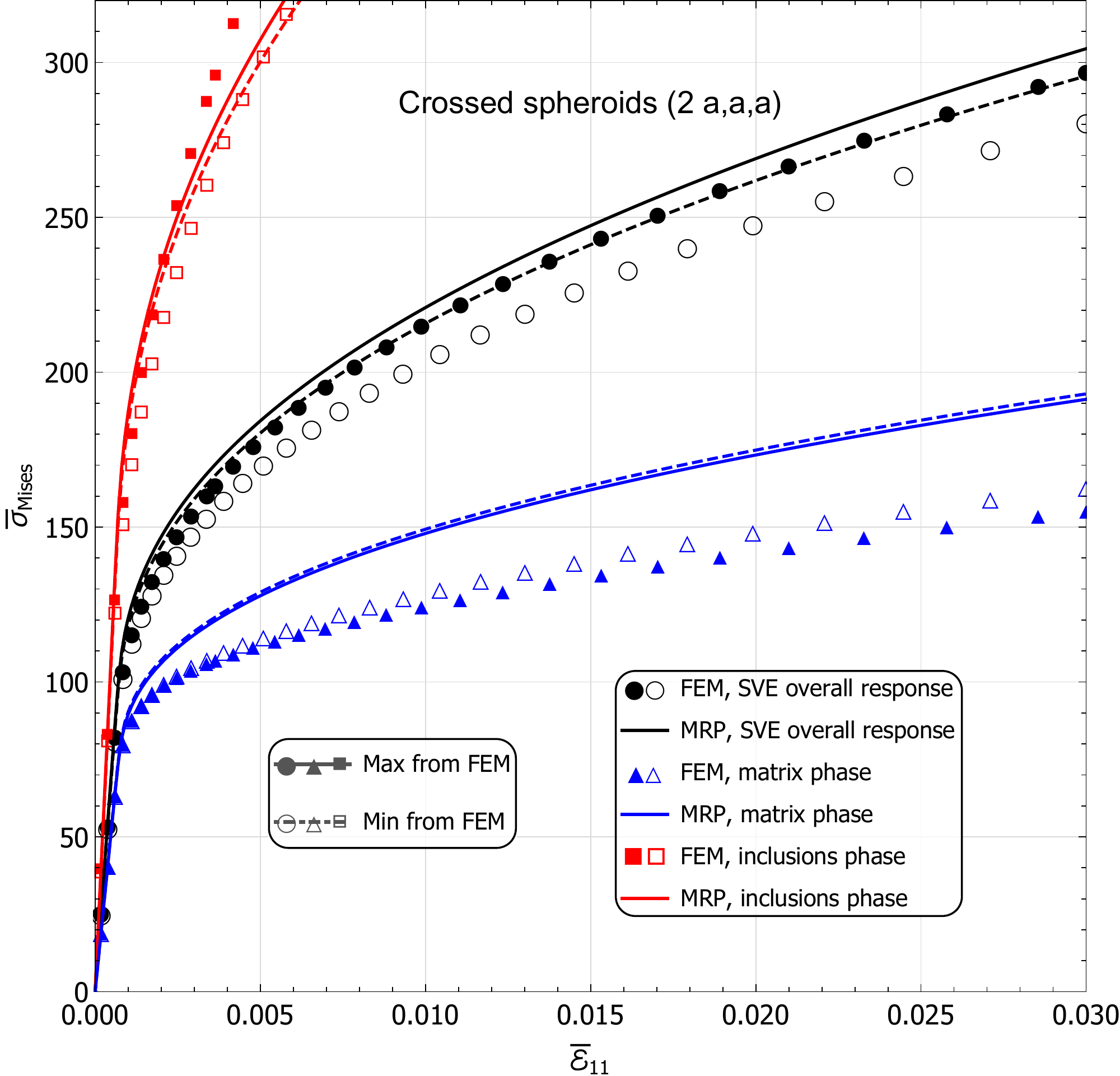}
\end{tabular}
\caption{
MRP and FE apparent elastic-plastic response of the MMC with $f_{\textup{i}}=0.30$ reinforced by ceramic particles of different shapes: (a) spheres, (b) prolate spheroids, (c) drilled oblate spheroids, and (d) crossed spheroids.
Results of the isochoric tension test: Huber-von Mises equivalent stress $\overline{\sigma}_{\textup{Mises}}$, vs. strain component $\overline{\varepsilon}_{11}$ in the direction of elongation.
For each shape 2 out of 30 random microstructures have been selected, which had the highest (abbr 'Max') and the lowest (abbr 'Min') overall Huber-von Mises stress at the strain magnitude $d=0.03$ in FEM simulations.
Fig.\ref{Fig:FEM_RVE_Shapes} presents structures with the smallest value of $\overline{\sigma}_{\textup{Mises}}$ for $\overline{\varepsilon}_{11}=0.03$.
Results for tangent linearization scheme are presented only in the case of MRP.
\label{Fig:Results_shapes_Mis_FEM}}
\end{figure}

\begin{figure}[H]
\centering
\begin{tabular}{c}
\includegraphics[angle=0,height=10cm]{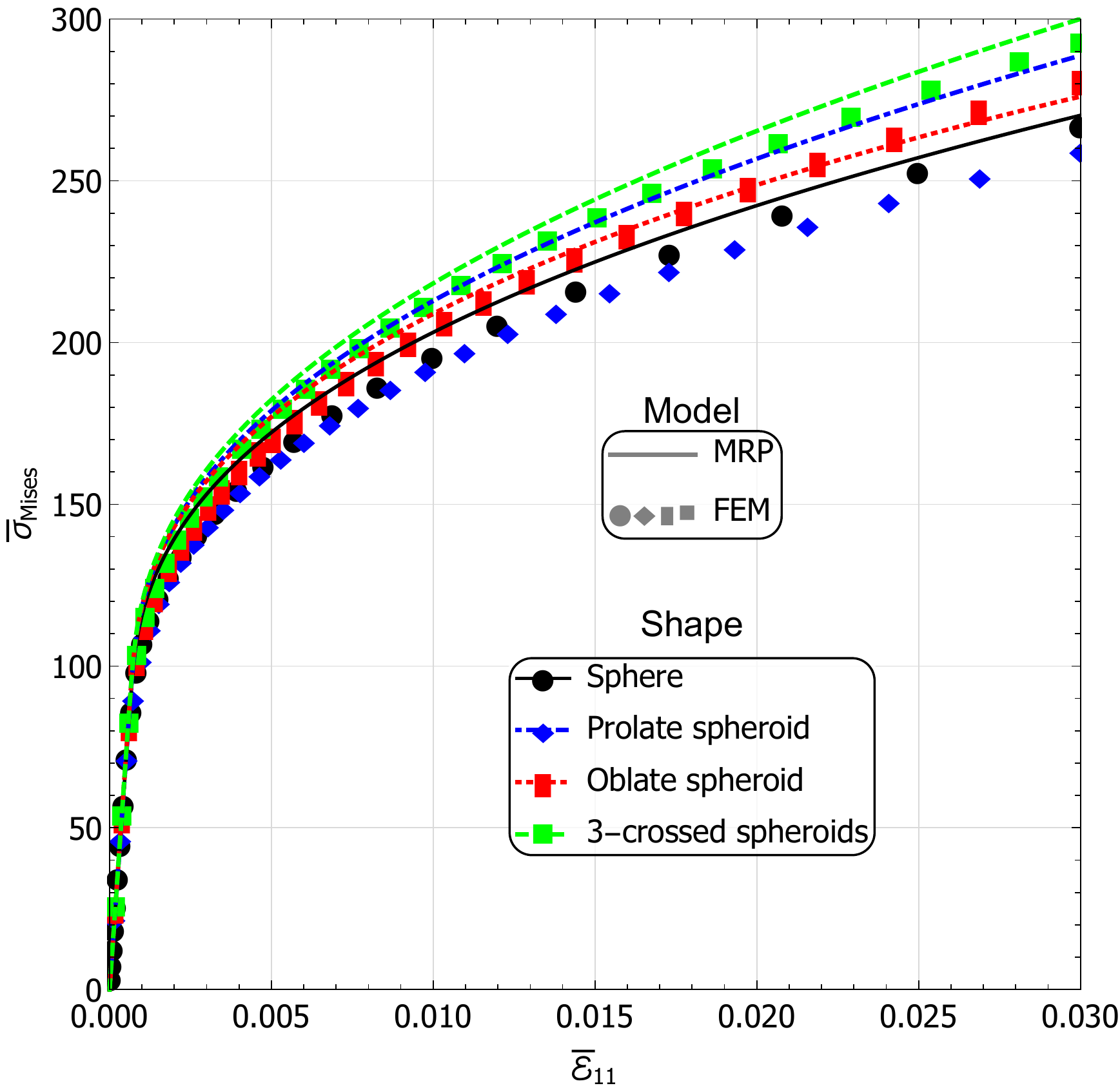}
\end{tabular}
\caption{
MRP estimates and results of FE simulation for the elastic-plastic effective response of the MMC with $f_{\textup{i}}=0.30$ reinforced by ceramic particles of different shapes: spheres, prolate spheroids, drilled oblate oblate spheroids, and crossed spheroids, according to the figure legend.
Results of the isochoric tension test: Huber-von Mises equivalent stress $\overline{\sigma}_{\textup{Mises}}$, vs. effective strain component $\overline{\varepsilon}_{11}$ in the direction of elongation.
The effective behaviour of the RVE is calculated as the average of the 30 SVEs apparent responses.
Tangent linearization scheme is employed.
\label{Fig:Results_shapes_Mis_FEM_eff}}
\end{figure}

Finally, the MRP model was verified by FE simulation of a fully random SVE.
The non-spherical particles are randomly placed and oriented.
The material parameters are listed in Tab.\ref{Tab:MMC}.
For each shape, thirty SVEs were generated and simulated using FEM.
For each shape two microstructures were selected from computational homogenisation, which had the highest (abbreviation 'max') and the lowest (abbreviation 'min') overall Huber-von Mises stress at the strain magnitude $d=0.03$ of the isochoric tension test $\mathbf{E}_1$ (Eq.\ref{Eq:BCTransStrainsText}).
Therefore, the Fig.\ref{Fig:Results_shapes_Mis_FEM} presents the apparent response of the MMC reinforced with ceramic particles of the same shape and size.
The effective behaviour of the MMC material is introduced in the Fig.\ref{Fig:Results_shapes_Mis_FEM_eff}.
The selected microstructures with the lowest stress 'min' are presented in Fig.\ref{Fig:FEM_RVE_Shapes}.
The mesh size was such that a subsequent mesh refinement gave a difference in the Huber-von Mises equivalent stress $\overline{\sigma}_{\textup{Mises}}$ of less than 4\%.
The selected microstructures with the highest stress denoted as 'max' have a higher overall matrix packing ratio than the microstructures with the lowest stress denoted as 'min'.
Fig.\ref{Fig:Results_shapes_Mis_FEM} presents the MRP estimates for the chosen 'max' and 'min' microstructures.
In all cases in Fig.\ref{Fig:Results_shapes_Mis_FEM}.a-d, the MRP evaluation of the matrix phase response is stiffer than that obtained in the respective FEM simulation.
The MRP model captures the fact that matrix stresses of 'min' microstructures are higher than those of 'max' microstructures (dashed lines are above solid lines in Fig.\ref{Fig:Results_shapes_Mis_FEM}).
The MRP estimates for the MMC reinforced by the prolate spheroids and the crossed spheroids are noticeably above the corresponding numerical results (Fig.\ref{Fig:Results_shapes_Mis_FEM}.b and d).
On the other hand, the apparent elastic-plastic behaviour of the MMC reinforced with spheres or drilled oblate spheroids is similar when modelled by the MRP scheme or computational homogenisation (Fig.\ref{Fig:Results_shapes_Mis_FEM}.a and c).
Except the prolate spheroids (Fig.\ref{Fig:Results_shapes_Mis_FEM}.b) the MRP model predicts a similar response in the inclusion phase as the numerical simulations (Fig.\ref{Fig:Results_shapes_Mis_FEM}.a, c and d).

In the previous figure, the apparent behaviour of the MMC is shown.
Fig.\ref{Fig:Results_shapes_Mis_FEM_eff} introduces the effective behaviour of the MMC with $f_{\textup{i}}=0.30$ reinforced by ceramic particles of different shapes: spheres, prolate spheroids, drilled oblate spheroids, and crossed spheroids.
The representative volume element response in Fig.\ref{Fig:Results_shapes_Mis_FEM_eff} is calculated as the average of the 30 statistical volume elements behaviour.
The MRP model well estimates the effective response of the MMC with 3-crossed spheroids, drilled oblate spheroids, and spheres.
The agreement between the MRP assessment and results of FE simulation of the MMC reinforced with the prolate spheroids could be improved.
Probably increasing the number of prolate spheroids in the volume element could improve results because from all studied shapes, the orientation of the prolate spheroids manifested the most significant influence on their response (Fig.\ref{Fig:Results_shapes_Mis_MRP_2}).

\begin{figure}[H]
\centering
\begin{tabular}{c}
\includegraphics[angle=0,height=10cm]{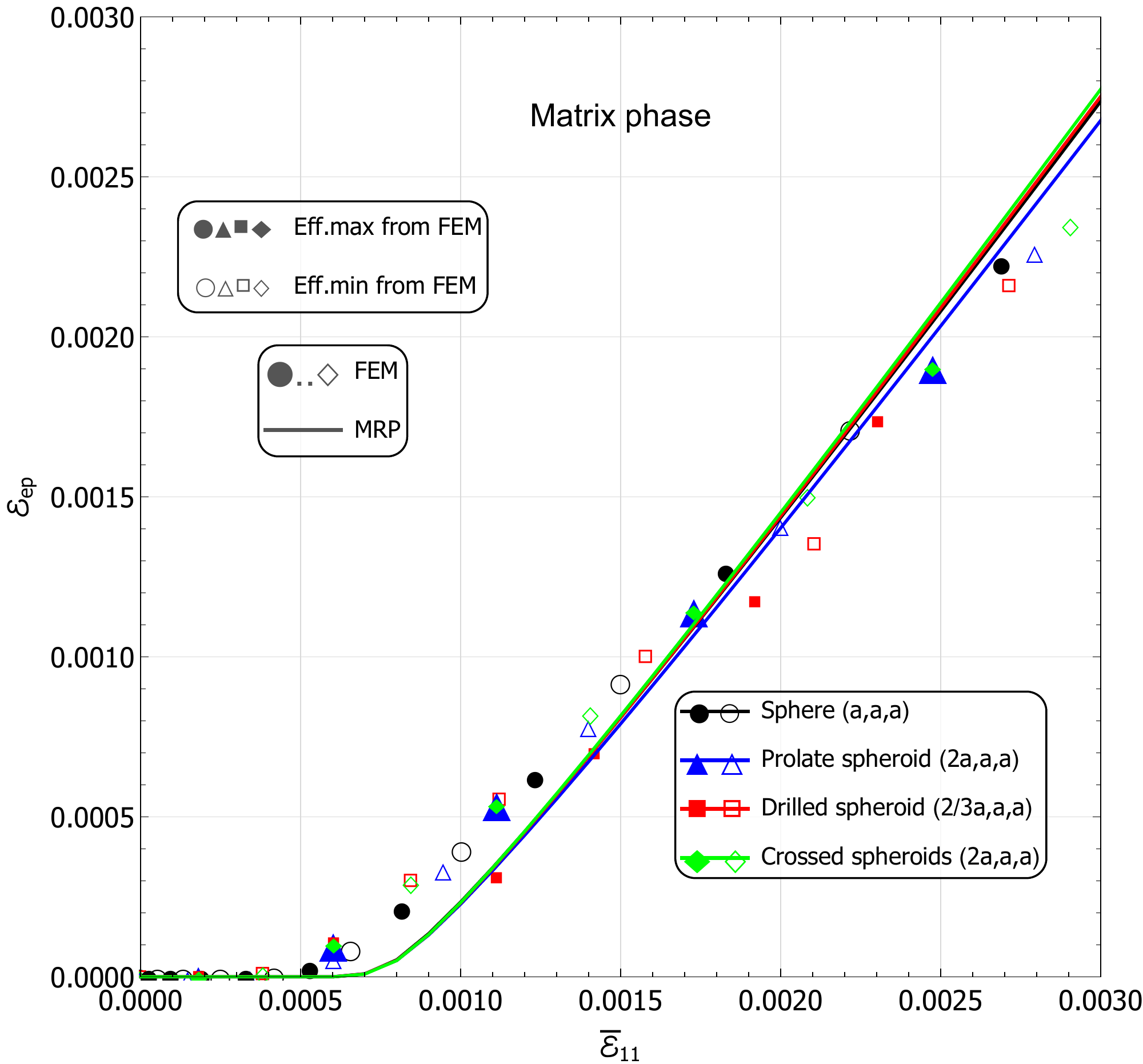}
\end{tabular}
\caption{
MRP estimates and FEM simulation of the elastic-plastic response of the MMC with $f_{\textup{i}}=0.30$ reinforced by different ceramic shapes: spheres, prolate spheroids, drilled oblate spheroids, and crossed spheroids.
Results of the isochoric tension test:
%(a) Huber-von Mises equivalent stress $\overline{\sigma}_{\textup{Mises}}$, (b) 
plastic strain $\varepsilon_{\textup{ep}}$ vs. strain component $\overline{\varepsilon}_{11}$ in the direction of elongation.
Min/max SVEs have been selected from 30 instances analyzed by FEM for each shape based on $\bar{\sigma}_{\textup{Mises}}$ at $\overline{\varepsilon}=0.03$
, Fig.\ref{Fig:FEM_RVE_Shapes} presents structures with the smallest value of $\overline{\sigma}_{\textup{Mises}}$ for $\overline{\varepsilon}_{11}=0.03$.
Only the tangent linearization scheme is presented.
\label{Fig:Results_shapes_phases_2_Mis_MRP}}
\end{figure}

\begin{figure}[H]
\centering
\begin{tabular}{ccccc}
&(a)&(b)&(c)&(d)\\
\includegraphics[angle=0,height=2cm]{UklWsp_2.jpg}&
\includegraphics[angle=0,height=3.5cm]{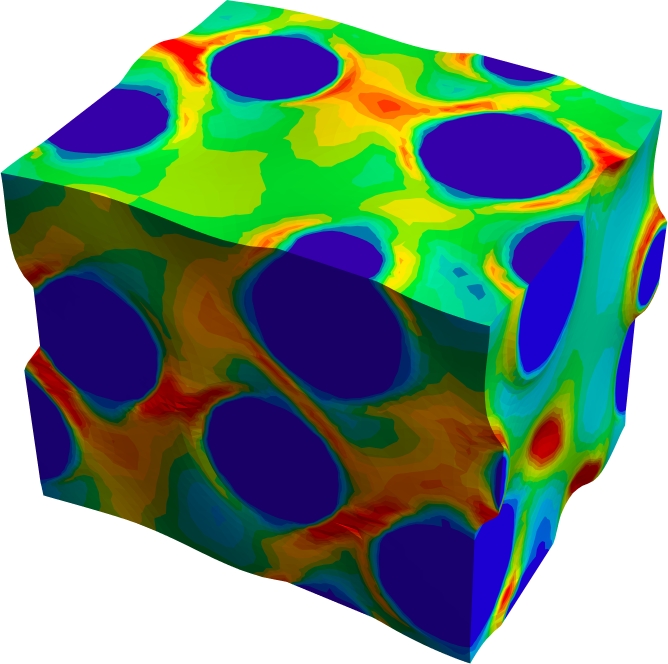}&
\includegraphics[angle=0,height=3.5cm]{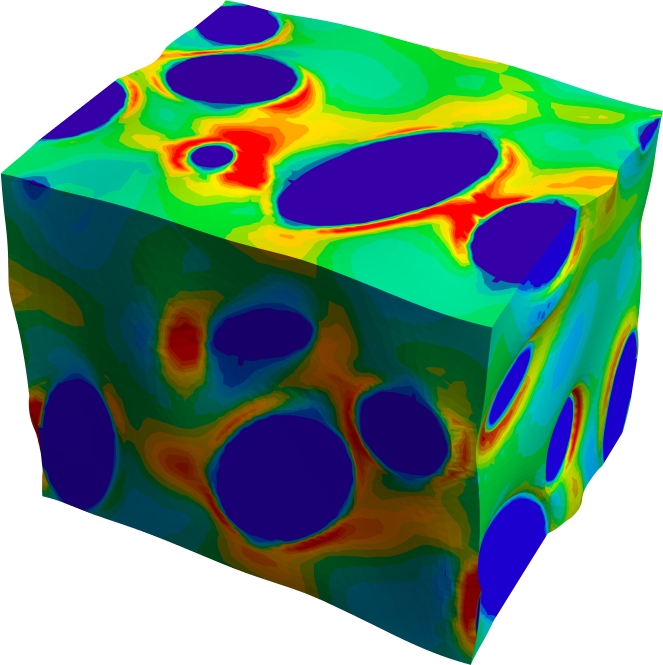}&
\includegraphics[angle=0,height=3.5cm]{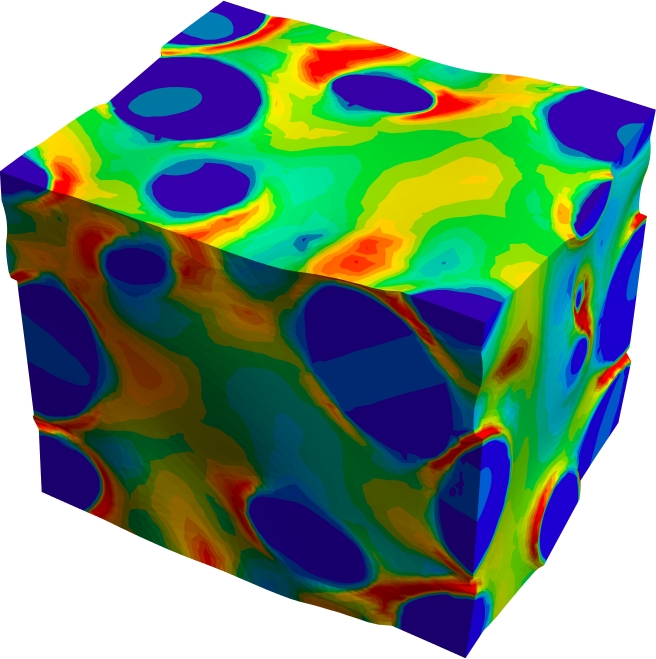}&
\includegraphics[angle=0,height=3.5cm]{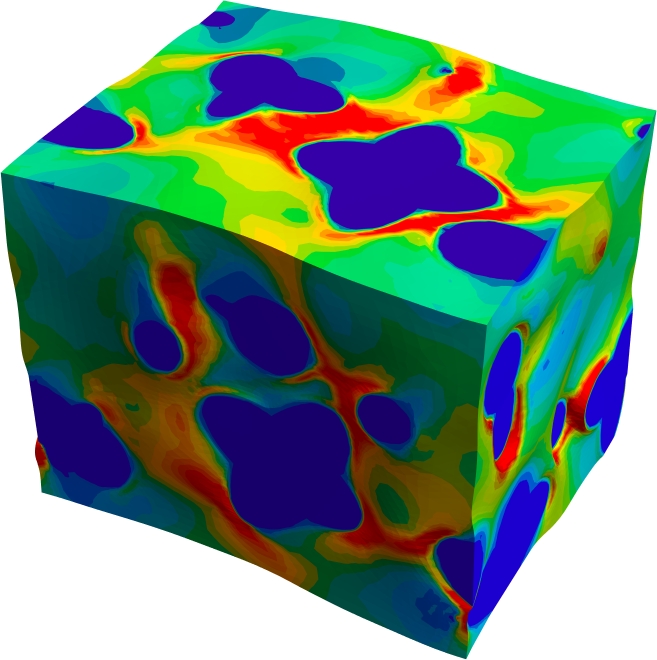}
\includegraphics[angle=0,height=3.5cm]{Legend01.jpg}
\end{tabular}
\caption{
Accumulated plastic deformation (legend 0-0.1) within the SVEs of a randomly generated structure (Fig.\ref{Fig:FEM_RVE_Shapes}) with boundaries fulfilling periodicity.
The actual deformation was enlarged five times to emphasize the periodic boundary conditions.
The volume fraction of the 10 inclusions in the volume is 30\%.
The SVE is subjected to isochoric tension in direction 1.
Countur maps are presented for the overall tensile strain $\varepsilon_{11}=0.03$.
\label{Fig:FEM_RVE_Shapes_Acc_plast}}
\end{figure}

Fig.\ref{Fig:Results_shapes_phases_2_Mis_MRP} presents the accumulated plastic strain $\varepsilon_{\textup{ep}}$ of the matrix phase in the studied SVEs (Fig.\ref{Fig:Results_shapes_Mis_FEM}),
and Fig.\ref{Fig:FEM_RVE_Shapes_Acc_plast} shows the accumulated plastic strain contour maps.
The MRP estimation of the plastic strain is similar for all shapes and both 'min' and 'max' microstructures (solid lines in Fig.\ref{Fig:Results_shapes_phases_2_Mis_MRP}).
The FE simulation results do not differ either.
The reason for the observed agreement could be that the average of the plastic strain over the whole volume of the matrix phase is presented.
To support this idea, note the presense of localization of the accumulated plastic deformation in Fig.\ref{Fig:FEM_RVE_Shapes_Acc_plast}, especially between particles along  the direction of elongation.
The current MRP approach, which relies on the mean-field approximation, does not account for this phenomenon.
These results indicate the course of future research.

\section{Conclusions}
%o modelu-MRP bazuje na RMTM i MT, jakie cechy mikrostr. opisuje
In this paper, a mean-field morphologically representative pattern approach is developed which is designed to estimate the elastic-plastic response of a Metal Matrix Composite reinforced by ceramic particles.
The MRP approach represents the MMC microstructure with the use of patterns.
Two models are employed to describe the MRP patterns: the Replacement Mori-Tanaka Model and the Mori-Tanaka (MT*) scheme.
The RMTM pattern describes the shape and orientation of an inhomogeneity in a computationaly efficient numerical manner.
In the RMTM, the dilute local concentration tensor is calculated using the finite element method.
The RMTM pattern used in the present work included composite inclusions made of a particle surrounded by two coatings: the first one defined by the minimum distances $\lambda_k$ between inclusions, the second one by the volume fraction of particles and the matrix outside the $\lambda$-coatings region.
Thus the minimum distances between inclusions specify the packing of the inclusions.
The inclusions' influence on the surrounding matrix is accounted for by the second coating.
In the MT* pattern, a medium of the composite inclusions surrounds the remaining matrix.
Therefore, the MT* pattern models the interaction between composite inclusions.

The incremental linearization proposed by Hill is employed in two variants: tangent and secant, depending on the definition of the current stiffness tensor.
The metal matrix is assumed to be a ductile material with linear elasticity and the Huber-von Mises yield function with the associated flow rule.
The inclusion phase is considered to be a linearly elastic material with parameters relevant to a ceramic.
As it is commonly observed \citep{Kursa2018, Majewski2020}, the mean-field model extended to elastoplasticity by the secant linearization predicts a stiffer response of the MMC than when applying the tangent linearization.

The mean-field models, including the MRP approach, were verified by computational homogenization performed using FEM.
Statistical volume elements with random distributions of ten particles and periodic boundary conditions were used.
Four different shapes of ceramic particles were studied: a sphere, a prolate spheroid, a drilled oblate spheroid filled with matrix phase, and three prolate spheroids crossing at right angles (Fig.\ref{Fig:Shapes}).

The shape of inclusions strongly affects the MMC response estimated by MRP or FEM.
As it was expected, the response of the MMC reinforced with prolate spheroids strongly depends on the orientation of particles (Fig.\ref{Fig:Results_shapes_Mis_MRP_2}.a).
Furthermore, the matrix in oblate spheroids is prevented by stiff ceramic particle from extensive plastic strain accumulation.
This phenomenon is observed in the FEM simulation (Fig.\ref{Fig:FEM_RVE_Shapes_Acc_plast}.c), and it corresponds to the MRP results of the MMC reinforced with oblate spheroids (Fig.\ref{Fig:Results_shapes_Mis_FEM}.c).
In the case of highly irregular shapes of the particles, the numerical simulations of the concentration tensor could be more challenging.
Especially when there is significant unevenness of the surface of the particle, which makes mesh quality an important factor.
On the other hand, it is substantially easier to simulate single irregular inhomogeneities than an RVE with many irregular inclusions.
In general, the novel MRP approach estimates well the impact of the shape of the particles on the effective MMC response (e.g. Fig.\ref{Fig:Results_shapes_Mis_FEM_eff}).

The packing effect was studied in detail for the spherical particles (Fig.\ref{Fig:Keff_Geff_elastic_packing} and Fig.\ref{Fig:Smises_plastic_packing}).
Although quantitatively the MRP results concerning the predicted Huber-von Mises stress are higher than the FEM outcomes (Fig.\ref{Fig:Results_spheres_Epl_Mis}.b), the packing impact in both approaches has the same character: smaller values of the packing ratio give a stiffer response of the MMC.

One may draw the inference from the presented model that the novel MRP approach improves classical micromechanical estimates in the case of moderate volume contents or non-spherical shapes of inclusions, especially for the non-linear response of the particulate composites.
The shape of particles may be used as a design parameter:
the optimal inclusions' shape can be found using the multi-objective optimization procedure presented in \citep{Kursa2014}.
Furthermore, it is foreseen that the MRP approach may be particularly efficient with regard to particulate composites with damage evolution.
The local fracture in ceramics and delaminations along the interface could be considered in the MRP approach by modifying the function approximating the concentration tensor.
The advantage of the presented scheme is analyzing each inclusion separately (each pattern represents a single inclusion).
With this approach, the appearance of local damage should be modelled by the presented MRP.
Therefore, an extension of the presented MRP approach to the non-linear regime with damage evolution is a subject of our current research.
In further research crack propagation at the inclusion-matrix interface is supposed to be accounted for as well.
\appendix
\section*{Declaration of competing interest}
The authors declare that they have no known competing financial interests or personal relationships that could have appeared to influence the work reported in this paper.

\section*{Data availability}
Data will be made available on request.

\section*{Funding}
The research was partially supported by the project No. 2017/25/N/ST8/01968 of the National Science Center (NCN), Poland.

\section*{Acknowledgements}
I would like to acknowledge my collaborators from IPPT PAN: prof. K. Kowalczyk-Gajewska and dr P. Holobut for providing language help.

\section{Model implementation\label{Ap:Model}}
Let us introduce the model implementation in a step-by-step manner on an example of the RVE with volume $V^{\textup{RVE}}$ and $N_{\textup{i}}$ inhomogeneities of the volume fraction $f_{\textup{i}}$.
The number of the MRP patterns $M$ is equal to $N_{\textup{i}}+1$ if each inhomogeneity is modelled separately.
Superscript $\alpha$ defines the MRP pattern $\alpha=1,...,M$.
In this publication, $\alpha=1,...,M-1$ are the RMTM patterns of the composite inclusions and $\alpha=M$ is the MT* pattern (e.g. Fig.\ref{Fig:MRP_10_spheres}).
$c^{\alpha}=V^{\alpha}/V^{\textup{RVE}}$ is the volume fraction occupied by the pattern $\alpha$.
Subscript $k$ defines a phase in the pattern:
\begin{itemize}
\item for $\alpha=1,...,M-1$: $k=1$ inclusion, $k=2$ the packing coating $\lambda$, and $k=3$ the inclusion influence coating $f_{\textup{i}}^\alpha \left(f_{\textup{m}}-f_{\lambda}\right)$, where $f_{\lambda}=\Sigma_{\alpha=1}^{M-1} f_{\lambda}^\alpha$,
\item for $\alpha=M$: $k=1$ matrix outside composites inclusions, $k=2$ the effective influence of the composites inclusions.
\end{itemize}
$f_k^{\alpha}=V_k^{\alpha}/V^{\alpha}$ and $\mathbb{L}_k^{\alpha}$ are the volume fraction and the stiffness tensor, respectively, of the phase $k$ in the subvolume $V^{\alpha}$.
In the paper, in the elastic regime, the stiffness tensors of phases are equal: $\mathbb{L}_1^{\alpha}=\mathbb{L}_{\textup{i}}$, $\mathbb{L}_2^{\alpha}=\mathbb{L}_{\textup{m}}$, $\mathbb{L}_3^{\alpha}=\mathbb{L}_{\textup{m}}$ for $\alpha=1,..,M-1$, and $\mathbb{L}_1^{M}=\mathbb{L}_{\textup{m}}$.
\begin{enumerate}
\item Calculate the composites inclusions $\alpha = 1, . . , M-1 $ (e.g. Fig.\ref{Fig:MRP_10_spheres}).
	\begin{enumerate}
	\item Use the Mori-Tanaka formulation for the composite:
	\begin{itemize}
	 \item inclusion: stiffness tensor $\mathbb{L}_1^{\alpha}$, volume fraction $f_1^{\alpha}/\left( f_1^{\alpha}+f_2^{\alpha}\right)$,
	 \item matrix: stiffness tensor $\mathbb{L}_2^{\alpha}$, volume fraction $f_2^{\alpha}/\left( f_1^{\alpha}+f_2^{\alpha}\right)$,
	\end{itemize}
	to calculate the concentration tensors: $\mathbb{A}_1^{\alpha}$, $\mathbb{A}_2^{\alpha}$, and the effective stiffness tensor $\mathbb{L}_{1+2}^{\alpha}$.
	\item Use the Mori-Tanaka formulation for the composite:
	\begin{itemize}
	 \item inclusion: stiffness tensor $\mathbb{L}_{1+2}^{\alpha}$, volume fraction $\left(f_1^{\alpha}+f_2^{\alpha}\right)$,
	 \item matrix: stiffness tensor $\mathbb{L}_3^{\alpha}$, volume fraction $f_3^{\alpha}$,
	\end{itemize}
	to calculate the concentration tensors: $\mathbb{A}_{1+2}^{\alpha}$, $\mathbb{A}_3^{\alpha}$, and the effective stiffness tensor $\mathbb{L}^{\alpha}$.
	\end{enumerate}
\item Calculate the MT* pattern $\alpha=M$, in which $c^{M}=1-\sum_{\alpha=1}^{\textup{M-1}}c^{\alpha}$.
	\begin{enumerate}
	\item Use the Mori-Tanaka formulation for the composite:
	\begin{itemize}
	 \item inclusion: stiffness tensor $\mathbb{L}_1^{M}$, volume fraction $f_1^{M}=\left(1-f_2^{M} c^{M}\right)/c^{M}$
	 \item matrix: stiffness tensor
	 $\mathbb{L}_2^{M}=\sum_{\alpha=1}^{\textup{M-1}}\left(\mathbb{L}^{\alpha} c^{\alpha}\right)/\left(1-c^{M} \right)$,\\
	 volume fraction	 $f_2^{M}=\sum_{\alpha=1}^{\textup{M-1}} \sum_{k=1}^{3} \left( f_{k}^{\alpha} c^{\alpha} \right)/\left( c^{M}  \right)$,
	\end{itemize}
	to calculate the concentration tensors: $\mathbb{A}_1^{M}$ and $\mathbb{A}_2^{M}$.
\end{enumerate}
\item Update the concentration tensors for $\alpha = 1,...,M-1 $:
	 $\mathbb{A}_k^{\alpha}=\mathbb{A}_k^{\alpha} \mathbb{A}_{1+2}^{\alpha} \mathbb{A}_2^{M}$ for $k=1,2$, and
	 $\mathbb{A}_3^{\alpha}=\mathbb{A}_3^{\alpha} \mathbb{A}_2^{M}$.
\item Calculate the MRP concentration tensors:
	 \begin{enumerate}
	 \item $\bar{\mathbb{A}}_k^{\alpha}=\mathbb{A}_k^{\alpha}\left[\sum_{\beta=1}^{\textup{M-1}}\sum_{l=1}^{3}f_l^{\beta}c^{\beta}\mathbb{A}_l^{\beta}+f_1^{M}c^{M}\mathbb{A}_1^{M}\right]^{-1}$ for $\alpha = 1,...,M-1 $ and $k=1,2,3$,
	 \item $\bar{\mathbb{A}}_1^{M}=\mathbb{A}_1^{M} \left[ \sum_{\beta=1}^{\textup{M-1}}\sum_{l=1}^{3}f_l^{\beta} c^{\beta}\mathbb{A}_l^{\beta}+f_1^{M}c^{M}\mathbb{A}_1^{M}\right]^{-1}$.
	 \end{enumerate}
	 \item The macroscopic stiffness tensor:\\
	 $\bar{\mathbb{L}}=\sum_{\alpha=1}^{\textup{M-1}}\sum_{k=1}^{3}f_k^{\alpha}c^{\alpha}\mathbb{L}_k^{\alpha}\bar{\mathbb{A}}_k^{\alpha}+f_1^{M}c^{M}\mathbb{L}_1^{M}\bar{\mathbb{A}}_1^{M}$
\end{enumerate}

\section{Concentration tensors\label{Ap:AppendixProjektory}}
The numerical strain concentration tensor $\mathbb{A}_{\textup{i}}^{\textup{NDil}}$ of a spherical inhomogeneity is isotropic,
for the crossed spheroids (Fig.\ref{Fig:Shapes}.d), $\mathbb{A}_{\textup{i}}^{\textup{NDil}}$ has cubic symmetry Eq.\ref{Eq:ACube}, and if the inhomogeneities are the prolate or drilled oblate spheroids (Fig.\ref{Fig:Shapes}.b and c), the symmetry group is that of transverse isotropy Eq.\ref{Eq:ATrans}

The projector tensors in spectral decomposition of $\mathbb{A}^{\textup{Cub}}$ (Eq.\ref{Eq:ACube}) are
\begin{equation}\label{Eq:PCube}
\mathbb{K}=\sum_{k=1}^3\mathbf{m}_k\otimes\mathbf{m}_k\otimes\mathbf{m}_k\otimes\mathbf{m}_k\,,\quad
\mathbb{I}^{\textup{P}}=\frac{1}{3}\mathbf{I}\otimes\mathbf{I}\,,
\end{equation}
where $\mathbf{m}_k$ are the main symmetry axes of the unit cell.

The orthogonal projectors $\mathbb{P}_K$ ($K=1,2,3$) of $\mathbb{A}^{\textup{Trans}}$ (Eq.\ref{Eq:ATrans}) are
\begin{equation}\label{Eq:PTransIso}
\begin{gathered}
\mathbb{P}_{1}= \mathbf{d}\otimes\mathbf{d}\,,\quad
\mathbf{d}=\frac{1}{\sqrt{6}}\left(3\mathbf{m}_1\otimes\mathbf{m}_1-\mathbf{I}\right)\,,
\end{gathered}
\end{equation}
\begin{equation}\label{Eq:PTransSShear}
\begin{gathered}
\mathbb{P}_{2}=
\frac{1}{2}
\left[
\left(\mathbf{m}_2\otimes\mathbf{m}_3+\mathbf{m}_3\otimes\mathbf{m}_2\right)
\otimes
\left(\mathbf{m}_2\otimes\mathbf{m}_3+\mathbf{m}_3\otimes\mathbf{m}_2\right)
+
\left(\mathbf{m}_2\otimes\mathbf{m}_2-\mathbf{m}_3\otimes\mathbf{m}_3\right)
\otimes
\left(\mathbf{m}_3\otimes\mathbf{m}_3-\mathbf{m}_2\otimes\mathbf{m}_2\right)\right]\,,
\end{gathered}
\end{equation}
\begin{equation}\label{Eq:PTransSShear2}
\begin{gathered}
\mathbb{P}_{3}=
\frac{1}{2}
\sum_{k=2,3}
\left(\mathbf{m}_1\otimes\mathbf{m}_k+\mathbf{m}_k\otimes\mathbf{m}_1\right)
\otimes
\left(\mathbf{m}_1\otimes\mathbf{m}_k+\mathbf{m}_k\otimes\mathbf{m}_1\right)\,,
\end{gathered}
\end{equation}
where $\mathbf{m}_k$ are the main symmetry axes of the unit cell.

\section{Periodic boundary conditions\label{Ap:AppendixPeriodic}}
In the numerical calculations we imposed overall strain $\mathbf{E}$ of the unit cell by sets of micro-periodic displacement boundary conditions on pairs of corresponding points A-B on the opposite faces of the unit cell, as follows
\begin{equation}\label{Eq:periodicBC}
\mathbf{u}_A-\mathbf{u}_B=\mathbf{E}\cdot(\mathbf{x}_A-\mathbf{x}_B)\,,
\end{equation}
where $\mathbf{u}_A$, $\mathbf{u}_B$, $\mathbf{x}_A$ and $\mathbf{x}_B$ are the initial positions and displacements of points $A$ and $B$, respectively.
$\mathbf{E}$ is equivalent to the local strain averaged over the SVE's volume $\mathbf{E}=1/V\int_V\boldsymbol{\varepsilon}\,dV$.
In presented studies unit dimensions of the unit cell were assumed.
To achieve boundary conditions (\ref{Eq:periodicBC}) in FE analysis, a particular multi-point constraint approach was employed.
Within this approach, displacements of a pair of nodes A and B are connected to the displacements of two nodes at the selected unit cell corners (Fig.\ref{Fig:UVRMTM}) as follows
\begin{equation}
\mathbf{u}_A-\mathbf{u}_B=\mathbf{u}_O-\mathbf{u}_{Xk}\,,\quad k=1,2,3\,.
\end{equation}
\section{Comparison between MRP and n-GSC models\label{Ap:nGSC}}
\begin{figure}[H]
\centering
\begin{tabular}{ccc}\\
(a) $f_{\lambda}/f_{\textup{m}}=0$&(b) $f_{\lambda}/f_{\textup{m}}=1/2$&(c) $f_{\lambda}/f_{\textup{m}}=1$\\
\includegraphics[angle=0,width=0.3 \textwidth]{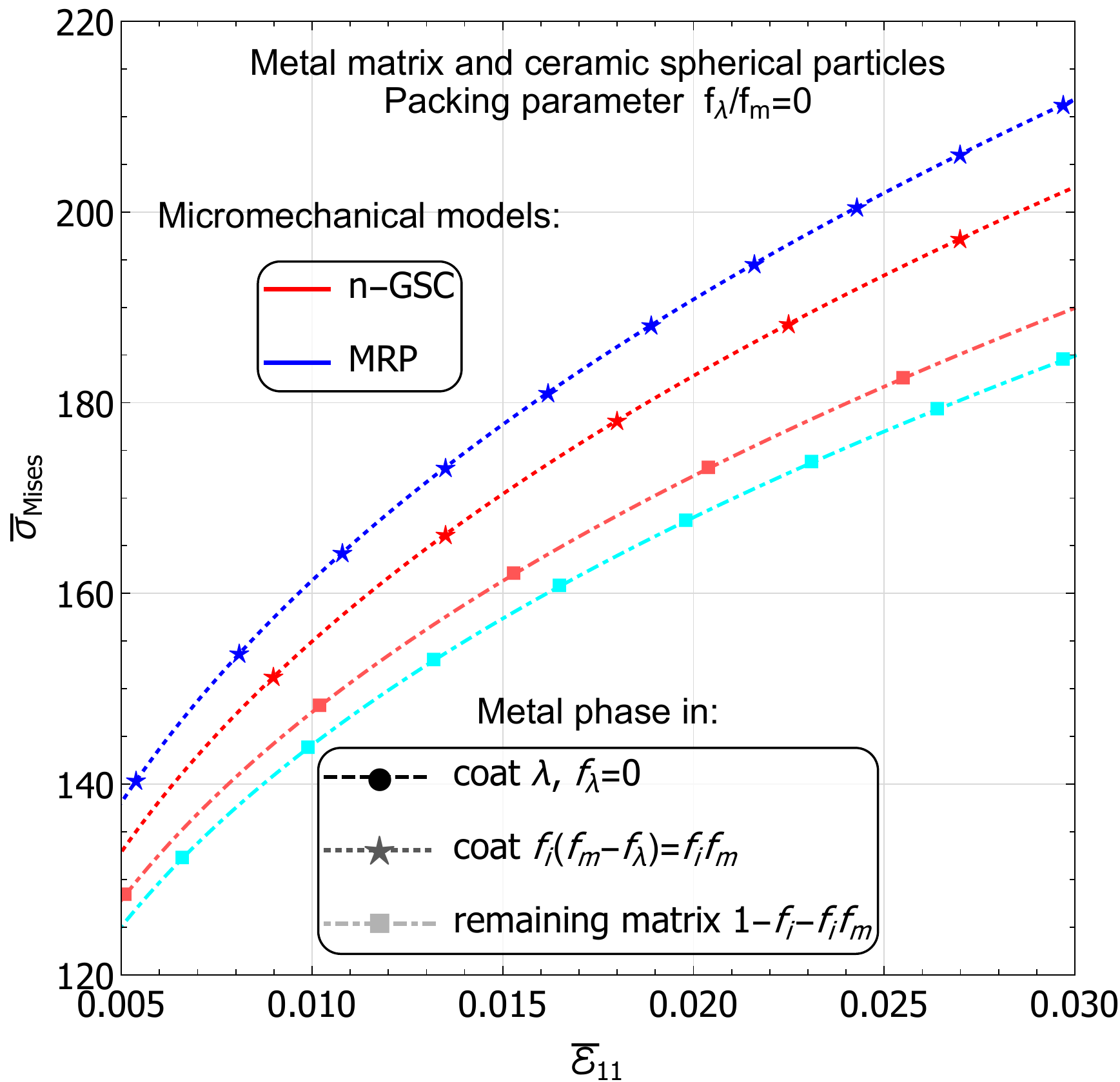}&
\includegraphics[angle=0,width=0.3 \textwidth]{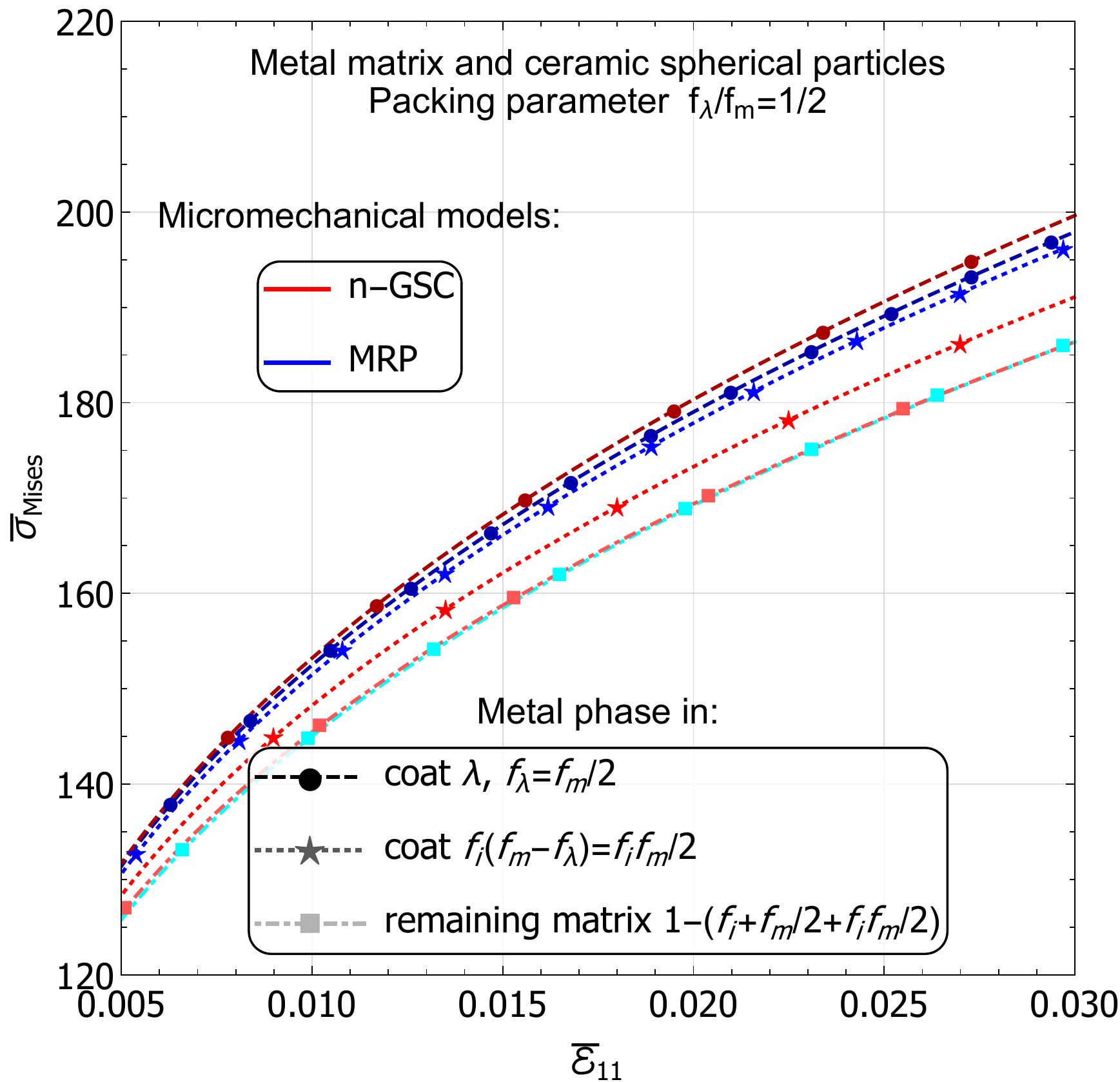}&
\includegraphics[angle=0,width=0.3 \textwidth]{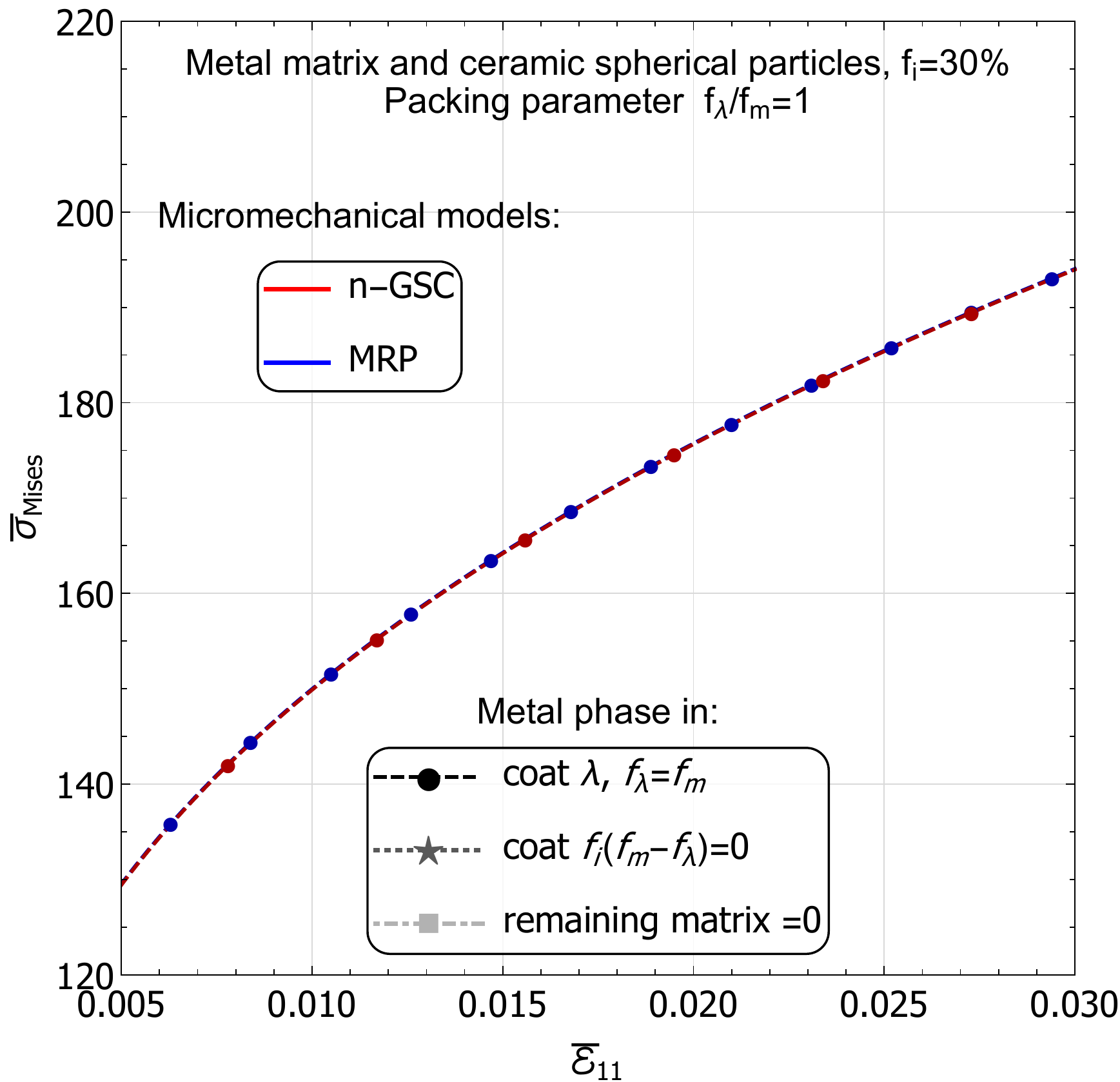}\\
\includegraphics[angle=0,width=0.3 \textwidth]{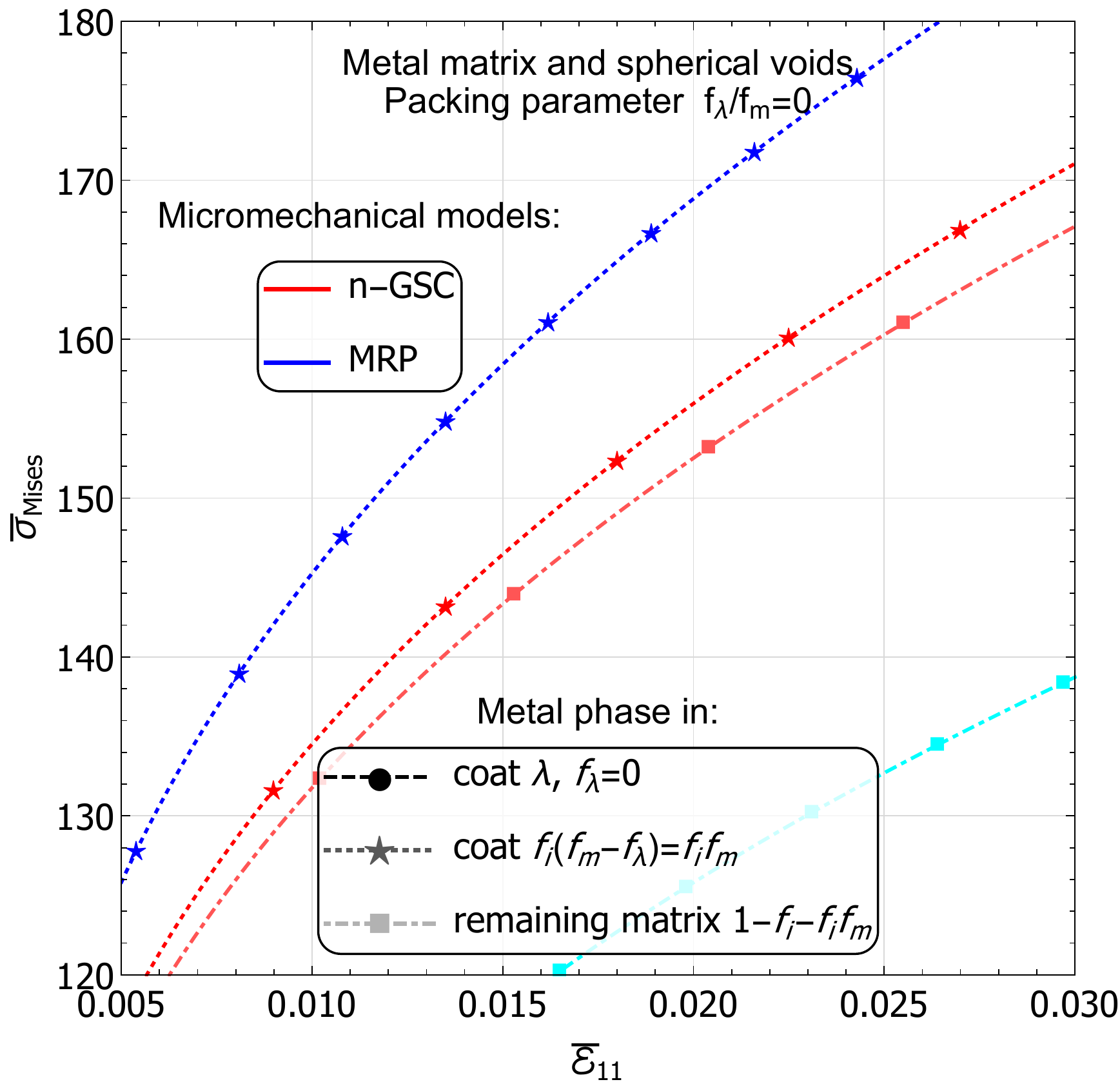}&
\includegraphics[angle=0,width=0.3 \textwidth]{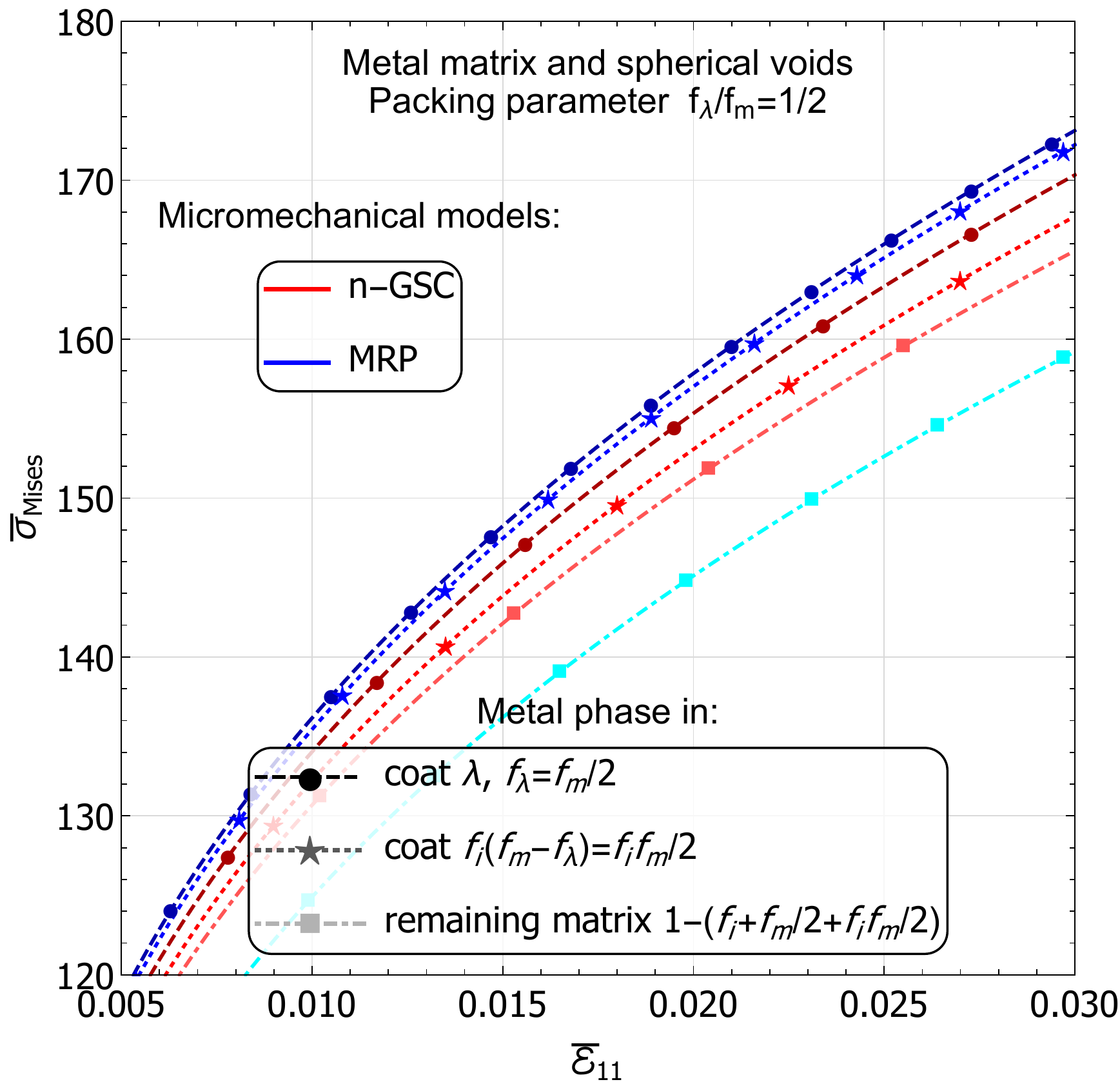}&
\includegraphics[angle=0,width=0.3 \textwidth]{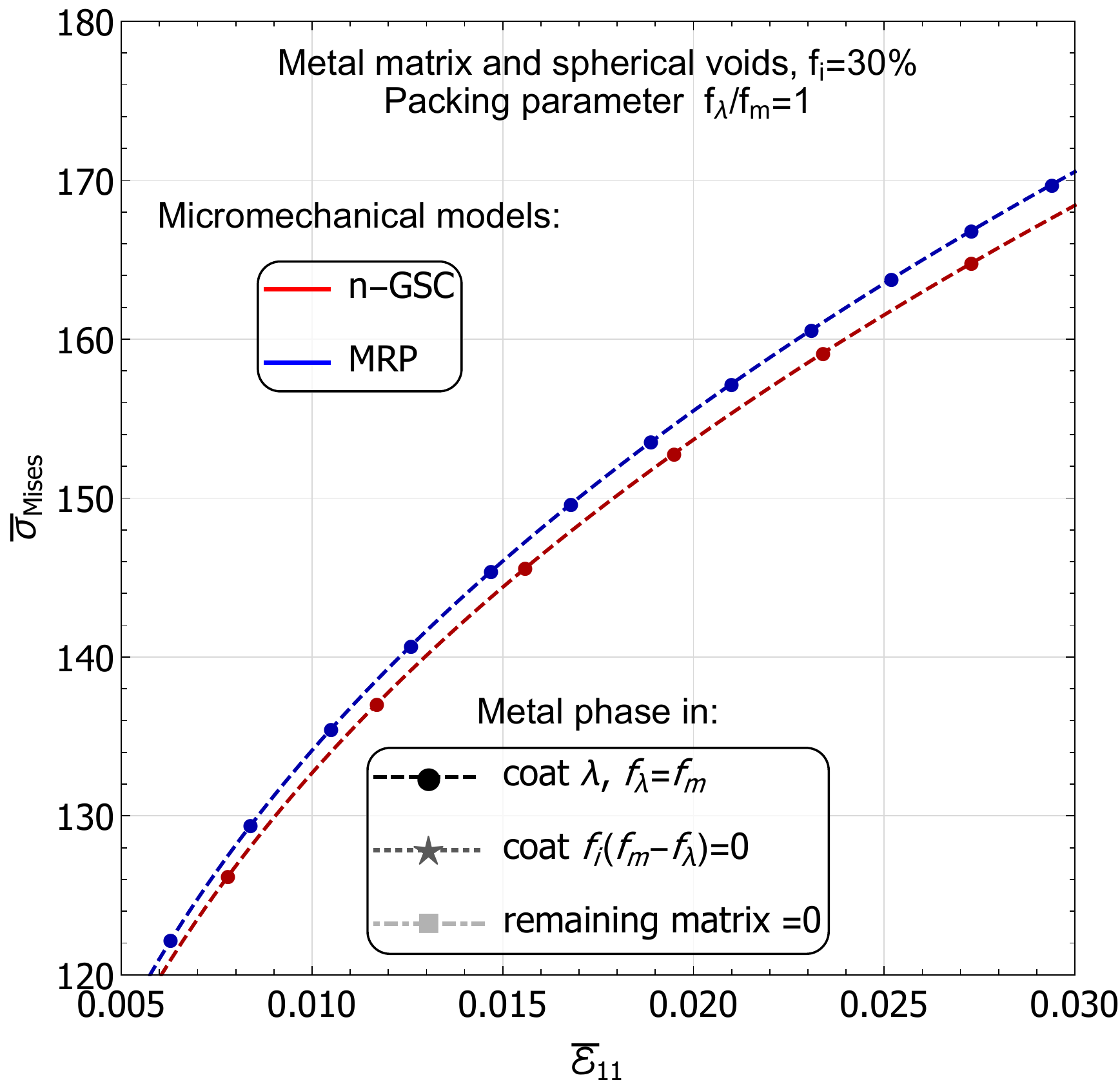}\\
\end{tabular}
\caption{
Comparison between the MRP model and the n-GSC scheme estimations of the elastic-plastic material composite with $f_{\textup{i}}=0.30$.
Huber-von Mises equivalent stress $\overline{\sigma}_{\textup{Mises}}$ vs. effective strain component $\overline{\varepsilon}_{11}$ in the direction of elongation of the isochoric tension test.
The first row, metal matrix reinforced by the ceramic spherical inclusions.
The second row, metal matrix with spherical voids.
The matrix packing parameter: (a) $f_{\lambda}/f_{\textup{m}}=0$, (b) $f_{\lambda}/f_{\textup{m}}=1/2$, and (c) $f_{\lambda}/f_{\textup{m}}=1$.
A tangent linearization scheme is used.
Only the metal phase response is presented in the plot.
\label{Fig:nGSC}}
\end{figure}

Material composites with spherical inclusions can be effectively estimated by the $n$-phase Generalized Self-Consistent (n-GSC) scheme \citep{Herve1993},
with the exact solution of the n-layer sphere.
In Fig.\ref{Fig:nGSC}, we compare the MRP model and the n-GSC scheme estimations of the elastic-plastic composite response.
Two variants were studied: the first row in Fig.\ref{Fig:nGSC} is a metal matrix reinforced by the ceramic spherical inclusions (Tab.\ref{Tab:MMC}),
the second row is a metal matrix with spherical voids.
Three matrix packing parameters were investigated: (a) $f_{\lambda}/f_{\textup{m}}=0$, (b) $f_{\lambda}/f_{\textup{m}}=1/2$, and (c) $f_{\lambda}/f_{\textup{m}}=1$.
The volume fraction of spherical inclusions (ceramic or voids balls) is $f_{\textup{i}}=0.30$.
In Fig.\ref{Fig:nGSC}, the Huber-von Mises equivalent stress $\overline{\sigma}_{\textup{Mises}}$ is plotted as a function of the effective strain component $\overline{\varepsilon}_{11}$ in the direction of elongation in the isochoric tension test.
A tangent linearization scheme is used for the non-linear material response.
The effective response $\overline{\sigma}_{\textup{Mises}}$ of the material composite is similar for both approaches.
Because the difference between n-GSC and MRP is visible for the metal phase in coats, in Fig.\ref{Fig:nGSC}, only the metal phase response is presented.
The plot range is limited for $\overline{\sigma}_{\textup{Mises}}>120$ and $\overline{\varepsilon}_{11}>0.005$ to highlight the outcomes contrast.
As one can notice, for $f_{\lambda}/f_{\textup{m}}=1$ (Fig.\ref{Fig:nGSC}.c) models MRP and n-GSC predict a similar response of the metal matrix.
For $f_{\lambda}/f_{\textup{m}}=1$, the MRP is equivalent to the MT solution, which gives similar results to the 2-phase GSC model.
When the matrix packing ratio is zero (Fig.\ref{Fig:nGSC}.a), $f_{\lambda}/f_{\textup{m}}=0$, 
the MRP estimations of the metal phases in MMC (Fig.\ref{Fig:nGSC}.a the first row) are near n-GSC results.
In the case of voids (Fig.\ref{Fig:nGSC}.a the second row) gap between the models is noticeable.
The difference in $\overline{\sigma}_{\textup{Mises}}$ between the matrix phases in RMTM and MT* patterns of the MRP is visible for the mentioned two cases (Fig.\ref{Fig:nGSC}.a and c).
In the third case $f_{\lambda}/f_{\textup{m}}=1/2$ (Fig.\ref{Fig:nGSC}.b), which gives 4-GSC and two coats in the RMTM pattern of the MRP model: the first layer $f_{\lambda}=f_{\textup{m}}/2$ and the second layer $f_{\textup{i}}f_{\textup{m}}/2$.
The difference between the two layers of the RMTM pattern is smaller than the exact solution of the n-GSC scheme.
Still, the MRP distinguishes the response of the matrix inside composite inclusions (RMTM patterns) and the remaining matrix (MT* pattern) in opposition to the n-GSC scheme.
\bibliography{Library}
\end{document}